\documentclass{style}
\usepackage{graphicx}
\usepackage{color}
\newcommand{\be}{\begin{equation}}
\newcommand{\ee}{\end{equation}}
\newcommand{\bea}{\begin{eqnarray}}
\newcommand{\eea}{\end{eqnarray}}

\def\vek#1{\mbox{\boldmath $#1$}}
\begin{document}
\title{Alpha Particle Clusters and their Condensation in Nuclear Systems}
\author{
Peter Schuck \\
\it Institut de Physique Nucl\'eaire, 91406 Orsay Cedex, France, \\ 
and Universit\'e Paris-Sud, Orsay, F-91505, France \\
and Laboratoire de Physique et de Mod\'elisation des Milieux Condens\'es,\\ 
CNRS et Universit\'e Joseph Fourier, UMR5493, 25 Av. des Martytrs, BP 166,\\ 
F-38042 Grenoble Cedex 9, France\\
Yasuro Funaki \\
\it School of Physics and Nuclear Energy Engineering and IRCNPC,\\ 
 Beihang University, Beijing 100191, China\\ 
and Nishina Center for Accelerator-Based Science, RIKEN, Wako 351-0198, Japan \\
Hisashi Horiuchi \\
\it Research Center for Nuclear Physics (RCNP), Osaka University, Osaka 567-0047, Japan \\
Gerd R\"opke \\ 
\it Institut f\"ur Physik, Universit\"at Rostock, D-18051 Rostock, Germany \\
Akihiro Tohsaki \\
\it Research Center for Nuclear Physics (RCNP), Osaka University, Osaka 567-0047, Japan \\
Taiichi Yamada \\
\it  Laboratory of Physics, Kanto Gakuin University, Yokohama 236-8501, Japan 
}


\date{}

\maketitle

\begin{abstract}
In this article we review the present status of $\alpha$ clustering in nuclear systems. An important aspect is first of all condensation in nuclear matter. Like for pairing, quartetting in matter is at the root of similar phenomena in finite nuclei. Cluster approaches for finite nuclei are shortly recapitulated in historical order. The $\alpha$ container model as recently been proposed by Tohsaki-Horiuchi-Schuck-R\"opke (THSR) will be outlined and the ensuing condensate aspect of the Hoyle state at 7.65 MeV in $^{12}$C investigated in some detail. A special case will be made with respect to the very accurate reproduction of the inelastic form factor from the ground to Hoyle state with the THSR description. The extended volume will be deduced. New developments concerning excitations of the Hoyle state will be discussed. After 15 years since the proposal of the $\alpha$ condensation concept a critical assessment of this idea will be given. Alpha gas states in other nuclei like $^{16}$O and $^{13}$C will be considered. An important aspect are experimental evidences, present and future ones. The THSR wave function can also describe configurations of one $\alpha$ particle on top of a doubly magic core. The cases of $^{20}$Ne and $^{212}$Po will be investigated.
\end{abstract}


\section{Introduction}

\begin{figure}
\begin{center}
\includegraphics[scale=0.3]{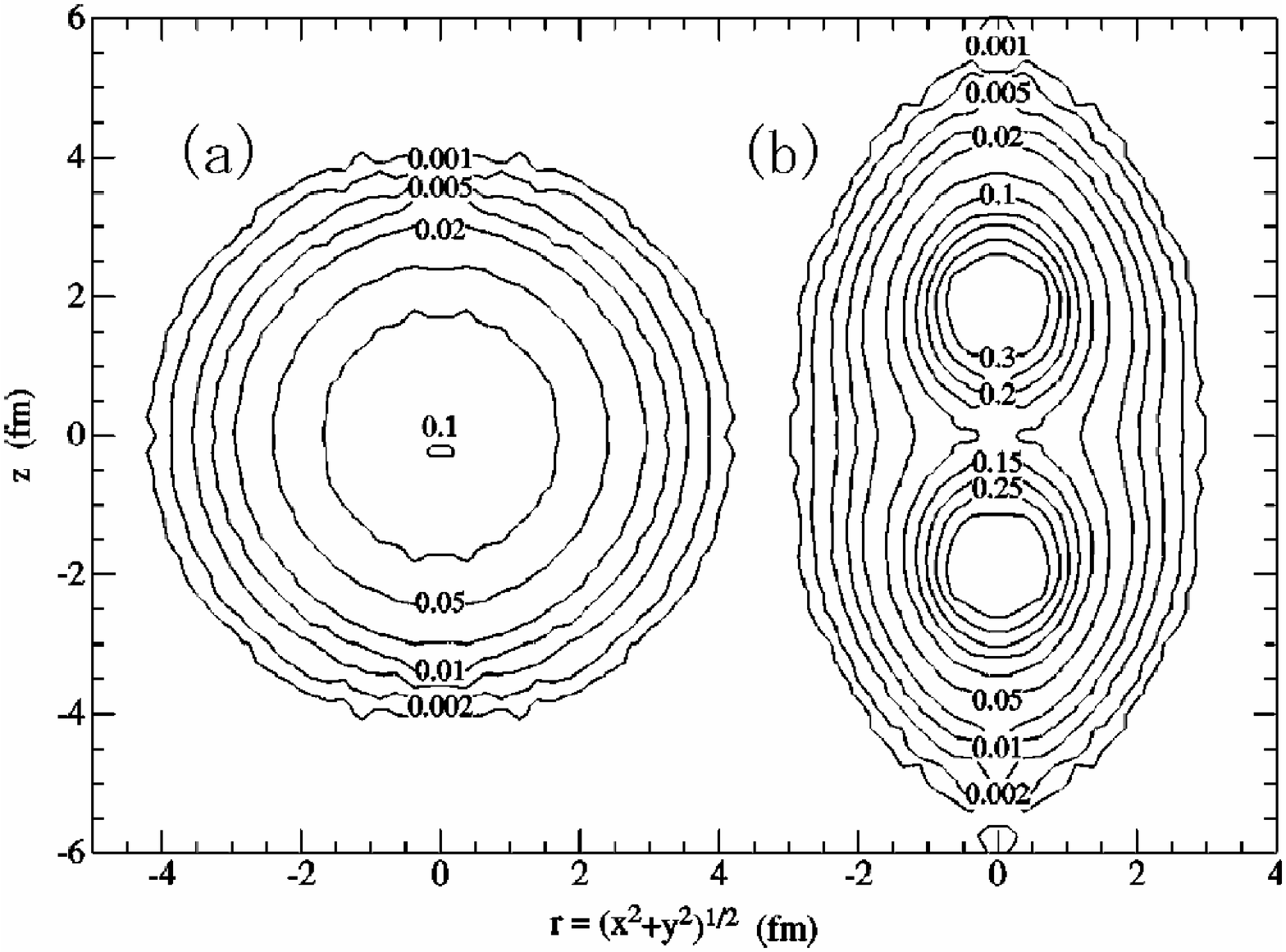}
\caption{\label{fig:gfmc_8be}
Green's function Monte Carlo results for $^8$Be. Left: laboratory frame; right: intrinsic frame. From \cite{Wiringa2000}}
\end{center}
\end{figure}

\begin{figure}
\begin{center}
\includegraphics[scale=0.5]{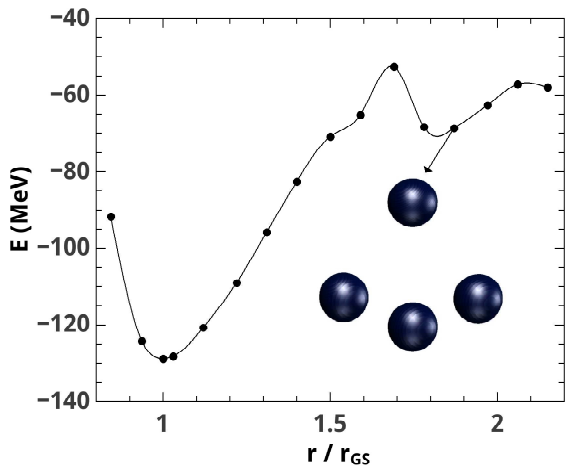}
\caption{\label{fig:16O_MF}
Mean-field energy of $^{16}$O as a function of its radius $r$. At a certain critical radius $^{16}$O clusters into four $\alpha$ particles (tetrahedron) \cite{Girod}. }
\end{center}
\end{figure}

Nuclei are very interesting objects from the many body point of view. They are self-bound droplets, i.e., clusters of fermions! As we know, this stems from the fact that in nuclear physics, there exist four different fermions: proton, spin up/down, neutron spin up/down. If there were only neutrons, no nuclei would exist. This is due to the Pauli exclusion principle. Take the case of the $\alpha$ particle described approximately by the  spherical harmonic oscillator as mean field potential: one can put two protons and two neutrons in the lowest (S) level, that is just the $\alpha$ particle. With four neutrons one would have to put two of them in the P-shell what is energetically very penalising, see however \cite{tetra}. Neutron matter is unbound whereas symmetric nuclear matter is bound. Of course, this is not only due to the Pauli principle. We know that the proton-neutron attraction is stronger than the neutron-neutron (or proton-proton) one. Proton and neutron form a bound state, the other two combinations not. The binding energy of the deuteron (1.1 MeV/nucleon) is to a large extent due to the tensor force. So is the  one of the $\alpha$ particle. The $\alpha$ particle is the lightest doubly magic nucleus with almost same binding per nucleon (7.07 MeV) as the strongest bound nucleus, i.e., Iron ($^{52}$Fe). The binding of the deuteron is about seven times weaker than the one of the $\alpha$ particle. The $\alpha$ is a very stiff particle. Its first excited state is at $\sim$ 20 MeV. This is factors higher than in any other nucleus. It helps to give to the $\alpha$ particle under some circumstances the property of an almost ideal boson. This happens, once the average density of the system is low as, e.g. in $^8$Be which has an average density at least four times smaller than the nuclear saturation density $\rho_0$. All nuclei, besides $^8$Be, have a ground state density around $\rho_0$ and can be described to lowest order as an ideal gas of fermions hold together by their proper mean field. $^8$Be is the only exception forming two loosely bound $\alpha$'s, see Fig.~\ref{fig:gfmc_8be}. For this singular situation exist general arguments but no detailed physical and numerical explanation (as far as we know). We will come to the discussion of $^8$Be later. However, radially expanding a heavier nucleus consisting of n$\alpha$ particles gives raise to a strong loss of its binding. At a critical expansion, i.e. low density, it is energetically more favorable that the nucleus breaks up into n$\alpha$ particles because each $\alpha$ particle can have (at its center) saturation. Of course, the sum of surface energies of all $\alpha$ particles is penalising but less than the loss of binding due to expansion. For illustration, we show in Fig.~\ref{fig:16O_MF} a pure mean field calculation of $^{16}$O which has broken up into a tetrahedron of four $\alpha$ particles at low density ~\cite{Girod}. Of course in this case the $\alpha$'s are fixed to definite spatial points and, thus, they form a crystal. In reality, however, the $\alpha$'s can move around lowering in this way the energy of the system greatly. We will come back to this in the main part of the review. Such $\alpha$ clustering scenarios are observed when two heavier nuclei collide head on at c.o.m. energies/particle around the Fermi energy. The nuclei first fuse and compress. Then decompress and at sufficiently low density the system breaks up into clusters. A great number of $\alpha$ particles is detected for central collisions, see \cite{Borderie}\cite{Raduta} and references in there. However, also in finite nuclei such low density n$\alpha$ systems can exist as resonances close to the n$\alpha$ disintegration threshold.

A very fameous example is the second $0^+$ state at 7.65 MeV in $^{12}$C, the so-called Hoyle state. Its existence was predicted in 1954 by the astrophysicist Fred Hoyle \cite{FHoyle} and later found practically at the predicted energy by Fowler {\it et al.} \cite{Fowler}. This state is supposed to be a loosely bound agglomerate of three $\alpha$ particles situated about 300 keV above the 3$\alpha$ disintegration threshold. As for the case of $^8$Be, this state is hold together by the Coulomb barrier. It is one of the most important states in nuclear physics because it is the gateway for Carbon production in the universe through the so-called triple $\alpha$ reaction ~\cite{Nomoto,NACRE,Ish13,OKK,Nguyen,Garrido,funaki-triple-a} and is, thus, responsible for life on earth. A great part of this article will deal with the description of the properties of this state. However, it is now believed that there exist heavier nuclei which show similar $\alpha$ gas states around the $\alpha$ disintegration threshold, for instance 
$^{16}$O around  14.4 MeV \cite{Fu08}. Alpha particles are bosons. If they are weakly interacting like e.g., in the Hoyle or other states, they may essentially be condensed in the 0S orbit of their own cluster mean field. We will dwell extensively on this 'condensation' aspect in the main part of the text.\\

Clustering and in particular $\alpha$ particle clustering has already a long history. The alpha particle model was first introduced by Gamow ~\cite{Gamow1, Gamow2}. Before the discovery of the neutron,  nuclei were assumed to be composed of $\alpha$ particles, protons and electrons. In 1937 Wefelmeier ~\cite{Wefel} proposed his well known model where  the n$\alpha$ particles are arranged in crystalline order in $Z=N$ nuclei. In the work of Hafstad and Teller in 1938 ~\cite{Hafstad}, the $\alpha$'s in a selfconjugate nucleus are arranged in close packed form interacting with nearest neighbors. The energy levels of $^{16}$O were discussed by Dennison ~\cite{Dennison} with a regular tetrahedron arrangement of the four $\alpha$'s. Other forerunners of $\alpha$ cluster physics with this kind of models were Kameny ~\cite{Kameny} and Glassgold and Galonsky ~\cite{Galonsky}. The latter discussed energy levels of $^{12}$C calculating the rotations and vibrations of an equilateral triangle arrangement of the three $\alpha$'s, see also a recent application of this idea in ~\cite{Gai} discussed below. Several works also tried to solve, e.g., the 3$\alpha$ system in considering the 3-body Schr\"odinger equation with an effective $\alpha$-$\alpha$ potential reproducing the $\alpha$-$\alpha$ phase shifts, see two recent publications ~\cite{Rimas, Ishikawa}. In the main part of the article we will discuss recent works of this type. In 1956 Morinaga came up with the idea that the Hoyle state could be a linear chain state of three $\alpha$ particles ~\cite{Morinaga}. This at that time somewhat spectacular idea found some echo in the community.  But in the 1970-ties first with the so-called Orthogonality Condition Method (OCM) Horiuchi ~\cite{OCM-Hori} and shortly later Kamimura {\it et al.}~\cite{Kami} and independently Uegaki {\it et al.} ~\cite{Uegaki} showed that the Hoyle state is in fact a weakly coupled system of three $\alpha$'s or in other words a gas like state of $\alpha$ particles in relative S-states. The emerging picture then was that the Hoyle state is of low density where a third $\alpha$ particle is orbiting in an S-wave around a $^8$Be-like object also being in an S-wave. Actually both groups in \cite{Kami, Uegaki} started with a fully microscopic 12 nucleon wave function where the c.o.m. part involving the c.o.m. Jacobi coordinates of the $\alpha$ particles was to be  determined by a variational 
Resonating Group Method (RGM) calculation in the first case and by a Generator Coordinate Method  (GCM) one in the second case ~\cite{Wh37, Hill}. Slight variants of the Volkov force ~\cite{Volkov} were used.

All known properties of the Hoyle state were reproduced in both cases with this parameter free calculations. Besides the Hoyle state several other states were predicted and agreement with experiment found. The second $2^+$ state was only confirmed very recently ~\cite{It11,Gai2+}. The achievements of these works were so outstanding and ahead of their time that--one is tempted to say--'as often after such an exploit' the physics of the Hoyle state stayed essentially dormant for roughly a quarter century. It was only in 2001 where a new aspect of the Hoyle state came on the forefront of discussion. Tohsaki, Horiuchi, Schuck, and R\"opke (THSR) proposed that the Hoyle state and other n$\alpha$ nuclei as, for instance, $^{16}$O with excitation energies roughly around the alpha disintegration threshold form actually an $\alpha$ particle condensate. They proposed a wave function of the (particle number projected) BCS type, however, the pair wave function replaced by a quartet one formed by a wide Gaussian for the c.o.m. motion and an intrinsic translationally invariant $\alpha$ particle wave function with a free space extension. The variational solution  with respect to the single size parameter of the c.o.m. Gaussian gave an almost 100 $\%$ squared overlap with Kamimura's wave function~\cite{Fu03}, thus, proving that implicitly the latter one has the more simplified (analytic) structure of the THSR wave function. Additionally it was later shown that THSR predicts a 70$\%$ occupancy of the three alpha's of the Hoyle state being in identical 0S orbit. This was rightly qualified as an $\alpha$ particle condensate. This interpretation of the Hoyle state and the prediction that in heavier n$\alpha$ nuclei similar $\alpha$ condensates may exist, triggered an immense new interest in the Hoyle state and $\alpha$ cluster states around it. Many experimental and theoretical articles have appeared since then including, e.g., 5 review articles on the subject ~\cite{rev1,rev2,rev3,rev4,rev5}. And the intensity of this type of studies does not seem to slow down.\\
For example very recently ab initio studies for cluster states appeared both using the nuclear effective-field-theory (EFT) approaches \cite{Meissner, Elhatisari1, Elhatisari2, Ulf2} and   the symplectic no-core shell model (NCSM) \cite{Tobin, Dreyfuss, Dytrych}  as we will discuss later in the main text.

Our article is organised as follows. In Sect.2 we show how in infinite nuclear matter, below a certain low critical density, $\alpha$ particle condensation appears. In Sects. 3-7, we recapitulate in condensed form the most important theoretical methods to treat $\alpha$ clustering in finite nuclei, that is the RGM, the OCM, the Brink and Generator Coordinate Method (Brink-GCM), the Antisymmetrised Molecular and Fermion Molecular Dynamics (AMD, FMD), and finally the wave function proposed by Tohsaki, Horiuchi, Schuck, R\"opke (THSR). In Sect. 8 and 9, the Hoyle state in $^{12}$C and its $\alpha$ condensate structure is discussed employing the THSR wave function. In Sect. 10 the spacial extension of the Hoyle state is investigated and in Sect.11 excited Hoyle states are studied. In Sect. 12, we will present the recent attempts to describe clustering from ab initio approaches. In Sect.13 we present an OCM study of the $0^+$ spectrum of $^{16}$O with the finding that only the 6-th $0^+$ state at 15.1 MeV can be interpreted as an $\alpha$ cluster condensate state. In Sect.14, a critical round up of the hypothesis of the Hoyle state being an $\alpha$ particle condensate is presented and the question asked: where do we stand after 15 years? In Sect.15 we show that also in the ground states of the lighter self conjugate nuclei non-negligeable correlations of the $\alpha$ type exist which can act as seeds to break those nuclei into $\alpha$ gas states when excited. In Sect.16 we treat the case what happens to the cluster states when an additional neutron is added to $^{12}$C. In Sect.17 we discuss the experimental situation concerning $\alpha$ condensation. In the next section 18 we come back to cases where the $\alpha$ is strongly present, even in the ground state. Such is the case for $^{20}$Ne where two doubly magic nuclei ($^{16}$O and $\alpha$) try to merge. In Sect.19, we point more in detail to the fact that the successful THSR description of cluster states sheds a new light on cluster dynamics being essentially non-localised in opposition to the old dumbell picture. Then in Sect.20, we come to another case of two merging doubly magic nuclei: $^{208}$Pb $+\alpha$ = $^{212}$Po. Finally, in Sect.21, we give an outlook and conclude.\\

\section{ Alpha particle Condensation in Infinite Matter}

The possibility of quartet, i.e., $\alpha$ particle condensation in
nuclear systems has only come to the forefront in recent years. First,
this may be due to the fact that quartet condensation, i.e.,
condensation of four tightly correlated fermions, is a technically much 
more difficult problem than is pairing. Second, as we will see,
the BEC-BCS transition for quartets is very different from the pair
case. As a matter of fact the analog to the weak coupling BCS like, long coherence
length regime does not exist for quartets. Rather, at higher densities
the quartets dissolve and go over into two Cooper pairs or a correlated four particle state.

Quartets are, of course, present in nuclear systems. In other fields
of physics they are much rarer. One knows that two excitons (bound states between an electron and a hole) in
semiconducters can form a bound state and the question has been asked
in the past whether bi-excitons can condense
~\cite{Noz}. In future cold atom devices, one may trap
four different species of fermions which, with the help of Feshbach
resonances, could form quartets (please remember that four different
fermions are quite necessary to form quartets for Pauli principle and,
thus, energetic reasons). Theoretical models of condensed matter have already been treated
and a quartet phase predicted ~\cite{Lechem}, see also ~\cite{Miyake}.

\begin{figure}
\begin{center}
\includegraphics[scale=0.7]{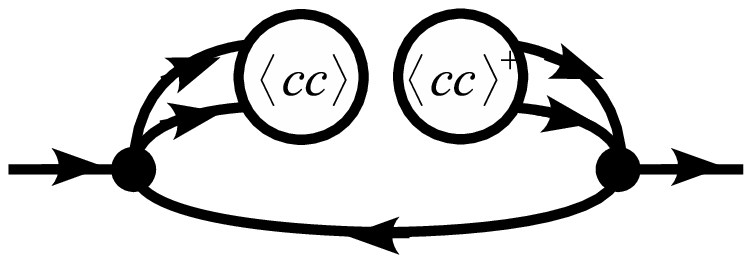}\hspace{2cm}
\includegraphics[scale=0.7]{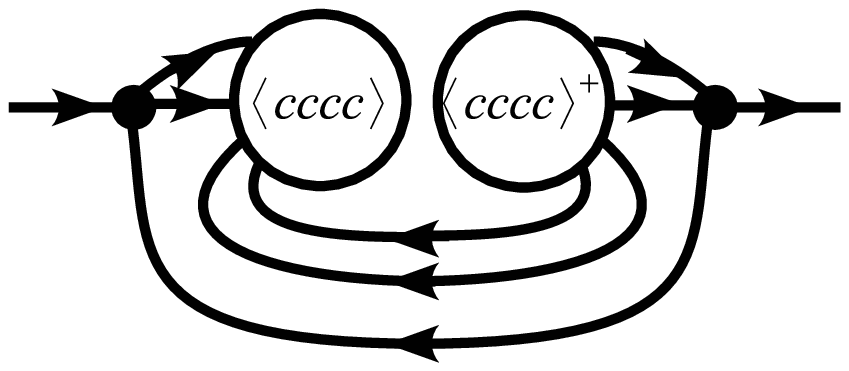}
\end{center}
\caption{\label{fig:Sogomassoperator} Single particle mass operator in
  case of pairing (upper panel) and quartetting (lower panel).}
\end{figure}

Let us start the theoretical description. For this it is convenient to
shortly repeat what is done in standard S-wave pairing. On the one hand,
we have the equation for the order parameter $\kappa(\vek{p}_1,
\vek{p}_2) = \langle c_{\vek{p}_1} c_{\vek{p}_2}\rangle$ where the brackets stand for expectation value in the BCS state and $c^+, c$ are fermion creation and destruction operators (we suppose S-wave pairing and suppress the spin dependence)
\begin{equation}
\kappa(\vek{p}_1, \vek{p}_2) = 
  \frac{1 - n(\vek{p}_1) - n(\vek{p}_2)}{e_{\vek{p}_1} + e_{\vek{p}_2} - 2 \mu } 
\sum_{\vek{p}'_1,\vek{p}_2'}
    \langle \vek{p}_1\vek{p}_2|v|\vek{p}_1'\vek{p}_2'\rangle 
    \kappa(\vek{p}_1', \vek{p}_2')
\end{equation}
with $e_k$ kinetic energy, eventually with a Hartree-Fock (HF) shift, and $\langle
\vek{p}_1 \vek{p}_2|v|\vek{p}_1'\vek{p}_2'\rangle =
\delta(\vek{K}-\vek{K}') v(\vek{q}, \vek{q}', \vek{K})$ the matrix element of
the force with $\vek{K},\vek{q}$ c.o.m. and relative momenta. One
recognises the in medium two-particle Bethe-Salpeter equation at $T=0$, taken at
the eigenvalue $E= 2\mu$ where $\mu$ is the chemical
potential. Inserting the standard BCS expression for the occupation
numbers
\begin{equation}
n(p) = \frac{1}{2}\bigg(1 - \frac{e_p - \mu}{2\sqrt{(e_p - \mu)^2 +
    \Delta^2}}\bigg)
\end{equation}
leads for pairs at rest, i.e., $\vek{K}=\vek{p}_1 + \vek{p}_2 =0$, to
the standard gap equation ~\cite{FW}. We want to proceed in an analogous
way with the quartets. In obvious short hand notation where we comprise in one number index momentum and spin, the in-medium
four fermion Bethe-Salpeter equation for the quartet order parameter
$K(1234) = \langle c_1c_2c_3c_4\rangle$ is given by \cite{Sogo1}
\begin{eqnarray}
&&(e_1 + e_2 + e_3 + e_4 - 4\mu)K(1234) = (1 -n_1-n_2) \nonumber \\
&& \times \sum_{1'2'}\langle 12|v|1'2'\rangle K(1'2'34)
   + \mbox{permutations}\,,
\label{quartetorderparametereqn}
\end{eqnarray}
We see that above equation is a rather straight forward extension of
the pairing case to the quartet one. The difficulty lies in the problem how
to find the single-particle (s.p.) occupation numbers $n_k$ in the quartet
case. Again, we will proceed in analogy to the pairing
case. Eliminating there the anomalous Green's function from the
$2\times 2$ set of Gorkov equations \cite{FW} leads to a
mass operator in the Dyson equation for the normal Green's function of
the form
\begin{equation}
M_{1,1'} = \frac{|\Delta_1|^2}{\omega + e_1 - 2\mu}\delta_{1,1'}\,.
\end{equation}
with the gap defined by
\begin{equation}
\Delta_1 = \sum_2 \langle 1\bar 1|v|2\bar 2\rangle \langle c_2c_{\bar 2}\rangle
\end{equation}
where '$\bar 1$' is the time reversed state of '$1$'.  Its graphical
representation is given in Fig.~\ref{fig:Sogomassoperator} (upper panel).
In the case of quartets, the derivation of a s.p. mass operator is
more tricky and we only want to give the final expression here (for
detailed derivation, see Appendix A and Ref.~\cite{Sogo1}):
\begin{equation}
M^{\rm{quartet}}_{1,1}(\omega ) = \sum_{234}\frac{\Delta_{1234}[\bar
    f_2\bar f_3\bar f_4 + f_2f_3f_4]\Delta^*_{1234}}{\omega +
  e_2+e_3+e_4 -4\mu}\,,
\label{Mquart}
\end{equation}
where $\bar f = 1-f$ and $f_i = \Theta(\mu - e_i)$ is the Fermi step
at zero temperature and the quartet gap matrix is given by
\begin{equation}
\Delta_{1234} = \sum_{1'2'}\langle 12|v|1'2'\rangle \langle c_{1'}c_{2'}c_3c_4\rangle
\label{eq:quartetmassoperator}
\end{equation}
This quartet mass operator is also depicted in Fig.~\ref{fig:Sogomassoperator} (lower panel).

Though, as mentioned, the derivation is slightly intricate, the final
result looks plausible. For instance, the three backward going fermion 
lines seen in the
lower panel of Fig.~\ref{fig:Sogomassoperator} give rise to the Fermi
occupation factors in the numerator of
Eq.~(\ref{Mquart}). This makes, as we will see, a
strong difference with pairing, since there with only a single fermion
line $\bar f + f =1$ and, thus, no phase space factor appears. Once we
have the mass operator, the occupation numbers can be calculated via
the standard procedure and the system of equations for the quartet
order parameter is closed.

\begin{figure*}[htbp]
\begin{center}
\includegraphics[scale=0.85]{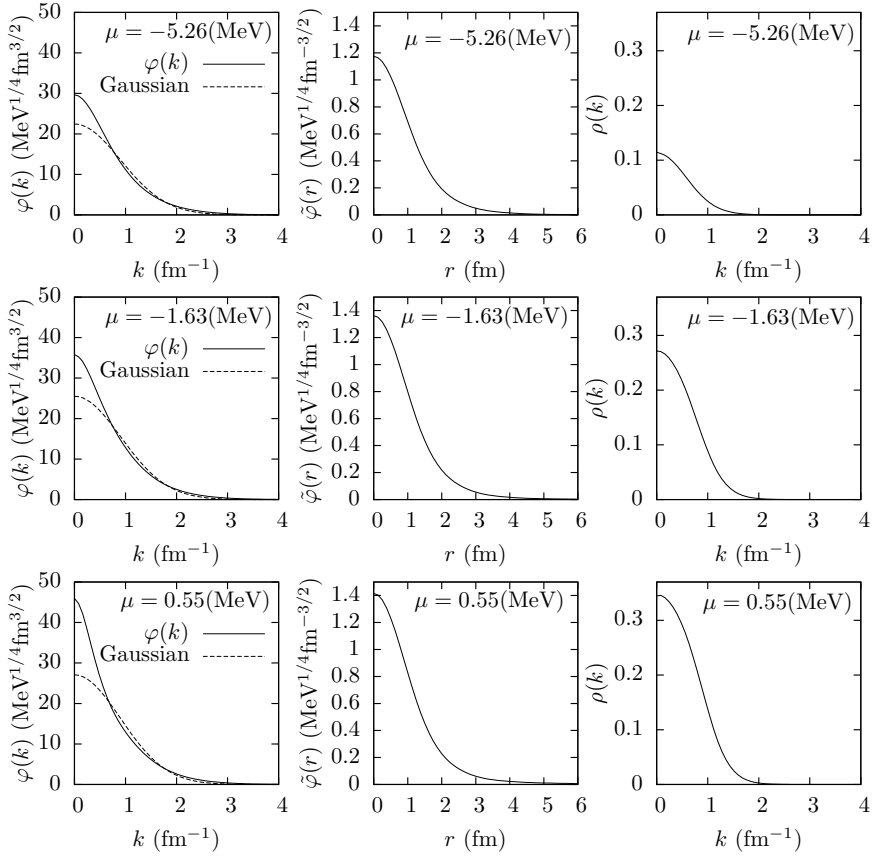}
\end{center}
\caption{\label{Sogofig6} Single particle wave functions $\varphi$ in
  momentum and position spaces (two left columns) and s.p. occupation
  numbers $\rho(k)$ (right column) \cite{Sogo1}.}
\end{figure*}

Numerically it is out of question that one solves this complicated
nonlinear set of four-body equations brute force. Luckily, there
exists a very efficient and simplifying approximation. It is known in
nuclear physics that, because of its strong binding, it is a good
approximation to treat the $\alpha$ particle in mean field as long as
it is projected on good total momentum. We therefore make the ansatz 
(see also ~\cite{Miyake})
\begin{equation}
\langle c_1c_2c_3c_4 \rangle \rightarrow 
  \varphi(\vek{k}_1)\varphi(\vek{k}_2)\varphi(\vek{k}_3)\varphi(\vek{k}_4) 
\delta(\vek{k}_1+\vek{k}_2+\vek{k}_3+\vek{k}_4)\,,
\label{eq:alphaprojectedmf}
\end{equation}
where $\varphi$ is a $0S$ single particle wave function in momentum
space. Again the scalar spin-isospin singlet part of the wave function has been suppressed. 
With this ansatz which is an eigenstate of the total momentum
operator with eigenvalue $\vek{K}=0$, the problem is still complicated
but reduces to the selfconsistent determination of $\varphi(\vek{k})$
what is a tremendous simplification and renders the problem
manageable. The reader should be aware of the fact that the approximation (\ref{eq:alphaprojectedmf}) is not a simple mean field ansatz. It is projected on good total momentum what induces strong correlations on top of the product of s.p. wave functions. Below, we will give an example where the high efficiency
of the product ansatz is demonstrated. Of course, with the mean field
ansatz we cannot use the bare nucleon-nucleon force. We took a
separable one with two parameters (strength and range) which were
adjusted to energy and radius of the free $\alpha$ particle. In
Fig.~\ref{Sogofig6},
we show the evolution with increasing chemical potential $\mu$
(density) of the single particle wave function in position and
momentum space (two left columns). We see that at higher $\mu$'s,
i.e., densities, the wave function deviates more and more from a
Gaussian. At slightly positive $\mu$ the system seems not to have a
solution anymore and selfconsistency cannot be achieved.

\begin{figure}[htbp]
\begin{center}
\includegraphics[scale=0.7]{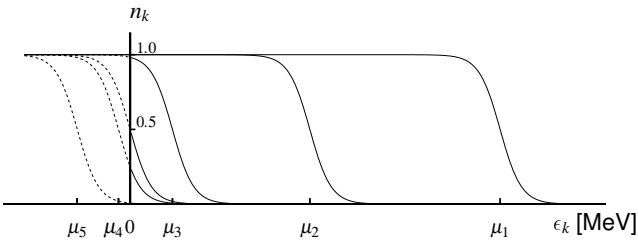}
\end{center}
\caption{Schematic (non-selfconsistent) view of BCS occupation numbers as the chemical potential varies from positive to negative (binding) values.}
\label{occs-BCS}
\end{figure}

Very interesting is the evolution of the occupation numbers $n_k 
(\equiv \rho(k))$
with $\mu$ (density) also shown in Fig.~\ref{Sogofig6} (right
column). It is seen that at slightly positive $\mu$ where the system
stops to find a solution, the occupation numbers are still far from
unity. The highest occupation number one obtains lies at around
$n_{k=0} \sim 0.35$. This is completely different from the BEC-BCS
cross-over in the case of pairing, where $\mu$ can vary from negative
to positive values and the occupation numbers saturate at unity when
$\mu$ goes well into the positive region, see Fig.\ref{occs-BCS}. 
We therefore see that, in the
case of quartetting, the system is still far from the regime of weak
coupling and large coherence length when it stops to have a
solution. One also sees from the extension of the wave functions that
the size of the $\alpha$ particles has barely increased. Before we
give an explanation for this behavior, let us study the critical
temperature where this breakdown of the solution is seen more clearly.

\begin{figure*}[htbp]
\begin{center}
\includegraphics[scale=1.]{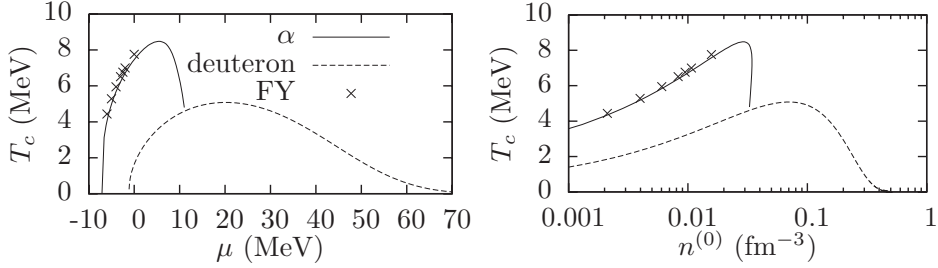}
\end{center}
\caption{\label{SogoTc} Critical temperatures for $\alpha$ particle
 (binding energy/nucleon $\sim$ 7.07 MeV) and deuteron (binding
  energy/nucleon $\sim$ 1.1 MeV) condensation as a function of $\mu$
  (left panel) and as a function of density (right panel)
  \cite{Sogo2}. The symbol FY stands for 'Faddeev-Yakubovsky' approach.}
\end{figure*}

In order to study the critical temperature for the onset of quartet
condensation, we have to linearise the equation for the order
parameter (\ref{quartetorderparametereqn}) in replacing the correlated
occupation numbers by the free Fermi-Dirac distributions at finite
temperature $n(p) \rightarrow
f(p)=[1+e^{(e_p-\mu)/T}]^{-1}$ with $e_p = p^2/(2m)$. Determining the temperature $T$ where
the equation is fullfilled gives the critical temperature
$T=T^{\alpha}_c$. This is the Thouless criterion for the critical
temperature of pairing \cite{Thouless} transposed to the quartet
case. In Fig.~\ref{SogoTc},
we show the evolution of $T^{\alpha}_c$ as a function of the chemical
potential (left panel) and of density (right panel)
\cite{Sogo2}, see also ~\cite{Sogo3} for the case of asymmetric matter. This figure shows very explicitly the excellent
performance of our momentum projected mean field ansatz for the
quartet order parameter. The crosses correspond to the full solution
of Eq.~(\ref{quartetorderparametereqn}) in the linearised finite
temperature regime with the rather realistic Malfliet-Tjohn bare
nucleon-nucleon potential \cite{Malfliet} whereas the
continuous line corresponds to the projected mean field solution. Both
results are litterally on top of one another (the full solution is
only available for negative chemical potentials). A non-projected mean field wave function would never give such a good agreement.  One clearly sees the
breakdown of quartetting at small positive $\mu$ (the fact that the
 critical $T_c$ breakdown occurs at a somewhat larger positive $\mu$ with respect to the full solution of the quartet gap equation with the ansatz (\ref{eq:alphaprojectedmf}) at $T$=0
  may be due to the fact that here we are at finite temperature, see also discussion below)
whereas n-p pairing (in the deuteron channel) continues
smoothly into the large $\mu$ region. It is worth mentioning that in
the isospin polarised case with more neutrons than protons, n-p
pairing is much more affected than quartetting (due to the much
stronger binding of the $\alpha$ particle) and finally loses against
$\alpha$ condensation \cite{Sogo3}. So, contrary to the pairing
case, where there is a smooth cross-over from BEC to BCS, in the case
of quartetting the transition to the dissolution of the $\alpha$
particles seems to occur quite abruptly and we have to seek for an
explanation of this somewhat surprising difference between pairing and
quartetting.

The explanation is in a sense rather trivial. It has to do with the
different level densities involved in the two systems. In the pairing
case, the s.p. mass operator only contains a single fermion (hole) line
propagator and the level density is given by

\begin{equation}
g_{1h}(\omega) = -\frac{1}{\pi}{\mbox{Im}}\, 
  \sum_p\frac{ \bar f(p) + f(p)}{\omega + e_p + i\eta}
  =\sum_p\delta(\omega+ e_p)
\end{equation}
\noindent

In the case of three fermions, as is the case of quartetting, we have for
the corresponding level density ( see also
\cite{Hiller})
\begin{eqnarray}
&& g_{3h}(\omega) = -\frac{1}{\pi}{\mbox{Im}}\, 
  {\mbox{Tr}}\frac{\bar f(p_1)\bar f(p_2)\bar f(p_3) + f(p_1)f(p_2)f(p_3)}
    {\omega + e_1 + e_2 + e_3 + i\eta} \nonumber \\
&&  = \mbox{Tr}[\bar f(p_1)\bar f(p_2)\bar f(p_3) + f(p_1)f(p_2)f(p_3)] \nonumber \\
&& \times    \delta(\omega + e_1 + e_2 + e_3)\,.
\end{eqnarray}

\begin{figure}[htbp]
\begin{center}
\includegraphics[scale=1.]{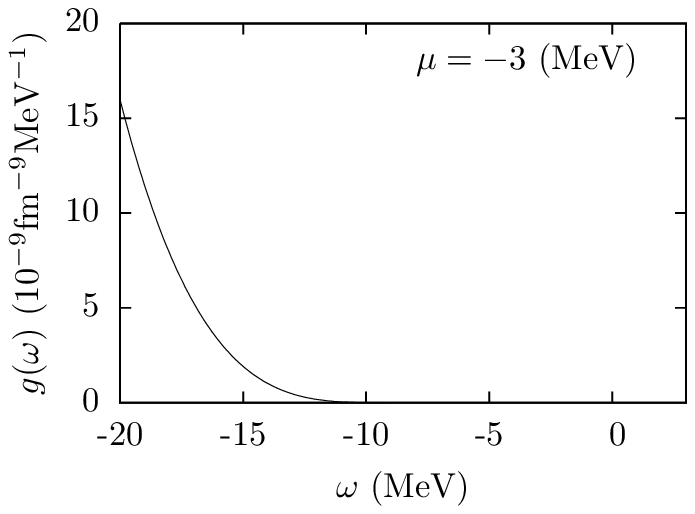}\hspace{5mm}
\includegraphics[scale=1.]{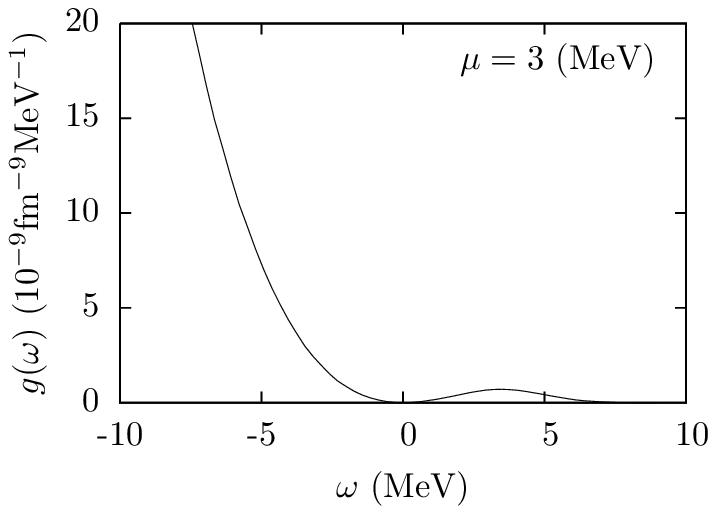}
\includegraphics[scale=1.]{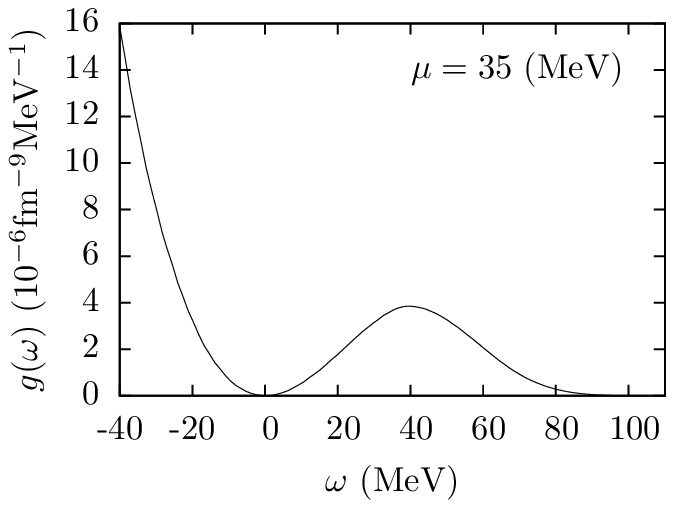}
\end{center}
\caption{\label{Sogoleveldensity} 3$h$-level density for negative (top)
  and two positive (bottom) chemical potentials \cite{Sogo1}.}
\end{figure}

In Fig.~\ref{Sogoleveldensity}, 
we give, for $T=0$, the results for negative and positive $\mu$. The
interesting case is $\mu > 0$. We see that phase-space constraint and
energy conservation cannot be fullilled simultaneously at the Fermi
energy and the level density is zero there. This is just the point
where quartetting should build up. Obviously, if there is no level
density, there cannot be quartetting. In the case of pairing there is
no phase space restriction and the level density is finite at the
Fermi energy. For negative $\mu$, $f(e_k)$ vanishes at zero
temperature and is exponentially small at finite $T$. Then there is no
fundamental difference between 1$h$ and 3$h$ level densities. This
explains the striking difference between pairing and quartetting in
the weak coupling regime. The same reasoning holds in considering the
in-medium four body equation (\ref{quartetorderparametereqn}). The
relevant in-medium four-fermion level density is also zero at $4\mu$
for $\mu > 0$ even for the quartet at rest. Actually the only case of an in-medium $n$-fermion
level density which remains finite at the Fermi energy is (besides $n$ =1) 
the $n=2$
case when the c.o.m. momentum of the pair is zero, as one may verify
straightforwardly. That is why pairing is such a special case,
different from condensation of all higher clusters. Of course, the level densities do no longer pass through zero, if we are at finite temperature. Only a strong depression may occur at the Fermi energy. This is probably, as mentioned,  the reason why the break down of the critical temperature is slightly less abrupt than at $T=0$. A well known example of zero level density at the Fermi energy may be familiar to the reader from Fermi liquid theory where the infinite mean free path of a fermion at the Fermi energy also is due to the fact that the 2p-1h level density (entering Fermi's golden rule) passes through zero at the Fermi energy.\\

In conclusion of this nuclear matter section concerning quartet condensation, we can say that for sure $\alpha$ particle condensation happens in low density nuclear matter. It may not be the only cluster which condenses because it is not the strongest bound even-even nucleus. In this context, the reader should always bare in mind that nuclei and nuclear sub-clusters only exist because there are four different fermions involved. In a dynamic process, the doubly magic $\alpha$ particle probably will be the first nucleus which condenses because of its small number of particles and its strong binding. A phenomenon of this type may happen in compact stars, as e.g., proto-neutron stars. Also neutron stars which are not completely cooled down may have in the outer crust a neutron gas between the nuclei, forming a Coulomb-lattice, with a good portion of protons. However, in which density, asymmetry, temperature range this may happen is an open question so far \cite{Lattimer, Shen, Typel, Horowitz, Gerd3}. Whereas the composition of nuclear matter at low densities and low temperatures is well investigated to give a partial density of $\alpha$ particles, correlation effects at higher densities may suppress the formation of a condensate. Further studies should be undertaken to better constrain this kind of phenomenon.

\begin{figure*}
\begin{center}
\includegraphics[width=12cm]{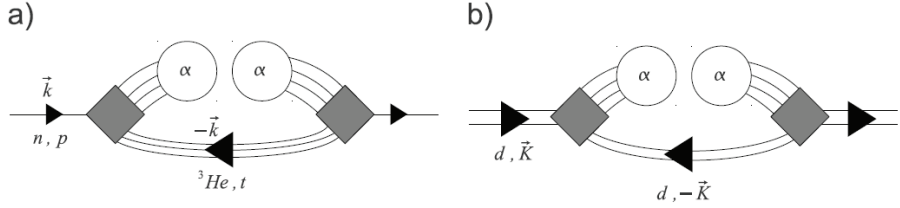}
\end{center}
\caption{Various additional processes leading to quartets.}
\label{d-d}
\end{figure*}

At this point, we should also mention that in the preceding quartet gap equation we considered an uncorrelated Fermi gas as background. In other words we treated a situation where four uncorrelated fermions directly collapse into the quartet order parameter. However, in nuclear physics with four different species of fermions, there may exist other processes leading to quartetting ($\alpha$ particle condensation). This stems from the fact that in a low density nuclear Fermi gas, there may, besides free nucleons, also deuterons, tritons, helions be present. Then two dimers and/or nucleons and trimers can form a quartet. Such processes are shown in Fig.\ref{d-d}. It remains as an important task for the future to treat a (hot) gas of nucleons, deuterons, tritons, helions, $\alpha$'s simultaneously with respect to mutual pairing and quartetting properties.

After all this, like with pairing, it is now very tempting to imagine that in finite nuclei precurser phenomena of $\alpha$ particle condensation are present. This will be the subject of the following sections.

\section{Resonating Group Method (RGM) for $\alpha$ particle clustered nuclei}

In finite nuclei special techniques have to be and have been developed to treat clustering, for instance $\alpha$ particle clustering.

The RGM is one of the most powerful microscopic cluster approaches for finite nuclei. It has been introduced by Wheeler~\cite{Wh37} and used by Kamimura {\it et al.} in 1977 in his famous work~\cite{Kami} to explain the cluster structure of ${^{12}{\rm C}}$ and, for instance, the Hoyle state. Let us shortly explain the method. \\

We will demonstrate the principle with the example of three $\alpha$ particles, the generalisation to other numbers of $\alpha$'s  being straightforward. The ansatz for the 3$\alpha$ RGM wave function has a very transparent form
\begin{equation}
\Psi_{\mbox{RGM}}({\vek r}_1....{\vek r}_{12}) \propto {\mathcal A}\chi({\vek \xi}_1,{\vek \xi}_2)\phi_{\alpha_1}\phi_{\alpha_2}\phi_{\alpha_3}
\label{rgm}
\end{equation}
Please note that in (\ref{rgm}), we suppressed the scalar spin-isospin part of the wave function of the $\alpha$ particle for brevity. Furthermore, we introduced the antisymmetriser ${\mathcal A}$, the Jacobi coordinates for the c.o.m. motion ${\vek \xi}_i$, and the intrinsic translationally invariant wave function for $\alpha$ particle number $i$
\begin{equation}
\phi_{\alpha_i} \propto \exp \bigg [-\sum_{k<l}({\vek r}_{i,k} - {\vek r}_{i,l})^2/(8b^2) \bigg ]
\label{a-wf}
\end{equation}

This $\alpha$ particle wave function contains the variational parameter $b$ leading to very reasonable $\alpha$ particle properties when used in modern energy density functionals (EDF's) \cite{PG}. Indeed, from the $\alpha$ particle wave function, one can construct the corresponding density matrix and local density which, when inserted into the EDF, yields an energy as a function of the width parameter $b$. Minimisation then leads to a definite $\alpha$ particle wave function. Because of the use of Jacobi coordinates, the total 3$\alpha$ wave function is translationally invariant. Given a microscopic hamiltonian $H$, the Schr\"odinger equation for the unknown function $\chi$ is given by
\begin{eqnarray}
&&\int d^3\xi'_1\int d^3 \xi'_2 {\mathcal H}({\vek \xi}_1{\vek \xi}_2,{\vek \xi}'_1{\vek \xi}'_2)\chi({\vek \xi}'_1,{\vek \xi}'_2) \nonumber \\
&& = E \int d^3\xi'_1\int d^3 \xi'_2   {\mathcal N}({\vek \xi}_1{\vek \xi}_2,{\vek \xi}'_1{\vek \xi}'_2)\chi({\vek \xi}'_1,{\vek \xi}'_2)
\label{rgm-eq}
\end{eqnarray}
\noindent
where
\begin{eqnarray}
&& \hspace{-0.5cm} {\mathcal H}({\vek \xi}_1{\vek \xi}_2,{\vek \xi}'_1{\vek \xi}'_2) 
 = \langle \Pi_{i=1}^2\delta({\vek \xi}_i-{\vek s}_i) \phi_{\alpha_1}\phi_{\alpha_2}\phi_{\alpha_3}| \nonumber \\ 
&& \hspace{1.5cm} (H-T_G){\mathcal A}|\Pi_{i=1}^2 \delta({\vek \xi}'_i - {\vek s}_i)\phi_{\alpha_1}\phi_{\alpha_2}\phi_{\alpha_3}\rangle \nonumber \\
\label{H-kernel}
\end{eqnarray}

\noindent
and analogously for the so-called norm kernel ${\mathcal N}$. The elimination of the c.o.m. kinetic energy $T_G$ is performed with substracting it from the Hamiltonian. In (\ref{H-kernel}) the ${\vek s}_i$ are again the Jacobi coordinates which have to be integrated over. The delta-functions in (\ref{H-kernel}) only serve for an easy book keeping of the Jacobi coordinates ${\vek \xi}$ at the end of the calculation. In order to make out of (\ref{rgm-eq}) a standard Schr\"odinger equation, we have to take the square root of the norm kernel, introduce a renormalised wave function $\tilde \chi = \sqrt{{\mathcal N}}\chi$ and divide ${\mathcal H}$ from left and right by $\sqrt{{\mathcal N}}$ what leads to $\tilde {\mathcal H} = {\mathcal N}^{-1/2}{\mathcal H}{\mathcal N}^{-1/2}$. Of course the division is only possible if we beforehand had diagonalised the norm kernel and eliminated the configurations belonging to zero eigenvalues. In the reduced space the Schr\"odinger equation then looks like
\begin{equation}
\tilde {\mathcal H}\tilde \chi = E \tilde \chi
\label{renorm-rgm}
\end{equation}
Since, for example in the case of two $\alpha$'s $({^{8}{\rm Be}})$ the nucleons occupy $0S$ orbits and in pure HO Slater approximation 4$\hbar \omega$ quanta are occupied from the four nucleons in the $P$-shell, the relative wave functions must at least accommodate 4$\hbar \omega$, if for overlapping configuration the Slater determinant shall be recovered. States with occupation lower than 4$\hbar \omega$ are so-called Pauli forbidden states. These Pauli forbidden states give rise to zero eigenvalues in the norm kernel and are thus automatically eliminated within the RGM formalism. On the other hand, the fact that the relative wave function must not have HO quanta smaller than four implies that it develops nodes in the region of overlap of the two $\alpha$'s. Since nodes generate kinetic energy, the amplitudes of oscillations at short distances will be small. This is precisely what we will find below when we treat ${^{8}{\rm Be}}$ in more detail. The above considerations can obviously be extended to any number of $\alpha$ particles. It should be mentioned, however, that the explicit evaluation of the antisymmetrisation is very complicated and the RGM equations have not been solved as they stand beyond the three $\alpha$ particles $({^{12}{\rm C}})$ case.\\
On the other hand, let us mention that RGM in the NCSM has recently successfully been used in describing scattering and reactions in light nuclei \cite{Navratil}.

\section{The Orthogonality Condition Model (OCM) and other Boson Models}

As we easily understand from eqs (\ref{rgm}) and (\ref{renorm-rgm}), the procedure to integrate out the internal coordinates of the $\alpha$'s leads to equations which are of bosonic type. It seems, therefore, natural to apply some further approximations to avoid the complexity with the antisymmetrisation. For example it can be shown that the eigenfunctions $u_F({\vek \xi})$ of the norm kernel which belong to the zero eigenvalues are just the Pauli forbidden states we discussed above. They satisfy the condition $ 
{\mathcal A}\{u_F({\vek \xi})\Pi_{i=1}^n\phi_{\alpha_i}\}=0$ for the case of  $n \alpha$ particles. This means that the antisymmetrised RGM wave function where $\chi$ is replaced by the Pauli forbidden $u_F$'s is exactly zero. This is a very strong boundary condition which is advised to incorporate into further approximation schemes. The idea of the OCM is, thus, the following: replace $\tilde {\mathcal H} = {\mathcal N}^{-1/2}{\mathcal H}{\mathcal N}^{-1/2}$ by an effective Hamiltonian $H^{(\mbox{OCM})}$ which contains effective phenomenological two and three body forces with adjustable parameters to mock up, e.g., the repulsion when two $\alpha$ particles come close
\begin{eqnarray}
&&\hspace{-1cm} H^{(\rm OCM)} = \sum_{i=1}^n T_i -T_G +\sum_{i<J=1}^n V^{\mbox{eff}}_{2\alpha}(i, j) \nonumber \\
&& \hspace{2.5cm} +\sum_{i<j<k=1}^n V^{\mbox{eff}}_{3\alpha}(i, j, k).
\label{V-OCM}
\end{eqnarray}

The effective local 2$\alpha$ and 3$\alpha$ potentials are presented as $V^{\mbox{eff}}_{2\alpha}(i, j)$ (including the Coulomb potential) and $V^{\mbox{eff}}_{3\alpha}(i, j, k)$, respectively. Then, the equation of the relative motion of the $n\alpha$ particles with $H^{(\mbox{OCM})}$, called the OCM equation, is written as
\begin{equation}
[H^{(\mbox{OCM})} - E]\Phi^{(\mbox{OCM})}_{n\alpha}=0
\label{OCM-eq}
\end{equation}
\begin{equation}
\langle u_F|\Phi^{(\mbox{OCM})}_{n\alpha}\rangle =0
\label{ocm-cond}
\end{equation}

\noindent
where $u_F$ represents the Pauli forbidden states as mentioned above. They have to be orthogonal to the physical states, a condition which is taken into account  in (\ref{ocm-cond}). Of course the wave function $\Phi^{(\mbox{OCM})}_{n\alpha}$ should be completely symmetrised with respect to any exchange of bosons. It has turned out that this approximate form of the RGM equations is very efficient and represents a viable approach for higher numbers of $\alpha$ particles. It has recently been successfully applied to the low lying spectrum of ${^{16}{\rm O}}$~\cite{Fu08} as we will discuss below.\\

Some authors go even further in the bosonisation of the problem. They discard the condition (\ref{ocm-cond}) completely and incorporate this in adjusting appropriately the effective forces. The two most recent ones are from i) Lazauskas 
{\it et al.}~\cite{Rimas} using the non-local Papp-Moszkowski potential~\cite{Pa08}. Good description of the ground state and Hoyle state positions was obtained. ii) Ishikawa~\cite{Ishikawa} obtained with local effective  two and three body forces a similar quality of the ${^{12}{\rm C}}$ spectrum. However, he, in addition, calculated also the decay properties of the Hoyle state concluding that the three body decay of three $\alpha$'s is so much hindered with respect to the sequential 2-body decay $\alpha + ^8$Be that its detection may be very difficult.

\section{Brink and Generator Coordinate Wave Functions}

The GCM was used by Uegaki {\it et al.} \cite{Uegaki} for the calculation of cluster states in ${^{12}{\rm C}}$. The GCM wave function is based on the so-called Brink wave function of the form
\begin{eqnarray}
&& \Psi_{\mbox{Brink}} \propto 
{\mathcal A}e^{-2({\vek R}_1 - {\vek S}_1)^2/b^2}e^{-2({\vek R}_2 - {\vek S}_2)^2/b^2} \nonumber \\
&&\hspace{2cm}\times e^{-2({\vek R}_3 - {\vek S}_3)^2/b^2} \phi_{\alpha_1}\phi_{\alpha_2}\phi_{\alpha_3}
\label{Brink}
\end{eqnarray}
with $\phi_{\alpha_i}$ as in (\ref{a-wf}).
The Brink wave function is in fact a perfect Slater determinant where always quadruples of 2 protons and 2 neutrons are placed on the same spatial position ${\vek S}_i$. This can be seen in noticing that a product of four Gaussians can be written as an intrinsic part $\phi_{\alpha}$ times a c.o.m. part. So, the Brink wave function places each $\alpha$ particle at a definite position and, thus, describes clustering as some sort of $\alpha$ particle crystal. Below, we will discuss the validity of this approach in more detail.

The corresponding GCM wave function is a superposition of Brink ones with a weight function $f$ which has to be determined from a variational calculation,
\begin{eqnarray}
&& \Psi_{\mbox{GCM}} \propto P_0\int d^3{\vek S}_1\int d^3{\vek S}_2\int d^3{\vek S}_3~f({\vek S}_1,{\vek S}_2,{\vek S}_3) \nonumber \\ 
&& \hspace{2cm} \times \Psi_{{\mbox{Brink}}}({\vek R}_1{\vek R}_2{\vek R}_3, {\vek S}_1{\vek S}_2{\vek S}_3).
\label{Uegaki}
\end{eqnarray}

It is clear that the GCM wave function is much richer than the single Brink one. Actually both wave functions, for practical use, have to be projected on good linear momentum (${\vek K} =0$) and on good angular momentum. To take off of the Brink wave function the total c.o.m. part is trivial because of the Gaussians in (\ref{Brink}) and is formally introduced by the projector $P_0$ in (\ref{Uegaki}). To project on good angular momentum needs usually some numerical calculation but, for example for the case of ${^{8}{\rm Be}}$ it can be done analytically ~\cite{8Be}. Let us remember that the projector on good angular momentum is given by
\begin{equation}
P^I_{MK}=\int d\Omega D^{J*}_{MK}(\Omega)R(\Omega).
\label{I-proj}
\end{equation}
where $D_{MK}^J$ are the Wigner functions of rotation and $R(\Omega)$ is the rotation operator ~\cite{RS}.\\
As mentioned, Uegaki {\it et al.} applied this technique at about the same time as Kamimura {\it et al.} with RGM to the cluster states of ${^{12}{\rm C}}$ with great success. We will present some details below.

\section{Antisymmetrised Molecular and Fermion Molecular Dynamics}

In 2007 the Hoyle state was also newly calculated by the practioners of Antisymmetrized Molecular Dynamics (AMD) (Kanada-En'yo {\it et al.} \cite{En12,Do97,En98,Ho91a,On92,En95,On04}) and Fermion Molecular Dynamics (FMD) (Chernykh {\it et al.} \cite{Chern}) approaches. In AMD one uses a Slater determinant of Brink-type of wave functions where the center of the packets ${\bf S}_i$ are replaced by complex numbers. This allows to give the center of the Gaussians a velocity as one easily realises. In FMD in addition the width parameters of the Gaussians are also complex numbers and, in principle, different for each nucleon. AMD and FMD do not contain any preconceived information of clustering. Both approaches found from a variational determination of the parameters of the wave function and a prior projection on good total linear and angular momenta that the Hoyle state has dominantly a 3-$\alpha$ cluster structure with no definite geometrical configurations. In this way the $\alpha$ cluster ansaetze of the earlier approaches were justified. 
As a performance, in \cite{Chern}, the inelastic form factor from the ground to Hoyle state was successfully reproduced in employing an effective nucleon-nucleon interaction V$_{\mbox{UCOM}}$ derived from the realistic bare Argonne V18 potential (plus a small phenomenological correction).


Kanada-En'yo {\it et al.} \cite{En98} pointed out that with AMD some breaking of the $\alpha$ clusters can and is taken into account. The Volkov  force \cite{Volkov} was employed in \cite{En98}. Again all properties of the Hoyle state were explained with these approaches. Like in the other works, the E0 transition probability came out $\sim$ 20 $\%$ too high. No bosonic occupation numbers were calculated, see Sect.9. It seems technically difficult to do this with these types of wave functions. However, one can suspect that if occupation numbers were calculated, the results would not be very different from the THSR results. This stems from the high sensitivity of the inelastic form factor (Sect.10) to the employed wave function.  Nonetheless, it would be important to produce the occupation numbers also with AMD and FMD.\\
In \cite{En98, Chern} some geometrical configurations of $\alpha$ particles in the Hoyle state are shown. No special configuration out of several is dominant. This reflects the fact that the Hoyle state is not in a crystal-like $\alpha$ configuration but rather forms to a large extent a Bose condensate. 

\section{THSR wave function and $^8$Be}

In 2001 Tohsaki, Horiuchi, Schuck, and R\"opke (THSR) proposed a new type of cluster wave function which has shed novel light on the dynamics of cluster, essentially $\alpha$ cluster motion in nuclei~\cite{To01}. The new aspect came from the assumption that for example the Hoyle state, but eventually also similar states in heavier $n\alpha$ nuclei, may be considered as a state of low density where the nucleus is broken up into $\alpha$ particles which move practically freely as bosons condensed into the same $0S$ orbit. This paper appeared after about a quarter century of silence about the Hoyle state and since then has triggered an enormous amount of new interest testified by a large amount of publications, both experimental as theoretical, see the review articles in ~\cite{rev1,rev2,rev3,rev4,rev5} and papers cited in there. However, before we consider ${^{12}{\rm C}}$ and the Hoyle state, we would like, for pedagogical reasons, to start out with ${^{8}{\rm Be}}$ which, as we know, is (a slightly unstable) nucleus with  strong 2$\alpha$ clustering, see Fig.1. In this case the THSR wave function reads
\begin{eqnarray}
\Psi_{\rm THSR} &\propto& {\mathcal A}\bigg \{ e^{ -\frac{2}{B^2}[({\vek R}_1-{\vek X}_G)^2 + ({\vek R}_2-{\vek X}_G)^2)]}\phi_{\alpha_1} \phi_{\alpha_2} \bigg \}
\nonumber\\
&\propto& {\mathcal A}\bigg \{ e^{ -\frac{r^2}{B^2}}\phi_{\alpha_1} \phi_{\alpha_2} \bigg \}    .
\label{thsr-2}
\end{eqnarray}
where the ${\vek R}_i$ are the c.o.m. coordinates of the two $\alpha$ particles , ${\vek X}_G = ({\vek R}_1 + {\vek R}_2)/2$ is the c.o.m. coordinate of the total system, and ${\vek r}= {\vek R}_1 - {\vek R}_2$ is the relative distance between the two $\alpha$ particles. The $\phi_{\alpha_i}$ are the same intrinsic $\alpha$ particle wave functions as in (\ref{a-wf}). We see that the THSR wave function is totally translationally invariant. Of course, (\ref{thsr-2}) is just a special case of the general THSR wave function for a gas of $n$ $\alpha$ particles
\begin{equation}
\Psi_{\rm THSR} \propto {\mathcal A}\psi_1\psi_2~~ ...~~ \psi_n \equiv {\mathcal A}|B\rangle
\label{thsr-n}
\end{equation}
with
\begin{equation}
\psi_i \propto e^{  -\frac{2}{B^2}({\vek R}_i-{\vek X}_G)^2}\phi_i
\label{thsr-1}
\end{equation}
and $X_G$ again the c.o.m. coordinate of the total system. Also, this $n\alpha$ wave function is translationally invariant and its c.o.m. part is usually expressed by the Jacobi coordinates.\\

As a technical point let us mention that for practical calculation the THSR wave function is rarely used in the form (\ref{thsr-2}). The point is that there has accumulated a lot of know-how in dealing with the Brink wave function and one wants to exploit this. To this purpose, the THSR wave function (\ref{thsr-n}) can be written as 

\begin{equation}
\Psi_{\rm THSR} \propto \int d^3R_1 ...d^3R_n\exp \bigg [-\sum_{i=1}^n\frac{R^2_i}{\beta^2}\bigg ]\Psi_{\rm Brink}
\label{thsr-brink}
\end{equation}
with the relation for the width $B^2 = b^2 + 2\beta^2$. Since the c.o.m. part of each $\alpha$ particle in the Brink wave function has a width $b$, the integral in (\ref{thsr-brink}) simply serves to transform the c.o.m. wave function of the $\alpha$ with a small width $b$ into one with a large width $B$. Otherwise there is of course strict equivalence of the two forms (\ref{thsr-n}) and (\ref{thsr-brink}) of the THSR wave function. As a side-remark, one may notice that if one worked in momentum space, the folding integral in (\ref{thsr-brink}) becomes simply a product of the c.o.m. and intrinsic $\alpha$ wave functions in momentum space. It remains to be seen whether this aspect can present some advantages in future studies.\\

From (\ref{thsr-n}), we see that the THSR wave function is analogous to a number projected BCS wave function in case of pairing and, therefore, suggests $\alpha$ particle condensation. However, for a wave function with a fixed number of particles, a bosonic type of condensation is not guaranteed and it has to be shown explicitly in how much the condensation phenomenon is realized. We will come to this point later.
The THSR wave function has two widths parameters $B$ and $b$ which are obtained from minimising the energy. The former describes the c.o.m. motion of the $\alpha$ particles which can extend over the whole volume of the nucleus and should, therefore, have a large width if the $\alpha$'s are well formed at low density. The width $b$ of the $\alpha$ particles should be much smaller and essentially stay at its free space value $b=1.36$ fm. However, if one squeezes the nucleus, the $\alpha$'s will strongly overlap and  quickly loose their identity, getting larger in size and finally, at normal nuclear densities dissolve completely into a Fermi gas. This happens for $b=B$. The mechanism which leads to this fast dissolution of the $\alpha$ particles was discussed in Section 2 in the case of infinite matter. One can show that the THSR wave function contains two limits exactly. For $b=B$ it becomes a pure Harmonic Oscillator Slater determinant~\cite{rev5}, whereas for $B>>b$ the $\alpha$ particles are so far apart from one another that the Pauli principle, i.e., the antisymmetriser, can be neglected leading to a pure product state of $\alpha$ particles, i.e., a condensate. These features of the THSR wave function show again the necessity to investigate to which end THSR is closer: to a Slater determinant or to an $\alpha$ particle condensate. We will study this in detail for the Hoyle state of $^{12}$C in the next section.\\

For the moment, let us continue with our study of $^8$Be. Of course, we know that $^8$Be is strongly deformed. It is straightforward to generalise the THSR wave function to deformed systems~\cite{8Be}. We suppose that the $\alpha$'s stay spherical and only their c.o.m. motion becomes deformed. This is easily achieved in adopting different width parameters $B_i$ in the different spatial directions. For example with $B_i^2 = b^2 + 2\beta_i^2$, we write for the c.o.m. part $\chi(r)$ of the THSR wave function 
\begin{equation}
\chi^{\rm THSR}({\vek r}) \propto \exp \bigg (-\frac{r_x^2 + r_y^2}{b^2 + 2 \beta_{\bot}^2} -\frac{r_z^2}{b^2 + 2\beta_z^2}\bigg )
\end{equation}


\begin{figure}
\begin{center}
\includegraphics[scale=0.65,angle=-90]{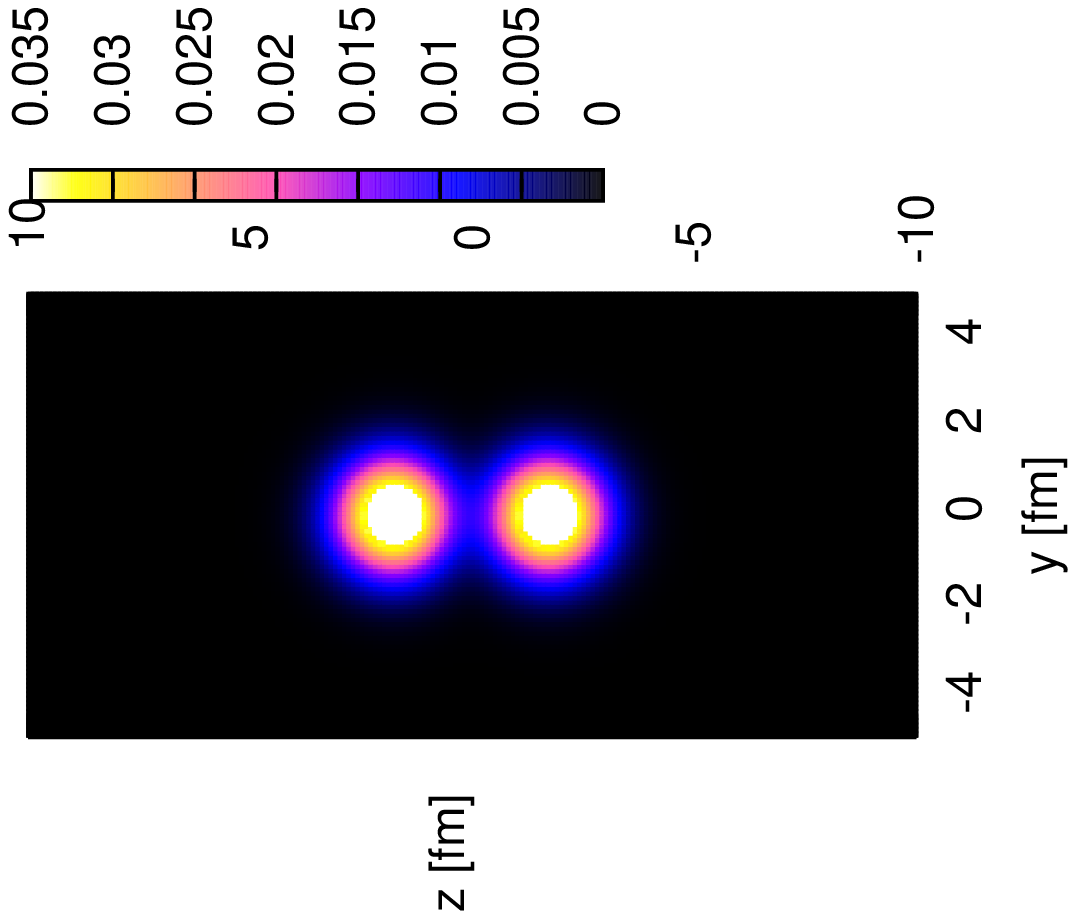}
\includegraphics[scale=0.65,angle=-90]{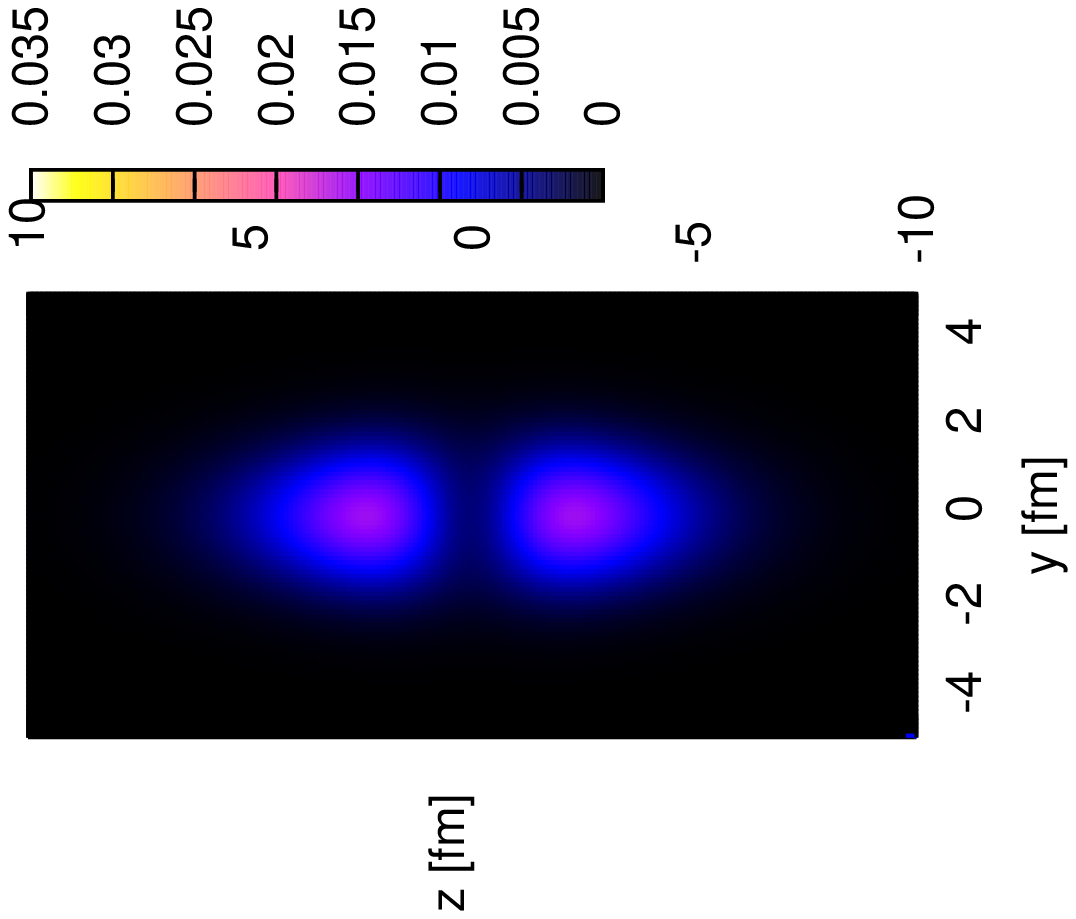}
\end{center}
\caption{ Comparison of single Brink and THSR densities for $^8$Be, from top to bottom.}
\label{densities8Be}
\end{figure}

\begin{figure}
\begin{center}
\includegraphics[scale=0.7]{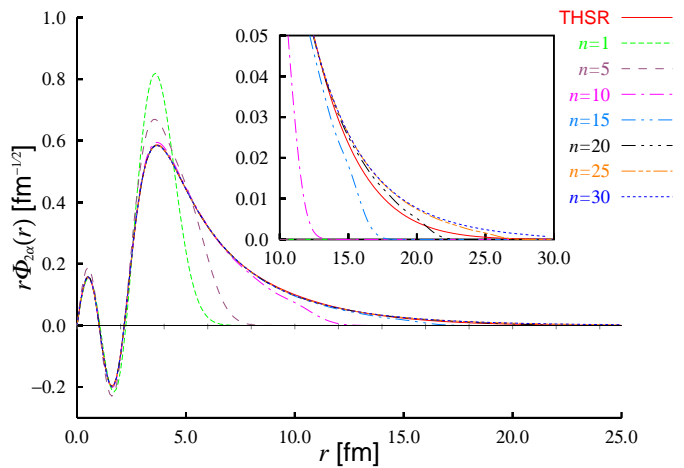}
\end{center}
\caption{Comparison of angular momentum projected ($J=0$) THSR wave function with a single ``Brink'' wave function for $^8$Be with $S=3.35$ fm (denoted by $n=1$). The convergence rate with the superposition of several ($n$) ``Brink'' wave functions \`a la Brink-GCM, is also shown. The line denoted by $n=30$ corresponds to the full RGM solution, see ~\cite{Funaki-WvO} for more details.}
\label{8Be-wf}
\end{figure}

Let us compare the deformed densities of $^8$Be one gets from a single Brink (\ref{Brink}) and the THSR wave functions. Using the Volkov force ~\cite{Volkov}, this is shown in Fig.~\ref{densities8Be}.  
We see that there is quite strong difference between the two distributions. The THSR one is much more diffuse  than the one obtained from a single Brink wave function. 
Actually this physical crystal-like dumbbell picture was the prevailing opinion of cluster physics before the introduction of the THSR wave function. We see that THSR offers a much more smeared out, quantal aspect of clustering. We will elaborate on this in more detail in Sect.19 where we discuss 'cluster localisation' versus 'delocalisation'.\\

One has to superpose many Brink wave functions (about 30) as done with the Brink-GCM approach to recover the quality of the single THSR wave function. In the laboratory frame, i.e. after angular momentum projection, the wave functions become almost identical~\cite{8Be,Funaki-WvO}. This is shown in Fig.~\ref{8Be-wf}. Actually, it is interesting that the angular momentum projection can be performed in this case analytically. With (\ref{I-proj}) we obtain
\begin{equation}
\hat P^{J=0}\chi^{\rm THSR}( r) \propto 
\frac{\exp(-r^2/B_{\bot}^2)}{ir}\mbox{Erf} \bigg (i \frac{ \sqrt{B_z^2 - B_{\bot}^2}}{B_{\bot}B_z}r \bigg )
\label{P-thsr}
\end{equation}
where Erf$(x)$ is the error function. With this projected $^8$Be THSR wave function, the results, e.g., for the ground state energy, are identical up to the 4th digit with the results from RGM ~\cite{Hill}.\\
 At this point let us also mention that $\alpha-\alpha$ scattering has recently been well described from an ab initio EFT calculation by Elhatisari {\it et al.} \cite{Elhatisari2} and that the structure of $^8$Be has been treated with the NCSM by Dytrych {\it et al.} \cite{Dytrych}.

After this relatively simple but instructive case of $^8$Be, let us move onward to $^{12}$C.


\section{The $^{12}$C nucleus and the Hoyle state}

Compared to the $^8$Be case, the situation in $^{12}$C is considerably more complex. First of all, the ground state of $^{12}$C is not a low density $\alpha$ cluster state as in $^8$Be. However, there exists a radially expanded state of about same low density as for $^8$Be which is a weakly interacting gas of three $\alpha$ particles forming a $0^+$ state at 7.65 MeV which is the famous Hoyle state already mentioned in the Introduction. 
The reason why the Hoyle state, in analogy to the case of $^8$Be is not the ground state of $^{12}$C is not absolutely clear. However, one may speculate that some sort of extra attraction acts between the three $\alpha$'s which makes the $\alpha$ gas state collapse to a much denser state of the Fermi gas type, that is to good approximation describable, as practically all other nuclei, by a Slater determinant. In $^{12}$C coexist, therefore, two types of quantum gases: fermionic ones and bosonic ones. We will see later that one can suppose that such is also the case in heavier self conjugate nuclei like $^{16}$O, etc.\\

As already pointed out in the Introduction, the Hoyle state and other states in $^{12}$C were explained in the 1970-ties by two pioneering works from Kamimura {\it et al.}~\cite{Kami} and Uegaki {\it et al.} ~\cite{Uegaki}. They used the RGM and Brink-GCM approaches, respectively. In 2001 the THSR wave function explained the Hoyle state with the $\alpha$ condensate type of wave function (\ref{thsr-n}) ~\cite{To01}. It was shown later that, taken the same ingredients, the THSR wave function has almost 100$\%$ squared overlap for the Hoyle state with the wave function of Kamimura {\it et al.} (and by the same token also of Uegaki {\it et al.}) ~\cite{Fu03,Uegaki}. Before we come, however, to a detailed presentation of the results, we have to explain how to use the THSR wave function in the case where the $\alpha$ gas state is not the ground state as in $^8$Be but an excited state. Two possibilities exist. Either one takes the large width parameter $B$ as Hill-Wheeler coordinate~\cite{Hill} and superposes a couple of THSR wave functions with different $B$-values leading to an eigenvalue equation which yields several eigen values including ground and Hoyle state ~\cite{To01}, or one minimises the energy under the condition that the excited state is orthogonal to the ground state ~\cite{Funaki-WvO}.

We will adopt the latter strategy because it has been shown, as already mentioned, that a {\it single} wave function of the THSR type is able to describe the Hoyle state with very good accuracy. \\

\begin{figure}
\begin{center}
\includegraphics[scale=0.7]{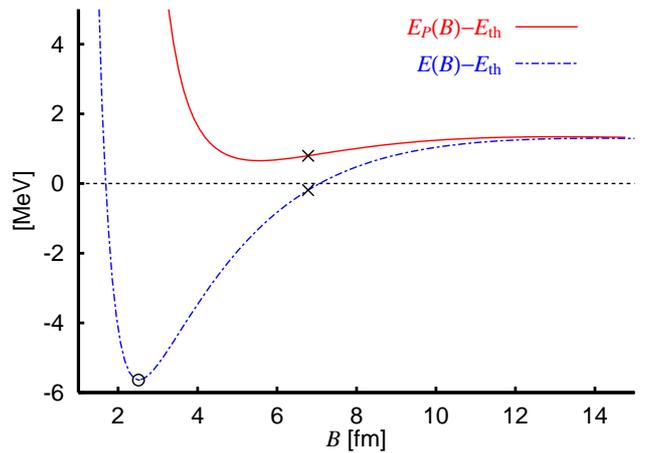}
\end{center}
\caption{Energy curve in the space orthogonal to the ground state, denoted by $E_P(B)$, together with the ground state $E(B)$. The values at the optimal $B$ values, $B_g$ and $B_H$ for the ground and Hoyle states, respectively, are marked by a {\it circle} and a {\it cross}.}
\label{energy-ortho-gs}
\end{figure}

\begin{figure}[h]
\begin{center}
\includegraphics[scale=0.65]{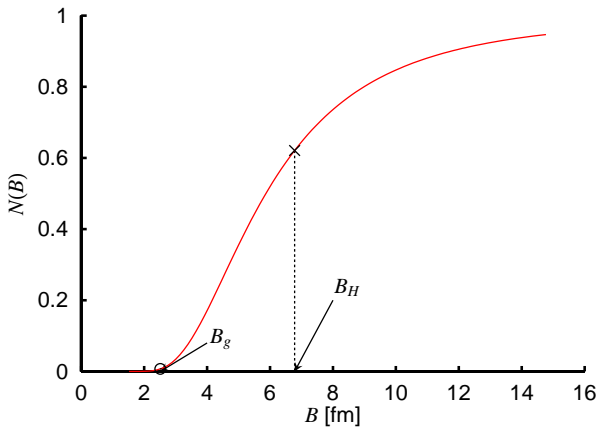}
\caption{Expectation value of the antisymmetrization operator in the product state $|B\rangle$. The value at the optimal $B$ values, $B_g$ for the ground state and $B_H$ for the Hoyle state, are denoted by a circle and a cross, respectively.}
\label{fig:anti}
\end{center}
\end{figure}

It is very interesting to consider the energy curves as a function of $B$-parameter for the ground state 
\begin{equation}
E(B) = \frac{ \langle \Psi_{\rm THSR}(B)|H|\Psi_{\rm THSR}(B)\rangle}
{\langle \Psi_{\rm THSR}(B)|\Psi_{\rm THSR}(B)\rangle}
\label{E-gs}
\end{equation}
and the first excited $0^+$ state in $^{12}$C
\begin{equation}
E_P(B) = \frac{\langle \hat P_{\bot}^{(\mbox{g.s.})}\Psi_{\rm THSR}(B)|H|\hat P_{\bot}^{(\mbox{g.s.})}\Psi_{\rm THSR}(B) \rangle}{\langle \hat P_{\bot}^{(\mbox{g.s.})}\Psi_{\rm THSR}(B)|\hat P_{\bot}^{(\mbox{g.s.})}\Psi_{\rm THSR}(B)\rangle}
\label{thsr-2nd}
\end{equation}
with $\hat P_{\bot}^{(\mbox{g.s.})} =1-|0_1^+ \rangle \langle 0_1^+ |$, the projector making the excited state orthogonal to the ground state.\\

The corresponding energy curves are shown in Fig.~\ref{energy-ortho-gs}. We see that the excited state has a minimum for a $B$-value almost three times as large as the one for the ground state. 

Actually, the ``optimal'' $B$ value is taken as the one giving the largest
squared overlap for the Hoyle state
between the solution of the Hill-Wheeler equation for th $|0_2^+>$ state, and the THSR wave
function, $|\mbox{THSR}(B)\rangle$.
 Figure 9 shows the energy surface of the Hoyle state in the
orthogonal space to the 
ground state (upper full line) and the minimum does not coincide with the optimal $B$ value.  On the other hand, if we define the ``optimal'' $B$ value as giving the largest
squared overlap between the
solution of the Hill-Wheeler equation and THSR wave function in the
orthogonal space to the ground state,
then the minimum in Fig.9 and the new ``optimal'' $B$ value become closer to
each other.

This study allows us to make a first investigation about the importance of the Pauli principle, i.e., of the antisymmetriser in the THSR wave function, in the ground state and in the Hoyle state. For this we define 
\begin{equation}
N(B) = \frac{ \langle B|{\mathcal A}|B\rangle}{\langle B|B\rangle}
\label{anti-sym}
\end{equation}
where $|B\rangle$ is the THSR wave function in (\ref{thsr-n}) without the antisymmetriser.
For $B \rightarrow \infty$ the quantity in (\ref{anti-sym}) tends to one, since, as already mentioned, the $\alpha$ particles are in this case so far apart from one another that antisymmetrisation becomes negligeable. The result for $N(B)$ is shown in Fig.~\ref{fig:anti} as a function of the width parameter $B$. We chose as optimal values of $B$ for describing the ground and Hoyle states, $B=B_g = 2.5$ fm and $B=B_H$= 6.8 fm, for which the normalised THSR wave functions give the best approximation of the ground state $0^+_1$ and the Hoyle state $0^+_2$, respectively (obtained by solving the Hill-Wheeler equation). We find that $N(B_H) \sim 0.62$ and $N(B_g) \sim 0.007$. These results indicate that the influence of the antisymmetrisation is strongly reduced in the Hoyle state compared with the ground state. This study gives us a first indication that the Hoyle state is quite close to the quartet condensation situation rather than being close to a Slater determinant. We will be more precise with this statement in the next section.

\section{Alpha particle occupation probabilities}

\begin{figure}[h]
\begin{center}
\includegraphics[scale=0.7]{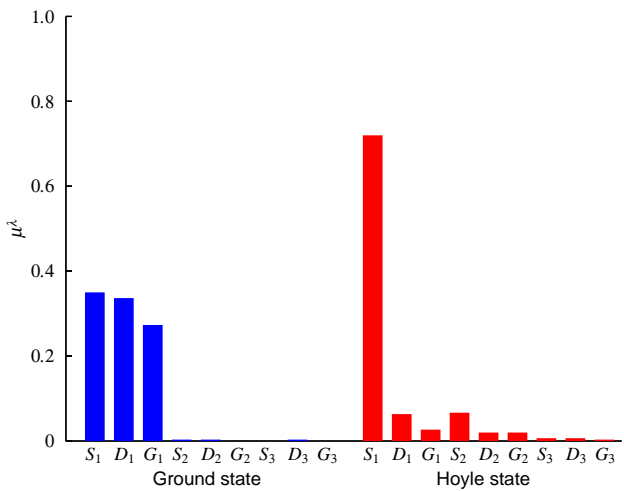}
\caption{Occupation of the single $\alpha$ orbitals of the Hoyle state of $^{12}$C compared with the ground state.}
\label{fig:3aocm_occupation}
\end{center}
\end{figure}

In the preceding section, we have seen a first indication that the influence of antisymmetrisation between the $\alpha$ particles in the Hoyle state is strongly weakened. A more direct way of measuring the degree of quartet condensation is to calculate the single $\alpha$ particle density matrix $\rho_{\alpha}({\vek \xi}_1, {\vek \xi}'_1)$ where from the density matrix formed with the fully translationally invariant THSR wave function all intrinsic $\alpha$ coordinates as well as all $\alpha$ c.o.m. Jacobi coordinates ${\vek \xi}_i$ besides one have been integrated out. A more detailed description of the procedure can be found in ~\cite{Yam09}. The eigenvalues of $\rho_{\alpha}$ correspond to the bosonic occupation numbers of the $\alpha$ particles. For example in the ideal boson condensate case, one will get for the Hoyle state one eigenvalue equal three (for the 0S state) and zero for all the others. However, we have seen that the Pauli principle, though being weak in the Hoyle state, it is not entirely inactive. This leads, besides from the action of the two body interaction, to a depletion of the lowest $\alpha$ state. Alpha particles are scattered out of the condensate, as one says. The calculation of this single $\alpha$ density matrix is technically complicated, even with the THSR wave function. There exist two older approximate (though quite reliable) calculations. The first one was performed by Suzuki {\it et al.} ~\cite{Suzuki}. In this work, the correlated Gaussian basis was used for the construction of the c.o.m. part of the RGM wave function. With an accurate approximation of the norm kernel, the amount of $\alpha$ condensation was calculated to be about 70$\%$. Afterwards, this problem was studied by Yamada {\it et al.} ~\cite{Yam05} using the 3$\alpha$ OCM approach. The result for the percentage of $\alpha$ condensation was equally about 70$\%$ and the distribution of the various occupation probabilities in the Hoyle state and the ground state of $^{12}$C is shown in Fig.~\ref{fig:3aocm_occupation}. We clearly see that the Hoyle state has a 0S occupancy of over 70$\%$ whereas all other occupancies are down by at least a factor of ten. On the other hand the occupancies of the ground state are democratically distributed over the configurations compatible with the shell model. We thus see that the Hoyle state is quite close to the ideal Bose gas picture. More recently Funaki {\it et al.} ~\cite{rev5} have achieved a calculation of the occupation numbers for the Hoyle state with the THSR wave function. It is found that the 0S wave is occupied with over 80$\%$. With the same technique the 0S occupation of the 15.1 MeV state in $^{16}$O was calculated to be over 60$\%$ ~\cite{Fu10}.\\
It is interesting to compare those numbers with typical fermionic occupation numbers. In this case a pure Slater determinant has fermion occupation numbers one or zero according to the Fermi step function. However, in reality measurements and also theories which go beyond the mean field approximation show that the occupancies are depleted and that the occupancies instead of being one are reduced to values ranging in the interval 0.7 to 0.8 ~\cite{Pandha, Gerd-Clark}. Therefore the nuclei in their ground states are as far away from an ideal Fermi gas as the Hoyle state (and possibly other Hoyle-analog states in heavier self-conjugate nuclei) is away from an ideal Bose gas. \\
It is also known from the interacting Bose gas that at zero temperature the Bose condensate is less than 100$\%$. For instance, in liquid $^4$He, the condensate fraction is less than 11$\%$. Calculations for $\alpha$ matter indicate a reduction of the Bose condensate with increasing density, see \cite{Gerd-Clark} where the suppression of the condensate fraction with increasing density is shown. In that paper, performing an artificial variation of the radius of the Hoyle state, the 0S occupancy is reduced with decreasing radius that indicates increasing density. This nice correspondence between the condensate fraction in homogeneous matter and single-state occupancy in nuclei underlines the analogy of $\alpha$ correlations in the Hoyle state with the Bose-Einstein condensate in homogeneous matter.

The above mentioned figure of 70-80$\%$ $\alpha$ condensate is confirmed by other less microscopic calculations which are based on a complete bosonisation of the three $\alpha$ problem with effective forces mocking up the Pauli principle. These approaches are also capable to find a quite good reproduction of the spectrum of $^{12}$C including the Hoyle state and they also result in a 70-80$\%$ realisation of the condensate. Such a study exists by Ishikawa  ~\cite{Ishikawa} . A similar study has been performed by Lazauskas {\it et al.} ~\cite{Rimas}. The latter authors only concluded (in agreement with Ishikawa ~\cite{Ishikawa}) that the $\alpha$ particles interact to 80 percent in relative 0S waves. However, there is a strong correlation between these numbers and the occupancies. This has been explicitly shown by Ishikawa who also calculated the bosonic occupation probabilities from his approach. He obtained 80\% of 0S state occupation \cite{Ishikawa}. We thus can conclude from all these studies that, indeed, the Hoyle state can be considered to be  in an $\alpha$ particle condensate for 70-80$\%$ of its time. The other 20-30\% contain residual $\alpha - \alpha$ correlations together with other configurations which empty the condensate to some extent. The picture that the three $\alpha$'s in the Hoyle state have a slightly  correlated two $\alpha$ state mostly in relative 0S state around which the third $\alpha$ is orbiting also mostly in a 0S state may be adequate. In a purely classical picture, one may say that two $\alpha$'s are moving in the lowest mode on their interconnecting straight line and the third $\alpha$ does the same on a straight line with respect to the c.o.m. of the first two. Of course, the orientations of the straight lines are not fixed in space and each one has to be averaged  over the whole volume.
Two $\alpha$ correlations and the Pauli principle are responsible for the fact that the Hoyle state is not entirely an ideal Bose condensate.

This situation  may be compared with a practically 100 percent occupancy in the case of cold bosonic atoms trapped in electro-magnetic devices. There the density is so low that the electron cloud of the atoms do not get into touch with one another and, therefore, an ideal Bose condensate state can be formed ~\cite{String}.

\section{Spacial Extension of the Hoyle state}

\begin{figure}[h]
\begin{center}
\includegraphics[scale=0.33]{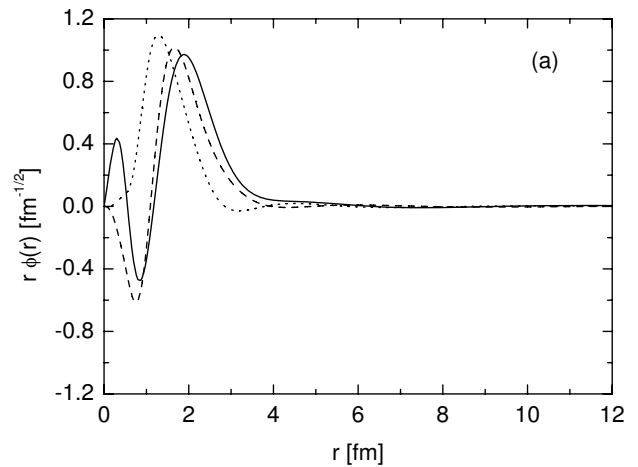}
\includegraphics[scale=0.33]{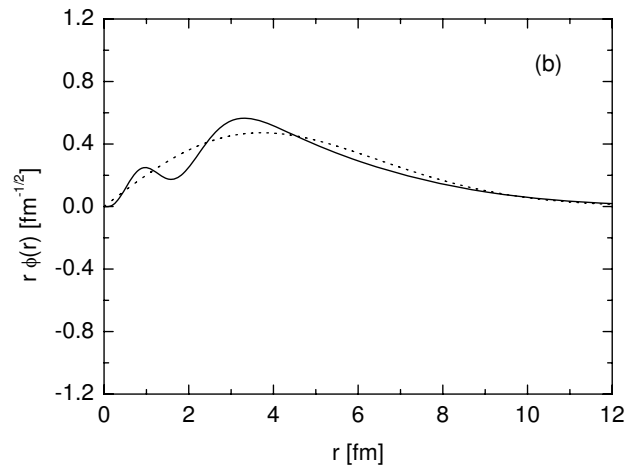}
\caption{Single $\alpha$ orbit in ground state (a) and in Hoyle state (b). In (a) the full line with two nodes corresponds to the S-wave, the broken line with one node to the P-wave and the dotted line with zero nodes to the G-wave.}
\label{fig:single_alpha}
\end{center}
\end{figure}

We have argued above that the Hoyle state has a similar low density as $^8$Be. Let us see what the THSR wave function tells us in this respect. First of all we give in Table~\ref{tab:thsr_12c} the rms radii of ground and Hoyle state calculated with THSR comparing it also with the RGM solution of Kamimura {\it et al.}~\cite{Kami} as well as with experimental data (the Hoyle state has a width of $\sim$ some eV and can, thus, be treated in bound state approximation what is implicitly done in those calculations).  We see that the rms radius of the Hoyle state is about 50$\%$ larger than the one of the ground state of $^{12}$C ($R_{\mbox{rms}} \sim 2.4$ fm). This leads to 3-4 times larger volume of the Hoyle state with respect to the ground state. In Fig.~\ref{fig:single_alpha}, we show the the single $\alpha$ 0S wave orbit corresponding to the largest occupancy of the Hoyle state. We see (lower panel) that this orbit is quite extended and resembles a Gaussian (drawn with the broken line, for comparison). There exist no nodes, only slight oscillations indicating that the Pauli principle is still active. There is no comparison with the oscillations in the ground state where the $\alpha$'s strongly overlap and effects from antisymmetrisation are very strong. Also the extension of the ground state orbits is much smaller than the one from the Hoyle state. In Table~\ref{tab:thsr_12c} are also given the monopole transition probabilities between Hoyle and ground state. Again there is agreement with the RGM result and also reasonable agreement with the experimental value though the theoretical values are larger by about 20$\%$. This transition probability is surprisingly large, a fact which can be explained with the Bayman-Bohr theorem  ~\cite{BB} and also from the fact that extra $\alpha$-like correlations are present in the ground state as will be discussed in section 14.\\

\begin{table*}[htbp]
\begin{center}
\caption{Comparison of the total energies, r.m.s. radii $(R_{\rm r.m.s.})$, and monopole strengths $(M(0_2^+\rightarrow 0_1^+))$ for $^{12}$C given by solving Hill-Wheeler equation~\cite{Hill} and from Ref.~\cite{Kami}. The effective two-nucleon force Volkov No.~2~\cite{Volkov} was adopted in the two cases for which the $3\alpha$ threshold energy is calculated to be $-82.04$ MeV.}
\label{tab:thsr_12c}
\begin{tabular}{ccccc}
\hline\hline
 &  & \hspace*{5mm}{THSR w.f.}\hspace*{5mm} & \hspace*{5mm}{\raisebox{-1.8ex}[0pt][0pt]{$3\alpha$ RGM \cite{Kami}}}\hspace*{5mm} & \hspace*{5mm}{\raisebox{-1.8ex}[0pt][0pt]{Exp.}}\hspace*{5mm} \\
 &  & (Hill-Wheeler) &  &  \\
\hline
\raisebox{-1.8ex}[0pt][0pt]{$E$~(MeV)} & $0_1^+$ & $-89.52$ & $-89.4$  & $-92.2$  \\
 & $0_2^+$ & $-81.79$ & $-81.7$  & $-84.6$  \\
\hline
\raisebox{-1.8ex}[0pt][0pt]{$R_{\rm r.m.s.}$~(fm)} & $0_1^+$ &   $\ \ \ 2.40$ &   $\ \ \ 2.40$ &   $\ \ \ 2.44$ \\
 & $0_2^+$ &   $\ \ \ 3.83$ &   $\ \ \ 3.47$ &  \\
\hline
$M(0_2^+\rightarrow 0_1^+)$~(fm$^2$) &  &   $\ \ \ 6.45$ &   $\ \ \ 6.7$ &   $\ \ \ 5.4$  \\
\hline\hline
\end{tabular}
\end{center}
\end{table*}

\begin{figure}[h]
\begin{center}
\includegraphics[width=7cm]{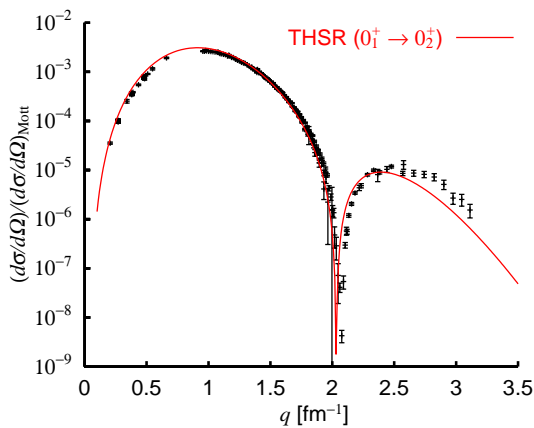}
\caption{Inelastic form factor  claculated with the THSR wave function (BEC) and the one of Kamimura {\it et al.}. The two results are on top of one another. Experimental data are from \cite{Chern}.}
\label{fig:fmfct}
\end{center}
\end{figure}

\begin{figure}
\begin{center}
\includegraphics[scale=0.75]{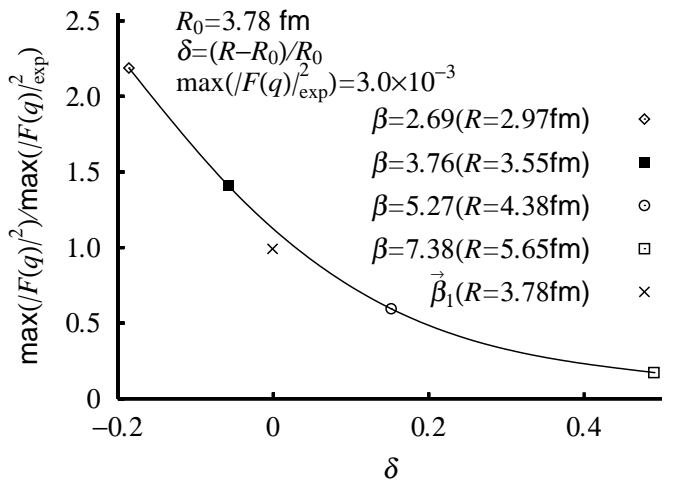}
\end{center}
\caption{Variation of the first maximum of the inelastic form factor with the radius of Hoyle state, see \cite{Fun06}.}
\label{inel-max}
\end{figure}

A very sensitive quantity is the inelastic form factor from ground to Hoyle state. In Fig.~\ref{fig:fmfct} we show a comparison of the result obtained with the THSR wave function and experimental data. We see practically perfect agreement. We want to stress that this result is obtained without any adjustable parameter what is a quite remarkable result for the following reason. Contrary to the position of the minimum, the absolute values of the inelastic form factor are very sensitive to the extension of the Hoyle state. In Fig.~\ref{inel-max} we show the dependence of the height of the first maximum as a function of an artificial variation of the radius of the Hoyle state ~\cite{Fun06}. We see that a 20$\%$ variation of the Hoyle radius gives a factor of two variation in the height of the maximum. A very strong sensitivity indeed! 


Let us also mention that the result of the RGM calculation \cite{Kami} in Fig. \ref{fig:fmfct} cannot be distinguished on the scale of the figure from the THSR one  demonstrating again the equivalence of both approaches. This strong sensitivity lends high credit to the THSR approach and to  all conclusions which are drawn from it concerning the Hoyle state. This concerns for instance the $\alpha$ particle condensation aspect discussed above. One is also tempted to conclude that any theory which reproduces this inelastic form factor describes implicitly the same properties as the THSR approach. One  recent very successful Green's function Monte Carlo (GFMC) calculation can also be interpreted in this way. We show in Fig.~\ref{QMC} in section 12 the result for the inelastic form factor from the GFMC  approach by Pieper {\it et al.}. We see that the agreement with experiment is practically the same as with the THSR one.

\section{Hoyle family of states in $^{12}$C}

In $^{12}$C, there exists besides the Hoyle state a number of other $\alpha$ gas states above the Hoyle state which one can qualify as excited states of the Hoyle state. For the description of those states it is indispensable to generalise the THSR ansatz. Indeed,  it is possible to make a natural extension of the 3$\alpha$ THSR wave function. The part of the 3$\alpha$ THSR wave function which corresponds to the c.o.m. motion of the $\alpha$ particles contains two Jacobi coordinates ${\vek \xi}_1$ and ${\vek \xi}_2$. To take account of $\alpha - \alpha$ correlations, that is, e.g., of the fact that two of the three $\alpha$'s can have a closer distance than the distance to the third $\alpha$ particle, it is possible to associate two different width parameters $B_1, B_2$ to the two Jacobi coordinates. In this case the translationally invariant THSR wave function has the following form (for the ground and Hoyle states, we recovered $B_1 = B_2$ to very good accuracy)
\begin{equation}
\Psi^{\rm THSR}_{3\alpha} = {\mathcal A}\bigg [ \exp \bigg ( -\frac{4}{3B_1^2}{\vek \xi}_1^2 - \frac{1}{B_2^2} {\vek \xi}_2^2\bigg )\phi_1\phi_2\phi_3 \bigg ]
\label{gen-thsr}
\end{equation}

\begin{figure}[h]
\begin{center}
\includegraphics[scale=1.]{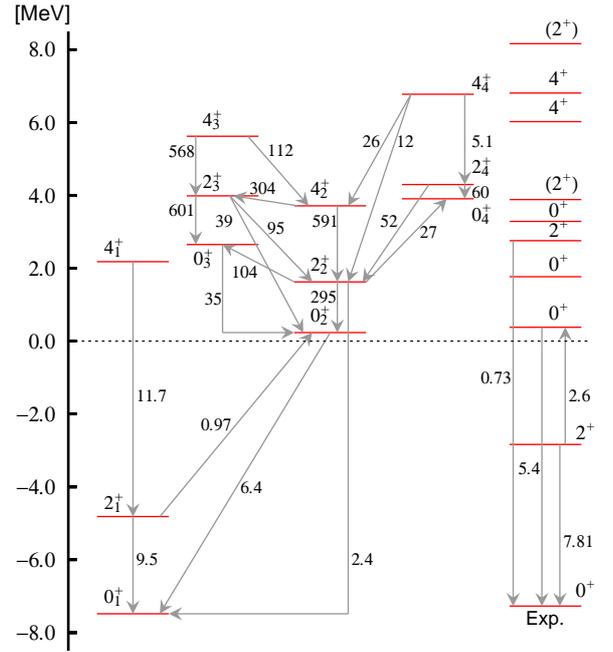}
\caption{Spectrum of $^{12}$C obtained from the extended THSR approach in comparison with experiment. The arrows indicate the transition strengths \cite{Funaki}.}
\label{fig:new_thsr}
\end{center}
\end{figure}

Of course, the $B_i$ may assume different values in the three spatial direction ($B_{i,x}, B_{i,y}, B_{i,z}$) to account for deformation and then the wave function should be projected on good angular momentum. With this type of generalised THSR wave function, one can get a much richer spectrum of $^{12}$C. In ~\cite{Funaki} by Funaki, axial symmetry has been assumed and the four $B$ parameters taken as Hill-Wheeler coordinates. In Fig.~\ref{fig:new_thsr}, the calculated energy spectrum is shown. One can see that besides the ground state band, there are many $J^{\pi}$ states obtained above the Hoyle state. All these states turn out to have large rms radii (3.7 $\sim$ 4.7 fm ), and therefore can be considered as excitations of the Hoyle state. The Hoyle state can, thus be considered as the 'ground state' of a new class of excited states in $^{12}$C. In particular, the nature of the series of states ($0_2^+, 2_2^+, 4_2^+$) and the $0_3^+$ and $0_4^+$ states have recently been much discussed from the experimental side. The $2_2^+$ state which theoretically has been predicted at a few MeV above the Hoyle state already in the early works of Kamimura {\it et al.} ~\cite{Kami}  and Uegaki {\it et al.} ~\cite{Uegaki} was recently confirmed by several experiments, see ~\cite{It11,Gai2+} and references in there. A strong candidate for a member of the Hoyle family of states with $J^{\pi}=4^+$ was also reported by Freer {\it et al.} ~\cite{Freer}. Itoh {\it et al.} recently pointed out that the broad $0^+$ resonance at 10.3 MeV  should be decomposed into two states: $0_3^+$ and $0_4^+$ ~\cite{It13, It11}. This finding is consistent with theoretical predictions where the $0_3^+$ state is considered as a breathing excitation of the Hoyle state~\cite{kurokawa05} and the $0_4^+$ state as the bent arm or linear chain configuration~\cite{En98}. \\

\begin{figure}[h]
\begin{center}
\includegraphics[width=7.5cm]{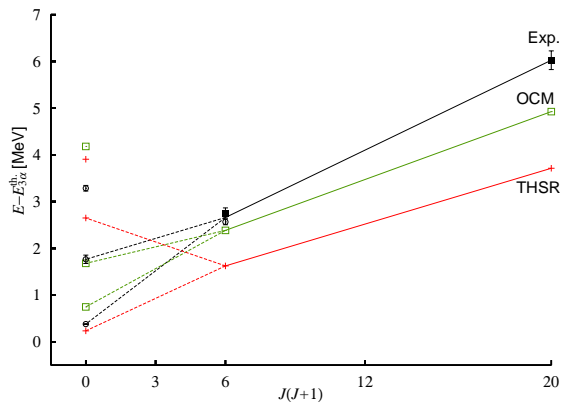}
\caption{So-called 'Hoyle' band as a function of $J(J+1)$ calculated from the  THSR approach (red) and from OCM (green) compared with experiment (black). The experimental values can be found in: $0^+_3, 0^+_4, 2^+_2$: \cite{It11}; $2^+_2$: \cite{Gai2+}; $4^+_2$: \cite{Freer}}
\label{fig:rota_new_thsr}
\end{center}
\end{figure}

In Fig.~\ref{fig:new_thsr}, the $E2$ transition strengths between $J$ and $J\pm2$ states and monopole transitions between $0^+$ states are also shown with corresponding arrows. We can note the very strong $E2$ transitions inside the Hoyle band, $B(E2; 4_2^+ \rightarrow 2_2^+)$ = 591 e$^2$fm$^4$ and $B(E2; 2_2^+ \rightarrow 0_2^+)$ = 295 e$^2$fm$^4$. The transition between the $2_2^+$ and $0_3^+$ states is also very large, $B(E2; 2_2^+ \rightarrow 0_3^+)$ = 104 e$^2$fm$^4$. In Fig.~\ref{fig:rota_new_thsr}, the calculated energy levels are plotted as a function of $J(J+1)$, together with the experimental data. There have been attempts to interpret this as a rotational band of a spinning triangle as this was successfully done for the ground state band \cite{Freer}. However, the situation may not be as straightforward as it seems. 

Because the two transitions $ 2_2^+ \rightarrow 0_2^+$ and $ 2_2^+ \rightarrow 0_3^+$    are of similar magnitude, no clear band head can be identified. It was also concluded in  Ref. \cite{Chern} that the states $0_2^+, 2_2^+, 4_2^+$ do not form a rotational band. The line which connects the two other hypothetical members of the rotational band, see Fig. \ref{fig:rota_new_thsr}, has  a slope which points to somewhere in between of the 0$^+_2$ and 0$^+_3$ states. However, to conclude from there that this gives raise to  a rotational band, may be premature. One should also realise that the $0_3^+$ state is strongly excited from the Hoyle state by monopole transition whose strength is obtained from the extended THSR calculation to be $M(E0;0_3^+ \rightarrow 0_2^+)$ = 35 fm$^2$. So, the $0_3^+$ state seems to be a state where one $\alpha$ particle has been lifted out of the condensate to the next higher S level with a node. This is confirmed in Fig. \ref{probas-0-2-4-hoyle} where the probabilities, $S^2_{[I,l]}$, of the third $\alpha$ orbiting in an $l$ wave around a $^8$Be-like, two $\alpha$ correlated pair with relative angular momentum $I$, are displayed. One sees that except for the $0_4^+$ state, all the states have the largest contribution from the $[0,l]$ channel. So, the picture which arises is as follows: in the Hoyle state, the three $\alpha$'s are all in relative 0S states with some $\alpha$-pair correlations (even with $I\ne 0$, see \cite{Rimas, Ishikawa}), responsible for emptying the $\alpha$ condensate by 20-30$\%$. This 0S-wave dominance, found by at least half a dozen of different theoretical 
works, see, e.g.,  \cite{OCM-Hori,Uegaki, Kami, Suzuki, Yam05,  Rimas, Ishikawa}, is incompatible with the picture of a rotating triangle. As mentioned, from the calculation the $0_3^+$ state is one where an $\alpha$ particle is in a higher nodal S state and the $0_4^+$ state is built out of an $\alpha$ particle orbiting in a D-wave around a (correlated) two $\alpha$ pair, also in a relative 0D state, see Fig. \ref{probas-0-2-4-hoyle}. The $2_2^+$ and $4_2^+$ states are a mixture of various relative angular momentum states (Fig. \ref{probas-0-2-4-hoyle}). Whether they can be qualified as members of a rotational band or, may be, rather of a vibrational band or a mixture of both, is an open question.
In any case, indeed,  they are very strongly connected by $B(E2)$ transitions: $B(E2;4_2^+ \rightarrow 2_2^+) = 560$ $e^2$fm$^4$. 
In this context, it should also be pointed out that Suhara {\it et al.} \cite{En'yo3} have recently investigated the effect of the possibility that the $\alpha$'s get polarized and/or deformed (the $\alpha$ breaking effect) in the $\alpha$ gas states. Apparently this has a substantial influence on the $\alpha$ gas states above the Hoyle state. For instance, it is claimed in that paper that the $0^+_3$ state is now the band head of the 'Hoyle band'. However, to validate this conclusion, one would like to see how well this approach reproduces the inelastic form factor to the Hoyle state.\\
Very recently, an interesting further contribution to the subject appeared \cite{Nakamura} where the authors reproduce some $\alpha$ gas states located just above the Hoyle state on grounds that the Hoyle state is an $\alpha$ condensate state. Only one adjustable parameter is involved. However, the used approach is novel and must further be tested before any firm conclusions can be drawn.

One may also wonder why, with the extended THSR approach, there is a relatively strong difference between the calculated and experimental, so-called Hoyle band? This may have to do with a deficiency inherent to the THSR wave function which so far has not been cured ( there may be ways to do it in the future). It concerns the fact that with THSR (as, by the way, with the Brink wave function), it is difficult to include the spin-orbit potential. This has as a consequence that the first $2^+$ and first $4^+$ states are quite wrong (too low) in energy because the strong energy splitting between $p_{3/2}$ and $p_{1/2}$ states is missing. This probably has a repercussion on the position of the second $2^+$ and $4^+$ states. This can be deduced from the OCM calculation by Ohtsubo {\it et al.} ~\cite{Ohtsubo} also shown in Fig. \ref{fig:rota_new_thsr} where the  $2^+$ and $4^+$ states of the ground state rotational band have been adjusted to experiment with a phenomenological force and, thus, the positions of the  $2^+$ and $4^+$ states of the so-called Hoyle-band are much improved. Additionally, this may also come from the fact that with this extended THSR wave function  a different force has to be adopted. Such investigations are under way.\\
One should  also mention that the excited $\alpha$ cluster states discussed above have a width much larger ($\sim$ 1 MeV) than the Hoyle state ($\sim$ 1 eV). Nevertheless, those widths are sufficiently small, so that the corresponding states can be treated in bound state approximation.\\

\begin{figure}
\begin{center}
\includegraphics[width=6cm]{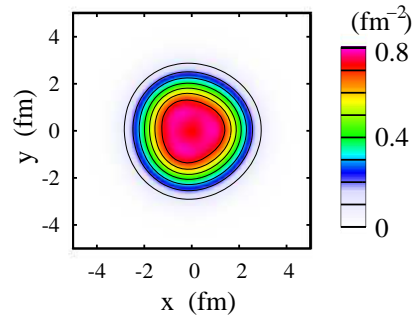}
\caption{\label{Yoshiko}
Intrinsic density distribution of the $^{12}$C ground state from a mean field calculation (we thank Y. Kanada-En'yo for providing this figure).}
\end{center}
\end{figure}

\begin{figure}
\begin{center}
\includegraphics[scale=0.6]{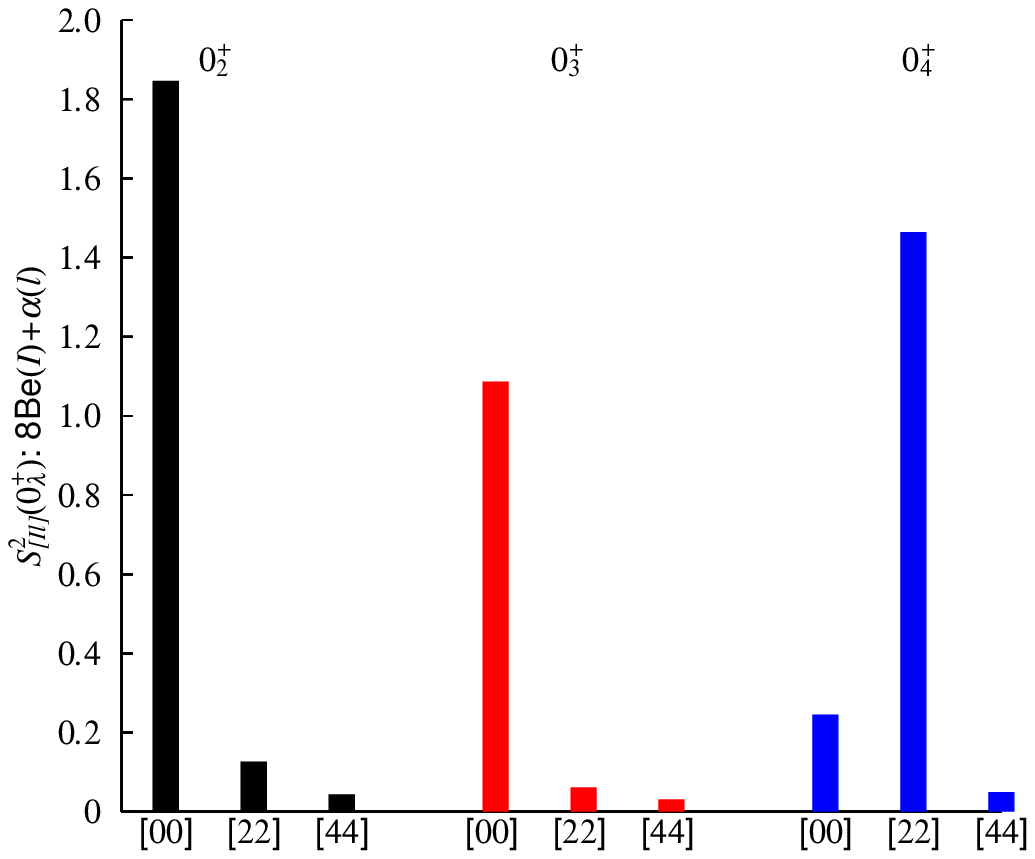}
\includegraphics[scale=0.6]{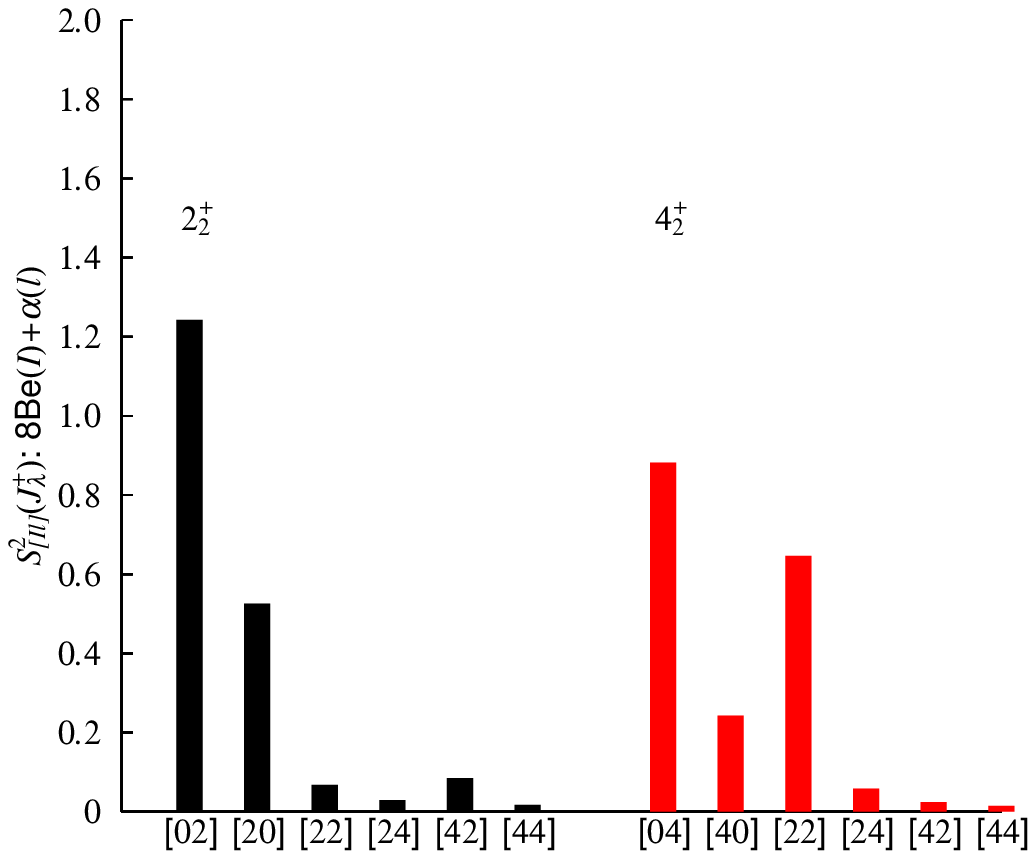}
\caption{\label{probas-0-2-4-hoyle}
Probability distributions for various components in the Hoyle and excitations of the Hoyle state (from \cite{Funaki}).}
\end{center}
\end{figure}

Let us dwell a little more on the ground state band.
In \cite{Gai, Freer} an algebraic model by  Iachello {\it et al.} \cite{Iachello}, originally due to Teller \cite{Teller}, was put forward and used on the hypothesis that the ground state of $^{12}$C has an equilateral triangle structure. The model then allows to calculate the rotational-vibrational (rot-vib) spectrum of three $\alpha$ particles. Notably a newly measured 5$^-$ state very nicely fits into the rotational band of a spinning triangle. This interpretation is also reinforced by the fact that for such a situation the 4$^+$ and 4$^-$ states should be degenerate what is effectively the case experimentally. In Fig. \ref{Yoshiko}, we show the triangular density distribution of the $^{12}$C ground state obtained from a pure mean field calculation.  This means a calculation without any projection on parity nor angular momentum. Therefore, symmetry is spontaneously broken into a triangular shape. The calculation is obtained under the same conditions as in \cite{Suhara}. However, in that work only figures with variation after projection are shown. This enhances the triangular shape. The Fig.\ref{Yoshiko} is unpublished. It must be said, however, that the broken symmetry to a triangular shape is very subtle and depends on the force used \cite{Suhara}. Mean field calculations with the Gogny force \cite{Girod2} and also with the relativistic  approach \cite{Ebran} do not show a spontaneous symmetry breaking into a triangular shape. It also should be mentioned that very recently Cseh {\it et al.} \cite{Cseh} have shown that the states of the ground state band can also be explained with U3 symmetry. So, the shape of the ground state of $^{12}$C is still an open question but  a triangular form seems definitely a possibility. 

\section{Ab initio and Quantum Monte Carlo (QMC) approaches to the Hoyle state}

Very recently a break through in the description of the Hoyle state was achieved by two groups \cite{Pieper, Meissner} using Monte Carlo techniques. In \cite{Meissner} Dean Lee {\it et al.} reproduced the low lying spectrum of $^{12}$C, including the Hoyle state, very accurately with a so-called ab initio  lattice QMC approach starting from effective chiral field theory \cite{Elhatisari1, Elhatisari2}. The sign problem has been circumvented exploiting the fact that SU(4) symmetry for the $\alpha$ particles is very well fulfilled. This parameter free first principle calculation is an important step forward in the explanation of the structure of $^{12}$C. 
On the other hand, all quantities which are more sensitive to details of the wave function have so far either not been calculated (e.g., inelastic form factor to the Hoyle state) or the results are in quite poor agreement with the results of practically all other theoretical approaches. This, for instance, is the case for the rms radius of the Hoyle state which in \cite{Meissner} is barely larger than the one of the ground state whereas it is usually believed that the Hoyle state is quite extended. The authors of \cite{Meissner} remark themselves that higher order contributions to the chiral expansion have to be included to account for the size of the Hoyle state. Concerning the shape of the Hoyle state, the authors in \cite{Meissner} obtain an obtuse triangular arrangement of the three $\alpha$'s. This seems to be in contradiction with the finding of many theoretical investigations of the Hoyle state where a relative 0S-wave dominance is found, see, e.g., 
\cite{OCM-Hori,Uegaki, Kami, Suzuki, Yam05,  Rimas, Ishikawa}. 

\begin{figure}
\includegraphics[width=6cm,angle=-90]{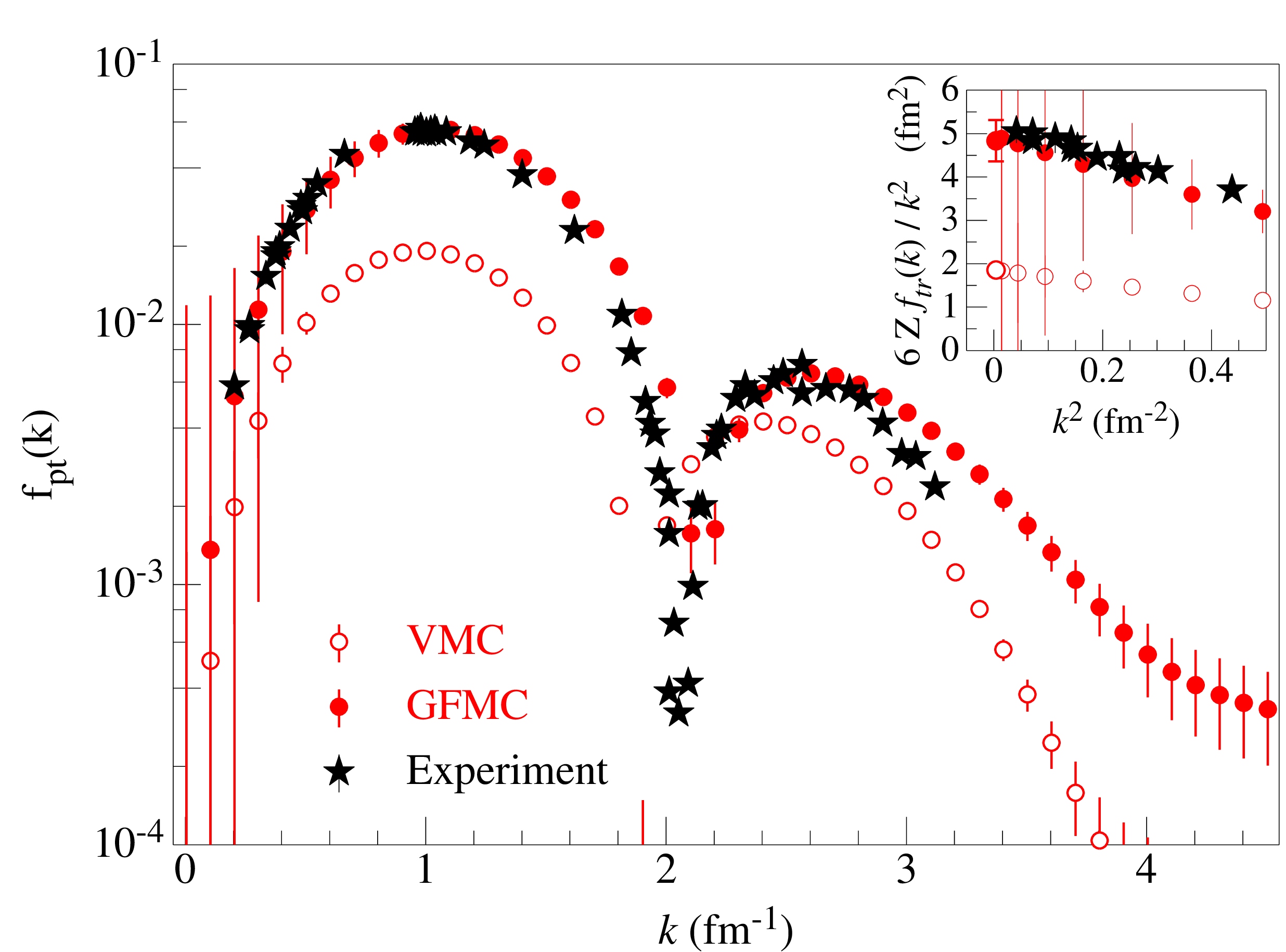}
\caption{\label{QMC}
Inelastic form factor from ground to Hoyle state from GFMC ~\cite{Pieper}, full circles.   
The open circles correspond to some approximate calculation, see \cite{Pieper}, and the black stars represent the experimental values \cite{Chern}.   }
\end{figure}

There also exist new Green's function Monte Carlo (GFMC)
results
with constrained path approximation using the Argonne v18 two-body and
Illinois-7 three-body forces, where the inelastic form factor for most of
the
experimental points is reproduced very accurately \cite{Pieper}, see Fig.~\ref{QMC}. In the insert of the upper panel, we see  that the rather precise experimental transition radius of 5.29$\pm$ 0.14 fm$^2$ given in \cite{Chern} is much better reproduced than in $\alpha$ cluster models (including the THSR model) which all yield an about 20\% too large value, see,e.g., \cite{To01}. This may also be the reason for the too slow drop off of the THSR density in the surface region, see Fig.\ref{Hoyle-density} below. The energy of the Hoyle state is with around 10 MeV in \cite{Pieper}  slightly worse than the one in \cite{Meissner}. In Fig.~\ref{Hoyle-density}, we compare the density of the Hoyle state (weighted with $r^2$) obtained with the THSR wave function and in \cite{Pieper}. We again see quite good agreement between both figures up to about 4 fm. For instance the kind of plateau between 1.5 and 4 fm seems to be very characteristic. It is, however, more pronounced in the GFMC calculation than from THSR. For a better appreciation, we repeat the results of THSR separately in the lower panel of Fig.\ref{Hoyle-density}. Beyond 4 fm, the density in \cite{Pieper} falls off more rapidly. As already mentioned, this may be due to the fact that the GFMC results are more accurate for small $q$-values. At any rate, the outcome of the three calculations in \cite{rev5, Kami, Pieper} is so close that it is difficult to believe that results for other quantities should be qualitatively different when calculated with the GFMC technique. This should, for instance, hold for the strong proportion of relative S-waves between the $\alpha$'s found with the other approaches discussed above.

\begin{figure}
\includegraphics[width=6.5cm]{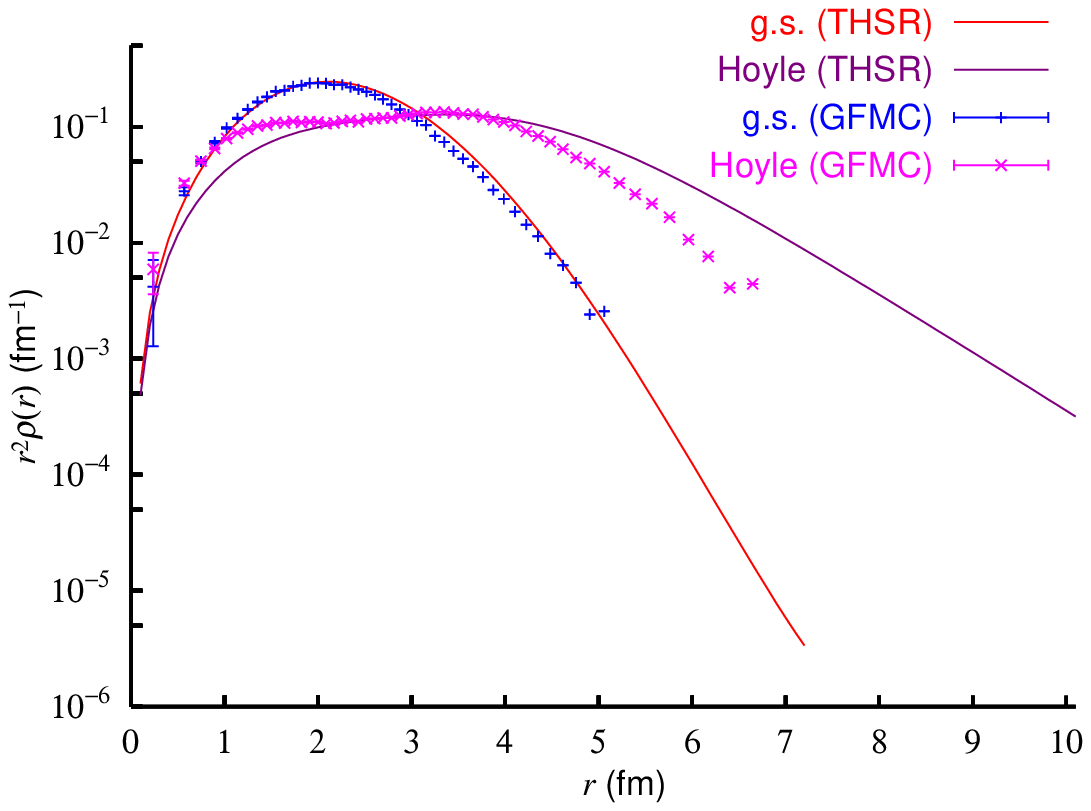}
\includegraphics[width=6.5cm]{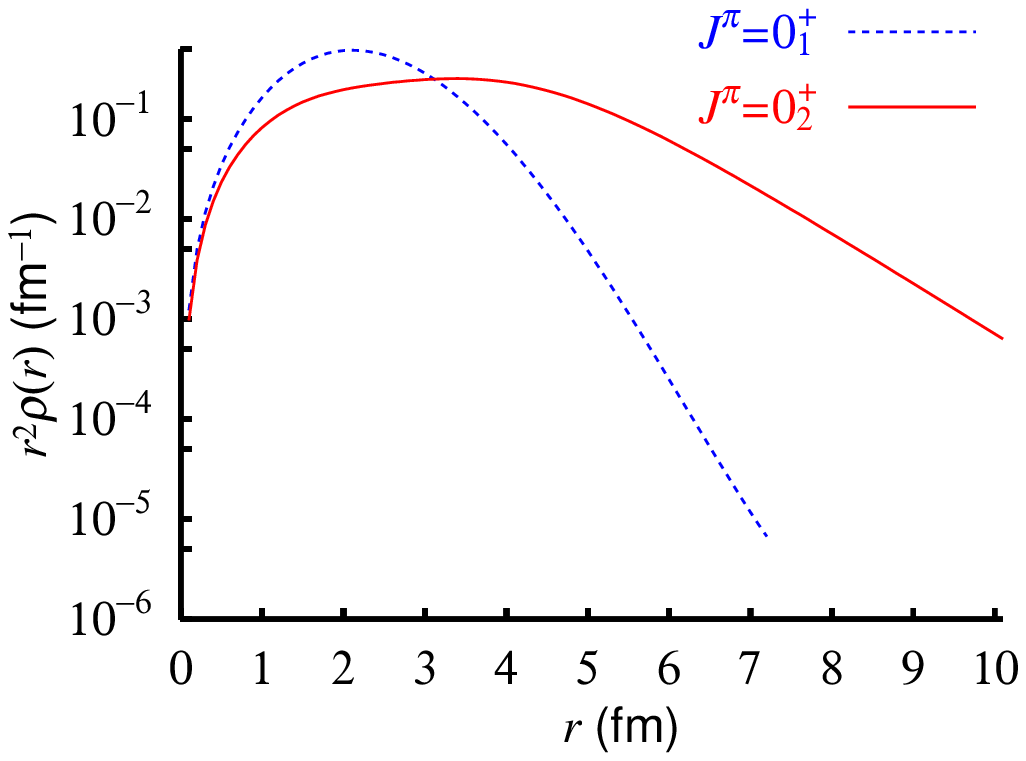}
\caption{Densities of the Hoyle state with GFMC ~\cite{Pieper} (magenta diamonds), and of ground state (blue crosses), upper panel, and with THSR, full lines. In the lower panel, the results with THSR are repeated for better appreciation. }
\label{Hoyle-density}
\end{figure}

Also with the symplectic no-core shell model (NCSM), there is now great progress in the description of cluster states including the Hoyle state \cite{Tobin, Dreyfuss, Dytrych}. The position of the Hoyle state and the second $2^+$ state in $^{12}$C are well reproduced in \cite{Dreyfuss}. The rms radius is with 2.97 fm on the lower side entailing a monopole transition which is quite a bit too low by about 40\%. Again what is missing is the inelastic form factor. As was pointed out several times, the very well measured inelastic form factor \cite{Chern} is highly sensitive to the ingredients of the wave function of the Hoyle state and it is mandatory that a theory reproduces this decisive quantity correctly.

\section{Alpha cluster states in $^{16}$O}

The situation with respect to $\alpha$ clustering was still relatively simple in $^{12}$C. There, one had to knock loose from the ground state one $\alpha$ particle to stay with $^8$Be which is itself a loosely bound two $\alpha$ state. So, immediately, knocking loose one $\alpha$ leads to the $\alpha$ gas, i.e., the Hoyle state. In the next higher self conjugate nucleus, $^{16}$O, the situation is already substantially more complex. Knocking loose one $\alpha$ from the ground state leads to $^{12}$C $+ \alpha$ configurations. Contrary to the situation in $^{12}$C, here the remaining cluster $^{12}$C can be in various compact states describable by the fermionic mean field approach before one reaches the four $\alpha$ gas state. Actually, as we will see, only the 6-th $0^+$ state in $^{16}$O is a good candidate for $\alpha$ particle condensation. This state is well known since long ~\cite{Selove} and lies at 15.1 MeV. The situation is, therefore, quite analogous to the one with the Hoyle state. The latter is about 300 keV above the 3$\alpha$ disintegration threshold. In $^{16}$O, the 4$\alpha$ disintegration threshold is at 14.4 MeV. Thus, the 15.1 MeV state is 700 keV above the threshold. Not so different from the situation with the Hoyle state. On the other hand the width of the Hoyle state is, like the one of $^8$Be, in the eV region, whereas the width of the 15.1 MeV state in $^{16}$O is 160 keV. This is large in comparison with the Hoyle state but still small considering that the excitation energy is about twice as high. It is tempting to say that the width is surprisingly small because the states to which it can decay, if we suppose that the 15.1 MeV state is an $\alpha$ condensate state, have radically different structure being either of the $^{12}$C$ + \alpha$ type with $^{12}$C in a compact form or other shell model states. Let us see what the theoretical approaches tell us more quantitatively.\\

\begin{figure}[h]
\begin{center}
\includegraphics[scale=1.]{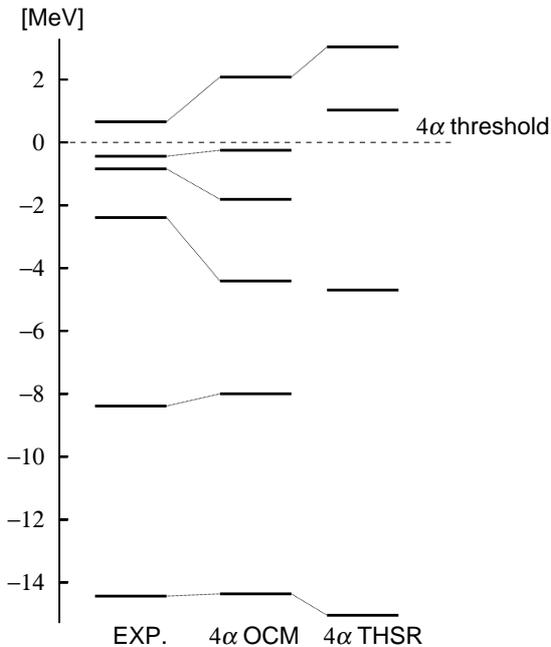}
\caption{Comparison of the $0^+$ spectra of $^{16}$O between experiment and theory (OCM, middle, and THSR, right).}
\label{fig:4a_levels}
\end{center}
\end{figure}

In the first application of THSR ~\cite{To01}, the spectrum was calculated not only for $^{12}$C but also for $^{16}$O. Four $0^+$ states were obtained. Short of two $0^+$ states with respect to the experimental situation if the highest state, as was done in ~\cite{To01} is interpreted as the 4$\alpha$ condensate state. Actually Wakasa, in reaction to our studies, has searched and found a so far undetected $0^+$ state at 13.6 MeV which in ~\cite{To01} was interpreted as the $\alpha$ condensate state. The situation with the missing of two $0^+$ states from the THSR approach is actually quite natural. In THSR the $\alpha$ particles are treated democratically whereas, as we just discussed, this is surely not the case in reality. The best solution would probably be, in analogy to the proposed wave function in (\ref{gen-thsr}), to introduce for each of the three Jacobi coordinates of the 4$\alpha$ THSR wave function a different $B$ parameter. This has not been achieved so far. As a matter of fact, the past experience with OCM is very satisfying. For example for $^{12}$C it reproduces also very well the Hoyle state, see Fig. 24 below. It was, therefore, natural that, in regard of the complex situation in $^{16}$O, first the more phenomenological OCM method was applied to obtain a realistic spectrum. This was done by Funaki {\it et al.} ~\cite{Fu08} . We show the spectrum of $^{16}$O obtained with OCM together with the result from the THSR approach and the experimental $0^+$ spectrum in Fig.~\ref{fig:4a_levels}. The modified Hasegawa-Nagata nucleon-nucleon interaction ~\cite{Hasegawa} has been used. We see that the 4$\alpha$ OCM calculation gives satisfactory reproduction of the first six $0^+$ states. Inspite of some quite tolerable discrepancies, this can be considered as a major achievement in view of the complexity of the situation. The lower part of the spectrum is actually in agreement with earlier OCM calculations ~\cite{Suzuki76, Suzuki2-76, Kato, Baye}. However, to reproduce the spectrum of the first six $0^+$ states, was only possible in extending considerably the configuration space with respect to the early calculations. Let us interpret the various states. The ground state is, of course, more or less a fermionic mean field state. The second state has been known since long to represent an $\alpha$ particle orbiting in an 0S wave around the ground state of $^{12}$C. In the third state an $\alpha$ is orbiting in a 0D wave around the first $2^+$ state in $^{12}$C. This $2_1^+$ state is well described by a particle-hole excitation and is, therefore, a non-clustered shell model state. The fourth state is represented by an $\alpha$ particle orbiting around the ground state of $^{12}$C in a higher nodal S-state. The fifth state is analysed as having a large spectroscopic factor for the configuration where the $\alpha$ orbits in a P wave around the first $1^-$ state in Carbon. The $0_6^+$ state is identified with the state at 15.1 MeV and as we will discuss, is believed to be the 4$\alpha$ condensate state, analogous to the Hoyle state.

\begin{figure}[h]
\begin{center}
\includegraphics[scale=0.7]{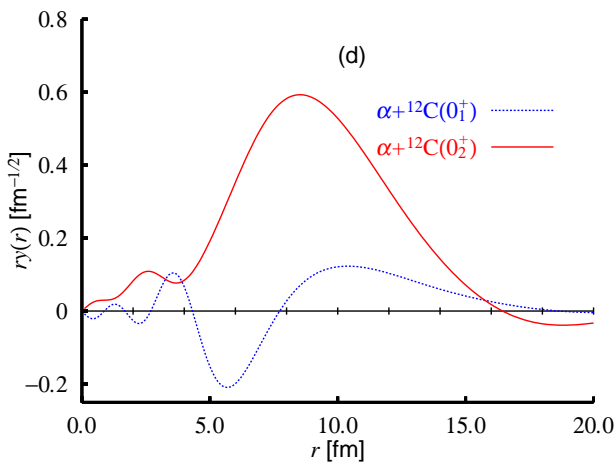}
\caption{Reduced width amplitudes in the two channels $\alpha + ^{12}$C($0^+_1$) (dotted curve) and $\alpha + ^{12}$C($0^+_2$) (solid curve) calculated with the THSR approach. }
\label{fig:4athsr_rwa}
\end{center}
\end{figure}

\begin{figure}[h]
\begin{center}
\includegraphics[scale=0.7]{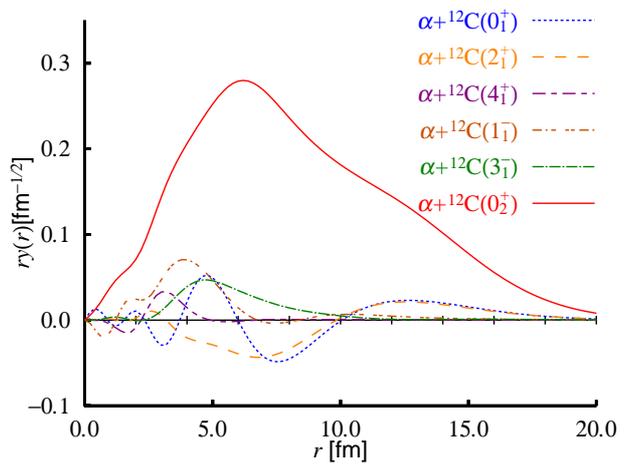}
\caption{ Reduced width amplitudes in the two channels $\alpha + ^{12}$C($0^+_1$) (dotted curve) and $\alpha + ^{12}$C($0^+_2$) (solid curve) calculated with the OCM approach. Also shown are the reduced width amplitudes of the other four $0^+$ states as indicated in the figure.}
\label{fig:4aocm_rwa}
\end{center}
\end{figure}

On the rhs of the spectrum we show in Fig.~\ref{fig:4a_levels}, the result with the THSR approach. As mentioned two states are missing. However, we will claim that at least the highest state and the lowest state, i.e., the ground state, have good correspondence between the OCM and THSR approaches. For this let us consider the so-called reduced width amplitude (RWA)
\begin{equation}
{\mathcal Y}_{[L, l]_J} = {\mathcal N}_{\mbox{RWA}}\langle [\frac{\delta(r'-r)}{r'^2}[\Phi_L(^{12}\mbox{C}), Y_l(\hat {\vek r}']_J\phi(\alpha)]|\Phi_J(^{16}\mbox{O})\rangle
\label{RWA}
\end{equation}
where $\Phi_L(^{12}\mbox{C})$ and $ \Phi_J(^{16}\mbox{O})\rangle $   are the states of $^{12}$C and $^{16}$O obtained by the THSR and OCM methods, respectively. The norm ${\mathcal N}_{\mbox{RWA}}$ is $\sqrt{\frac{16!}{12!4!}}$ and $\sqrt{\frac{4!}{3!1!}}$ for THSR and OCM, respectively. These RWA amplitudes are very close to spectroscopic factors and tell to which degree one state can be described as a product of two other states. In Fig.~\ref{fig:4athsr_rwa} and Fig.~\ref{fig:4aocm_rwa}, we show these amplitudes for the highest state with THSR and with OCM, respectively. Besides an overall factor of about two, we notice quite close agreement. The large spatial extension of the highest state in both calculations can qualify this state of being the 4 $\alpha$ condensate state. The other two states in THSR may describe the intermediate states in some average way.\\

\begin{table*}[htbp]
\caption{Binding energies $E$ measured from the $4\alpha$ threshold energy, r.m.s. radii $R_{\rm rms}$, monopole matrix elements $M(E0)$, and 
$\alpha$-decay widths $\Gamma$, in units of MeV, fm, fm$^2$, and MeV, respectively.}\label{tab:4athsr}
\begin{tabular}{cccccccccccccc}
\hline\hline
 &  & \multicolumn{2}{c}{THSR} &   &   &  \multicolumn{2}{c}{$4\alpha$ OCM} &  &  & \multicolumn{2}{c}{Experiment} &  \\
 & $E$ & $R_{\rm rms}$ & $M(E0)$ & $\Gamma$ &  $E$ & $R_{\rm rms}$ & $M(E0)$ & $\Gamma$ &   $E$ & $R_{\rm rms}$ & $M(E0)$ & $\Gamma$ \\
\hline
$0_1^+$ & $-15.1$ & 2.5 &  &   &   $-14.4$ & 2.7 &  &   &  $-14.4$ & $2.71$ &  &  \\
$0_2^+$ & $-4.7$ & 3.1 & 9.8 &    &  $-8.00$ & 3.0 & 3.9 &   &   $-8.39$ & &  $3.55$ &  \\
$0_3^+$ &  &  &  &  &  $-4.41$ & 3.1 & 2.4 &  &    $-2.39$& &  $4.03$ &  \\
$0_4^+$ & $1.03$ & 4.2 & 2.5 & 1.6 &   $-1.81$ & 4.0 & 2.4 & $\sim 0.6^a$ &   $-0.84$ &  &  & 0.6 \\
$0_5^+$ &  &  &  &   & $-0.25$ & 3.1 & 2.6 & $\sim 0.2^a$ &   $-0.43$ &  & $3.3$ & 0.185 \\
$0_6^+$ & $3.04$ & 6.1 & 1.2 & 0.14 &   $2.08$ & 5.6 & 1.0 & $\sim 0.14^a$  & $0.66$ &  &  & 0.166 \\
\hline\hline
\end{tabular} \\
a: Present calculated values taken from Ref.~\cite{yamada12} are larger than those shown in Ref.~\cite{Fu10} by a factor $\sqrt{4!/(3 !1 !)}^2=4$, 
since the factor $\sqrt{4!/(3 !1 !)}$, which should be added to the RWA for the $4\alpha$ OCM, was missing in Ref.~\cite{Fu10}.
\end{table*}

We show in Table~\ref{tab:4athsr} the comparison of energy $E$, rms radius $R_{\mbox{rms}}$, monopole matrix element to the ground state $M(E0)$, and $\alpha$-decay width $\Gamma$ between THSR, OCM, and experimental data. The $\alpha$-decay width is calculated based on the $R$-matrix theory. The most striking feature in Table~\ref{tab:4athsr} is the fact that the decay-widths of the highest state agree perfectly well between theory and experiment. For instance, the two theoretical approaches are quite different. So, this good agreement, very likely, is not an accident and shows that the physical content of the corresponding wave functions is essentially correct, that is a very extended gas of 4$\alpha$ particles. The result for the occupation probability given in ~\cite{Fu08, Fu10} shows that again the 15.1 MeV state can, to a large percentage, interpreted as a 4$\alpha$ condensate state, i.e., as a Hoyle analog state. There are also experimental indications that the picture of Hoyle excited states which we have discussed above repeats itself, to a certain extent, for $^{16}$O ~\cite{It14}. If all this will finally be firmly established by future experimental and theoretical investigation, this constitutes a very exciting new field of nuclear physics.

 It should also be mentioned that recently a calculation with the AMD approach by Kanada En'yo \cite{En'yo2} has mostly confirmed the results 
given in \cite{Fu08}, see also \cite{yamada12}. In \cite{Ulf2} the ground state and first excited $0^+$ states of $^{16}$O has been calculated with the lattice QMC approach with good success concerning the position. However, no excited states around the $\alpha$ disintegration threshold have been obtained as yet.

\section{Summary of approaches to Hoyle and Hoyle-analog states: the $\alpha$ condensation picture, where do we stand after 15 years?}

As already mentioned several times, the hypothesis that the Hoyle state and other Hoyle-like states are to a large extent $\alpha$ particles condensates has started with the publication of Tohsaki et al. in 2001 ~\cite{To01}. It got a large echo in the community. After 15 years, it is legitimate to ask the question what remains from this hypothesis. To this end, let us make a compact summary of all the approaches which are dealing or have dealt in the past with the Hoyle or Hoyle like states discussing the for or contra of the $\alpha$ condensate picture.\\

The first correct,  nowadays widely accepted point of view, has been given in the work of Horiuchi {\it et al.} in 
1974 ~\cite{OCM-Hori}. For the first time, employing the OCM approach, it was concluded that the Hoyle state is a state of three $\alpha$ particles interacting weakly in relative 0S-states. From there to jump to the idea of the Hoyle state being a condensate, it is only a small step. Next came the fully microscopic approaches solving RGM  respectively GCM equations by Kamimura {\it et al.} ~\cite{Kami} and Uegaki {\it et al.} ~\cite{Uegaki} for $^{12}$C. Concerning the Hoyle state the conclusions were the same as the one of Horiuchi. We cite from Uegaki {\it et al.} ~\cite{Uegaki}: {\it In a number of excited states which belong to the new ``phase'', the $^{12}$C nucleus should be considered to dissociate into 3 $\alpha$ clusters which interact weakly with each other and move almost freely over a wide region}. And further: {\it The $0_2^+$ state is the lowest state which belongs to the new ``phase'', and could be considered to be a finite system of $\alpha$-boson gas}. This is practically the same language as we use now with the THSR approach, only that we use the more modern term of 'Bose-condensation' which came much into vogue after the advent of Bose-Einstein Condensation (BEC) in cold atom physics ~\cite{String}. We want to stress the point that these early OCM, RGM, and GCM approaches are not just any sketchy model calculations for $\alpha$ clustering. On the contrary, they are very powerful and even to day not by-passed theories for the 12 nucleon problem of $^{12}$C. The point of THSR is that a more direct ansatz of the $\alpha$ particle condensation type is made which at the same time makes the numerics less heavy and what allowed to confirm the $\alpha$ condensate picture. Otherwise, as it was mentioned above, the {\it squared} overlap of a THSR wave function with, e.g., the one of Kamimura {\it et al.} is close to 100 $\%$! The merit of THSR also is that the hypothesis of $\alpha$ gas states being a general feature in self-conjugate nuclei has been advanced for the first time.\\

Let us continue with the enumeration of theoretical descriptions of the Hoyle state. In 2007 Kanada-En'yo achieved to confirm the Hoyle state interpretation of ~\cite{Kami, Uegaki} with the AMD method which does not contain any preconceived element of $\alpha$ clustering.
Later  in 2007 Chernykh {\it et al.} achieved the same with a variant of AMD, the so-called ' fermionic molecular dynamics' (FMD) ~\cite{Chern}. Also the inelastic form factor was calculated with reasonable success. With not much risk to be wrong, one may say that any microscopic theory which reproduces without adjustable parameters the inelastic form factor (see our discussion about this in Sect.10), implicitly deals with a wave function which has the same or very close properties as the one of Kamimura {\it et al.} and Uegaki {\it et al.} and, thus, as the one of THSR. There are also the pure bosonic approaches which put all the antisymmetrisation and Pauli principle effects into an effective boson-boson interaction. The most recent approaches of this type are the ones of Lazauskas {\it et al.} \cite{Rimas} and of Ishikawa {\it et al.} ~\cite{Ishikawa}. Both studies reproduce the Hoyle state quite well. In ~\cite{Ishikawa} also the bosonic occupations have been calculated with about 80$\%$ occupancy of the 0S state, similar to what was obtained earlier in ~\cite{Yam05}. Lazauskas {\it et al.} \cite{Rimas} did not calculate the bosonic occupation numbers but concluded that in the Hoyle state wave function pairs of bosons are to 80 $\%$ in a relative two boson 0S configuration. Since there is a strong correlation between relative 0S states of two bosons in the three boson wave function and the 0S bosonic occupancy, one may say that the works of Lazauskas {\it et al.} and the one of Ishikawa {\it et al.} give mutually consistent results. Ishikawa {\it et al.} also calculate the simultaneous 3$\alpha$ decay versus the two body decay into $\alpha + ^8$Be. They find, in agreement with other estimates ~\cite{Freer2, Kirse} that the three body decay with respect to the two body one is suppressed by a factor of at least 10$^{-4}$. This, however, does not speak against the $\alpha$ condensation interpretation of the Hoyle state. It only states that the three body decay is much suppressed with respect to two body decay what is a quantity difficult to calculate from first principles. As mentioned, very recently there exists a GFMC result from  Pieper {\it et al.} with very good reproduction of the inelastic form factor ~\cite{Pieper}. The quality thereof is comparable to the one obtained by RGM ~\cite{Kami}, GCM ~\cite{Uegaki}, and THSR ~\cite{To01}, besides for the limit of small momenta where GFCM yields about 20\% better results. As we mentioned already, if one could calculate with GFMC the bosonic occupancies, the results very likely would be in agreement with the ones mentioned above. Lattice QMC calculations give excellent results for the low lying part of the spectrum of $^{12}$C but a calculation of the inelastic form factor is missing \cite{Meissner}.\\

 An algebraic approach put forward by Iachello et al. ~\cite{Iachello}, originally due to Teller {\it et al.},    was published recently by Freer {\it et al.} ~\cite{Gai, Freer}. The model is based on the assumption that the $^{12}$C ground state is an equilateral triangle formed by three $\alpha$ particles and that this configuration can undergo coupled rotational-vibrational excitations. Indeed the model can very well explain the ground state band. This interpretation of equilateral triangle is reinforced by the fact that for such a situation the 4$^+$ and 4$^-$ states should be degenerate what is effectively the case experimentally ~\cite{Freer}. On the other hand Cseh {\it et al.} \cite{Cseh} showed that also a U3 symmetric model can equally well describe the states of the  ground state band. The authors of \cite{Freer} then tried to repeat their reasoning tentatively for the 'rotational' band with the Hoyle state as band head. However, as we discussed in Sect. 11, the fact that the Hoyle state forms a band head is not at all established since the $0_3^+$ could  be the band head as well in view of the fact that its $B(E2)$ transition to the $2_2^+$ state is of the same order as the one from the Hoyle state. So no well defined band head exists. The $2_2^+$ and $4_2^+$ states lie as a function of $J(J+1)$ on a straight line which points to somewhere in between the $0^+_3$ and $0^+_4$ states. Also the inelastic form factor to the Hoyle state is underestimated by an order of magnitude \cite{Bijker} with the algebraic approach. In \cite{En'yo3} it is claimed that including '$\alpha$ breaking effects', the $0^+_3$ state becomes the band head of the Hoyle band. Thus, it seems to us that the rotational band interpretation of a hypothetical 'Hoyle band' is on uncertain grounds. More theoretical and experimental investigations are certainly necessary to fully elucidate the situation.\\

\begin{figure}
\includegraphics[width=7.5cm]{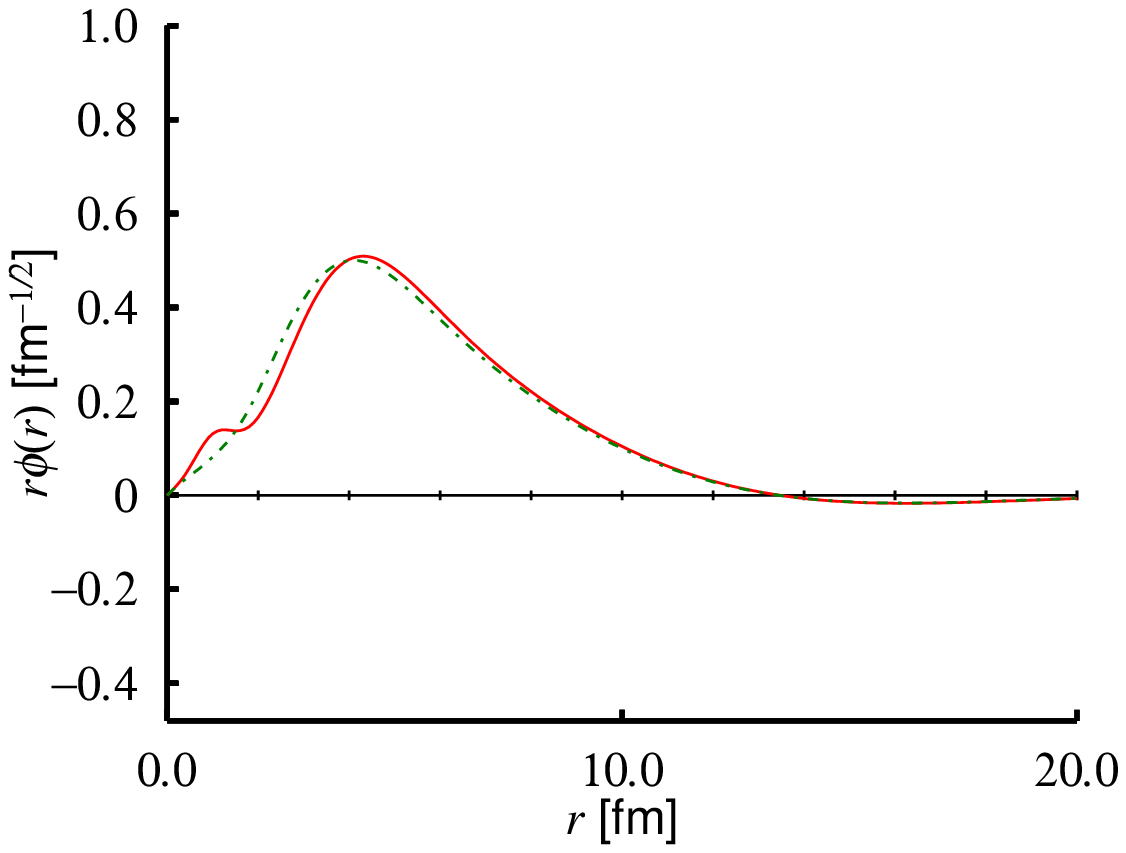}
\includegraphics[width=7.5cm]{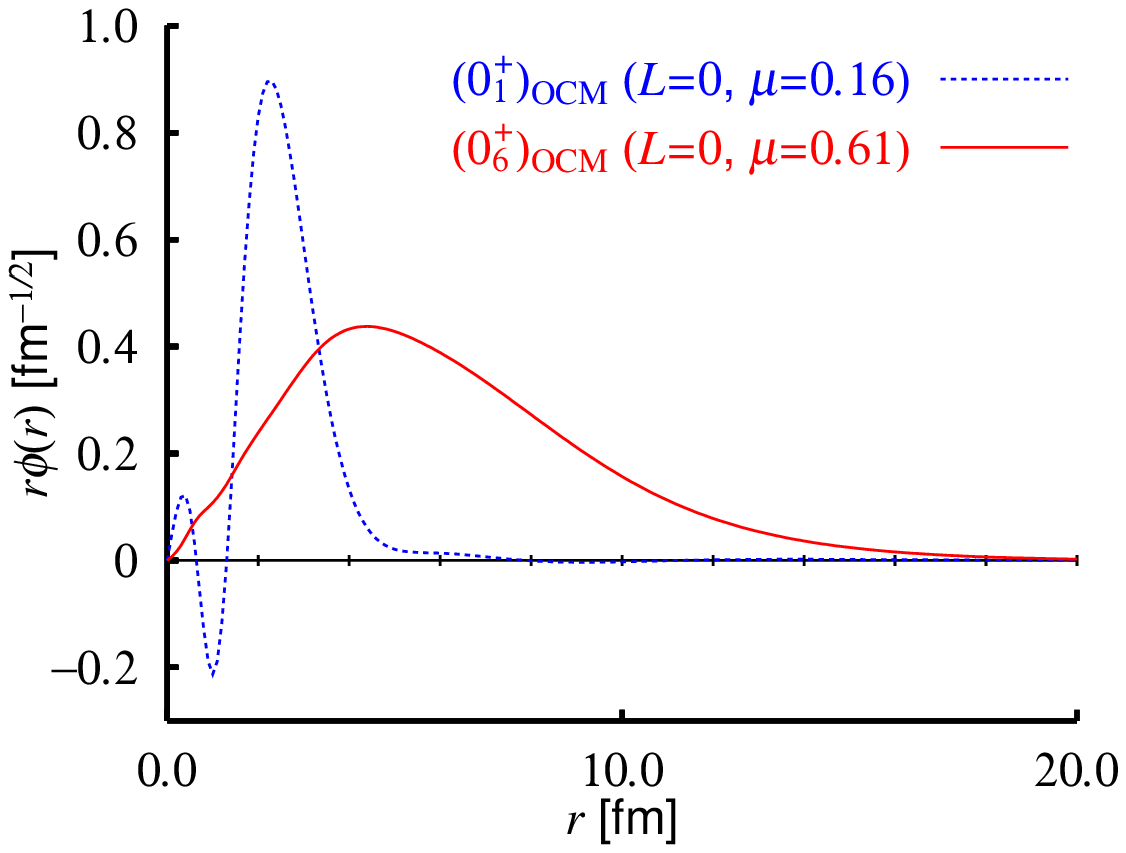}
\caption{\label{alpha-wavefcts} Comparison of single $\alpha$ particle wave functions in the condensate states of $^{12}$C (upper panel) and $^{16}$O (lower panel). One should remark the similarity of both wave functions (up to a scale factor). The dotted line in the upper panel is a best fit of a Gaussian to the calculated curve (full line). In the lower panel, the dotted line represents the single $\alpha$ particle wave function in the ground state. The figures are taken from \cite{16O}.}
\end{figure}

Last but certainly not least, let us mention $\alpha$ condensation in nuclear matter. Nobody contests the fact that infinite matter at low density becomes unstable with respect to cluster formation. A good candidate for a cluster phase is certainly given by $\alpha$ particles. As a matter of fact, as with pairing, quartetting has started in infinite matter with the work of R\"opke {\it et al.} ~\cite{Sogo1, Sogo2}. We have learned in Section 2 about the particular features of quartet condensation versus pair condensation. Most importantly we should remember that quartet condensation, contrary to what happens with pairing, only exists at low density where the chemical potential is still negative, i.e., the quartets are still bound (BEC). There does not exist a long coherence length, weak coupling phase for quartets. In other words quartet condensation only exists for densities where they do not overlap strongly. Therefore it is legitimate to think that the low density Hoyle and other Hoyle-like states are just a finite size manifestation of what happens in infinite matter. The dense ground states of nuclei cannot be considered as an $\alpha$ condensate.\\
The existence of $\alpha$ cluster condensation was critisized by Zinner and Jensen ~\cite{Zinner}. The main arguments are that the De Broglie wave length is too short and that the $\alpha$ condensates decay too fast (have a too large width). However, in ~\cite{Funaki-WvO} it was demonstrated that the de Broglie wave length is by factors larger than the extension of, e.g., the Hoyle state and in what concerns the decay of $\alpha$ condensate states, we argued already that the life times (widths) of those states are much longer 
(smaller) than what could be expected from their excitation energy. The width of a condensate merely is small because of the exotic structure of the condensate states having very  little overlap with states underneath. A good example is the sixth $0^+$ state at 15.1 MeV in $^{16}$O which has a width of only 160 keV (see \cite{16O} for an accurate estimate of this width). In \cite{Zinner} it was also stated that a condensate wave function should cover the whole nuclear volume what is actually the case with the large values of the $B$ parameters in the THSR wave functions. Another criterion namely that, besides a trivial norm factor, the condensate wave function should not change from one nucleus to the other is also very well fulfilled \cite{Funaki-WvO}, see Fig.\ref{alpha-wavefcts}.  The bosonic occupation numbers which in our mind constitute the best signature of condensate states were not calculated in ~\cite{Zinner}.\\

In conclusion of this section, we can say that several consistent and reliable microscopic approaches are in favor of the $\alpha$ condensate interpretation of the Hoyle state, either from direct calculations of the 
occupancies ~\cite{Suzuki} \cite{Yam05} or from the fact that wave functions, obtained from different approaches, are mutually consistent in as far as their squared overlaps approach the 100 $\%$. We again stress the point that a good reproduction of  the inelastic form factor of the Hoyle state (and other Hoyle-like states, to be measured in the future) is absolutely necessary for a a theory to be reliable. We do not see any work which clearly speaks against the condensate interpretation. $\alpha$ particle condensation is therefore a very useful new concept. Should the condensate picture be further confirmed by future studies, e.g., by the Monte Carlo approaches, this constituted a very exciting and rich novel feature of nuclear physics revealing that both Bose and Fermi gases can exist, at least in self conjugate nuclei, on equal footing.

\section{Alpha-type of correlations in ground states.}

\begin{figure*}
\begin{center}
\includegraphics[scale=0.42]{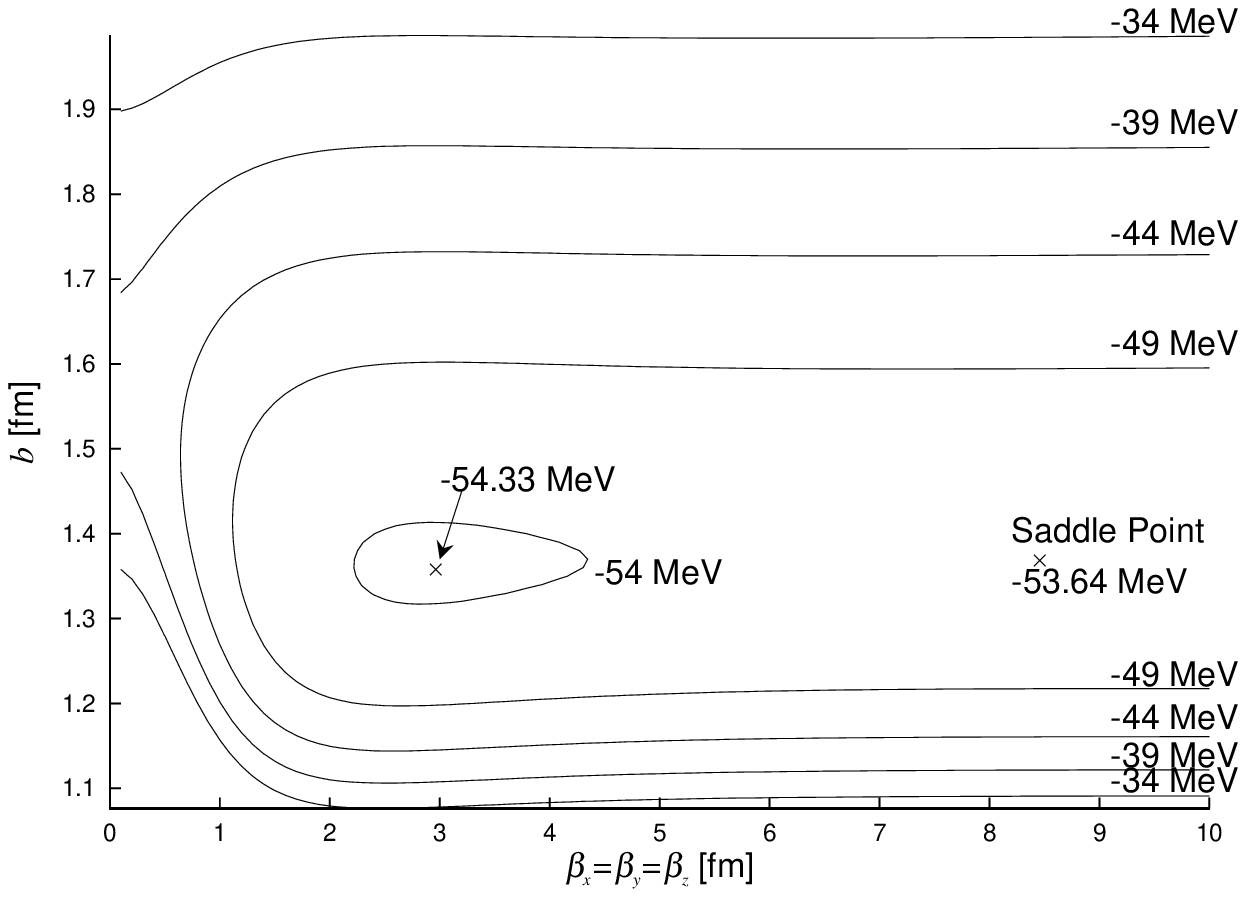}
\includegraphics[scale=0.44]{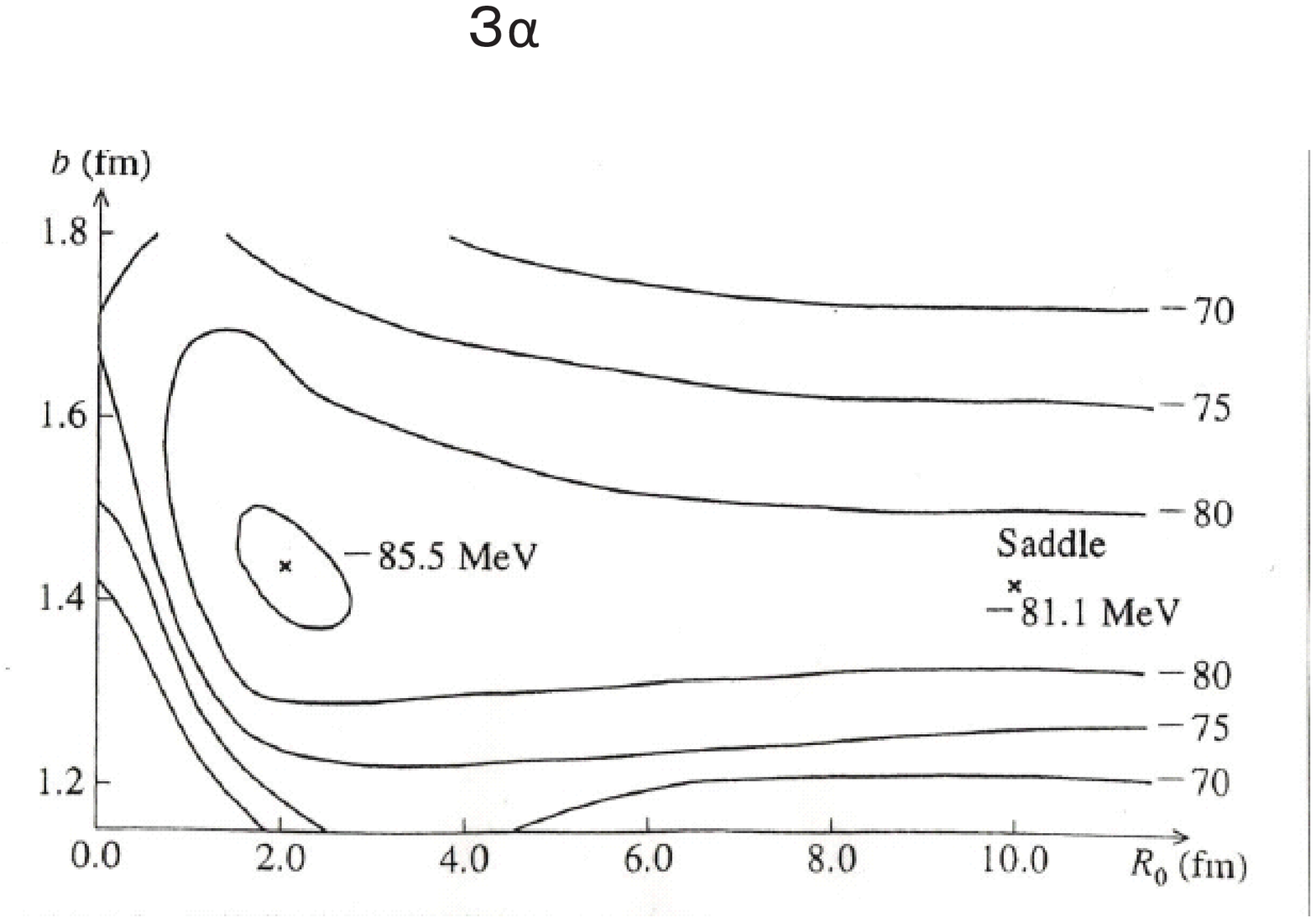}
\end{center}
\hspace{0.5cm}
\begin{center}
\includegraphics[scale=0.41]{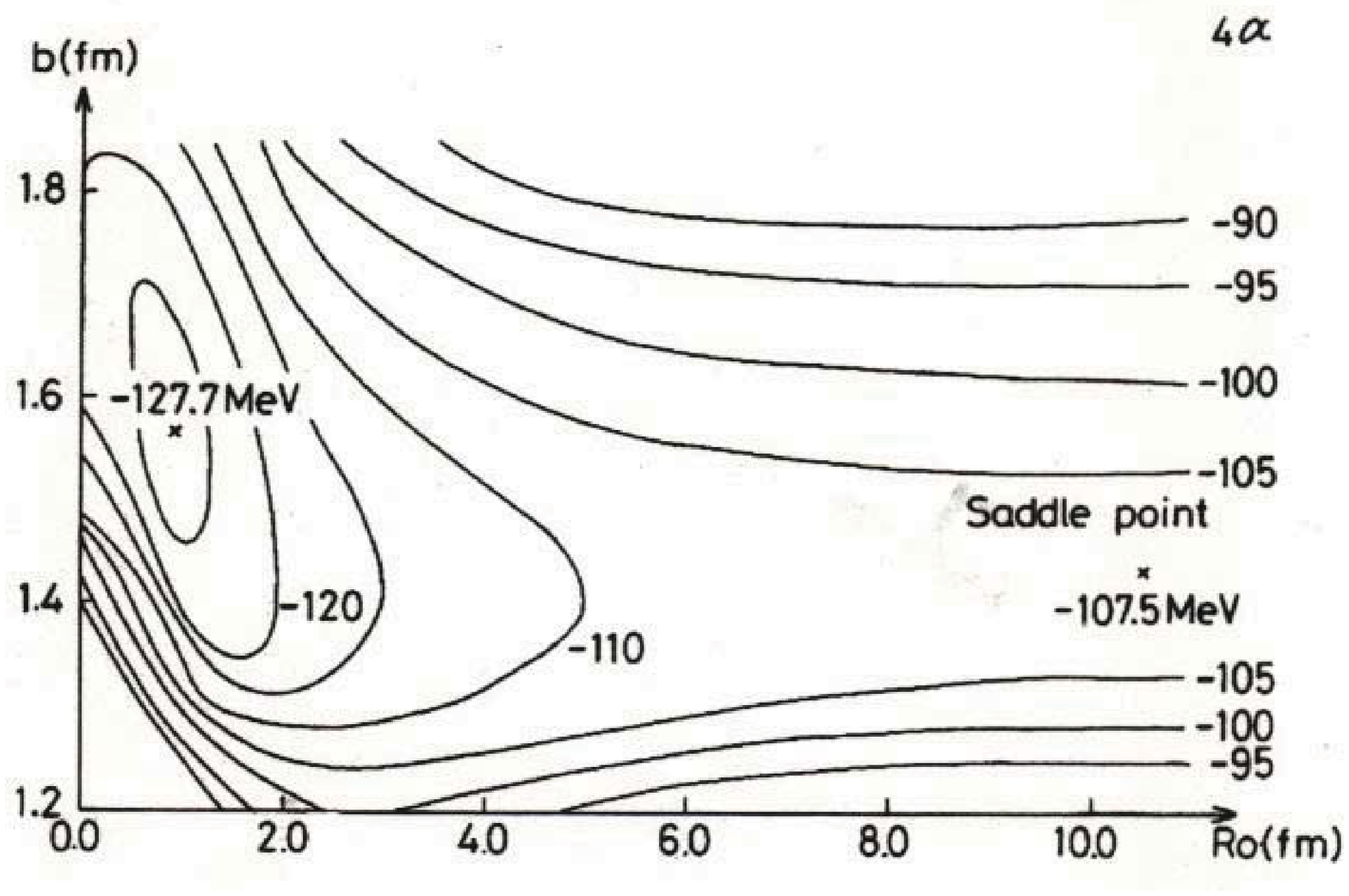}
\includegraphics[scale=0.41]{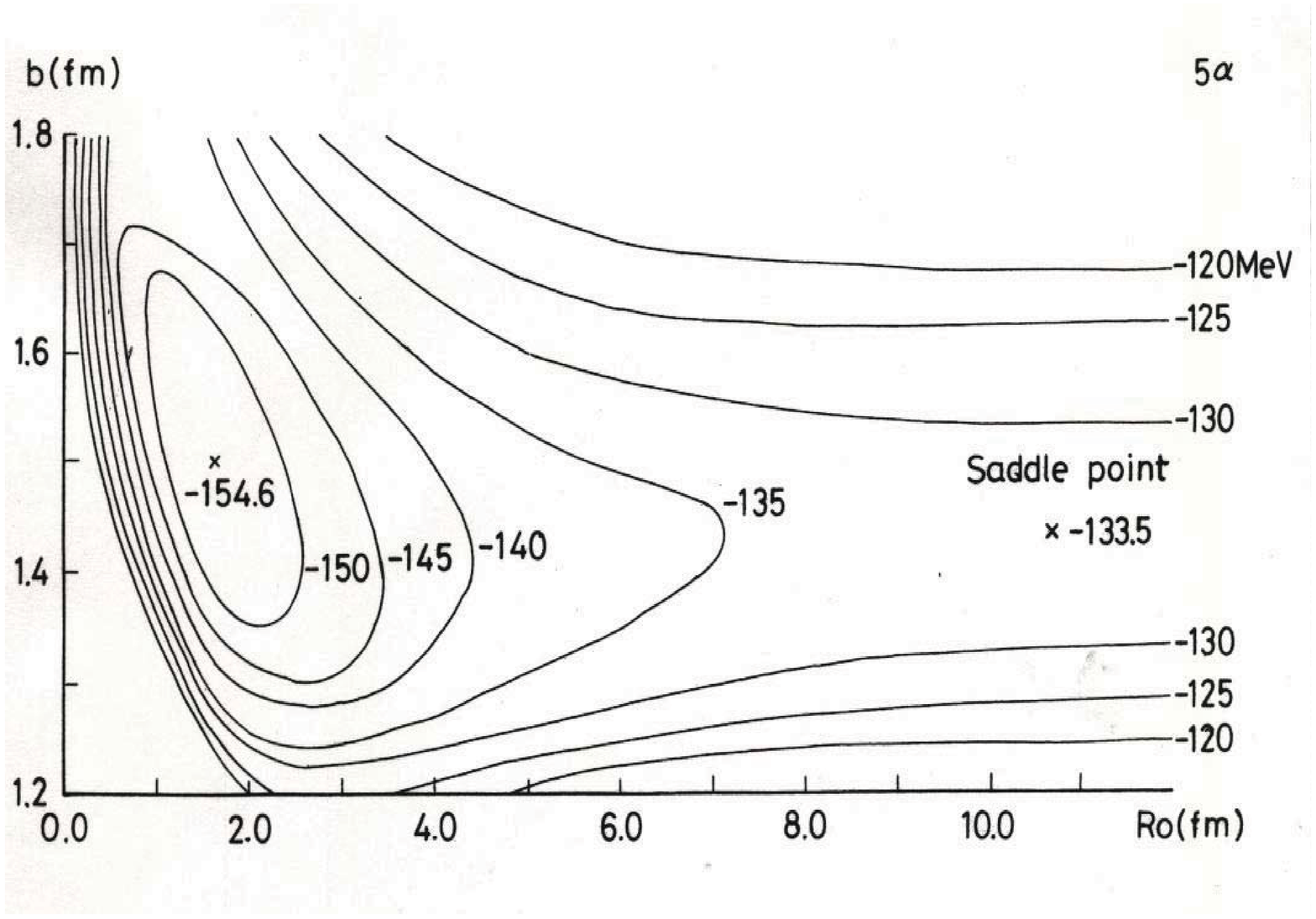}
\end{center}
\caption{Energy contourlines of $^8$Be, $^{12}$C, $^{16}$O, $^{20}$Ne in the space of the two width parameters $b, B$ of the THSR wave function $(B^2 = b^2 + 2R^2_0)$.}
\label{contour}
\end{figure*}

So far, we mostly have considered strong $\alpha$ correlations in the Hoyle-family of states. However, even in the ground states of the lighter self-conjugate nuclei non-negligeable $\alpha$-type of correlations are revealed from the calculations. Since in the ground states the $\alpha$'s strongly overlap, they are much deformed and extended and one cannot talk of real $\alpha$ particles anymore. Also these deformed entities, contrary to the pairing case as we have learned from the infinite matter section 2, cannot condensate because the 4-body in medium level densities pass through zero at the Fermi energy where the correlations should build up. Nevertheless that 4-body correlations are there can best be seen with our two parameter ($b$ and $B$) THSR wave function tracing the energy $E(b, B)= \langle \mbox{THSR}|H|\mbox{THSR}\rangle/ \langle \mbox{THSR}|\mbox{THSR}\rangle $ as a function of those width parameters. We recall that for $B=b$ we are in the Slater determinant limit, whereas for $B >> b$ a pure Bose condensate appears. \\
In Fig.~\ref{contour}, we show the contour maps of the energy landscape of $^8$Be,  $^{12}$C, $^{16}$O, and $^{20}$Ne (the $B$ parameter is related to $R_0$ by $B^2 = b^2 + 2R^2_0$). We clearly see that
energies are not minimal at the Slater limit $b = B$. Rather substantial energy is gained in going to higher $B$ values. For example for $^{16}$O, we have an energy gain from $\sim$ -120 MeV down to $\sim$ -126 MeV. That corresponds to  a gain of binding of $\sim$ 5$\%$. In $^{12}$C the gain in energy is somewhat stronger. However, we should not forget that this is a spherical calculation whereas $^{12}$C is deformed in its ground state. In $^{20}$Ne the situation is even more exaggerated because this nucleus has already a pronounced $^{16}$O $+ \alpha$ structure in its ground state which is not at all accounted for in a spherical calculation. Unfortunately, it is difficult to go to heavier $n\alpha$ nuclei with the THSR wave function because the explicit antisymmetrization becomes more and more difficult. In any case one can say that substantial $\alpha$ like correlations are present in the ground states of these nuclei which should not be neglected, e.g., when one establishes nuclear mass tables. In looking at those figures, one should be aware of the fact that $^8$Be is an exception in the series. It is the only nucleus which already in its ground state has a pronounced $\alpha$ particle configuration with low average density, similar to the one of the Hoyle state. We added $^8$Be for completeness but for considerations of systematics one should exclude this nucleus. It is an open but important question how these $\alpha$-like correlations evolve with mass number and/or with asymmetry. From Fig.\ref{contour}, it is difficult to draw any conclusions because their number is too small and also because $^{12}$C and $^{20}$Ne are deformed whereas those nuclei are constrained to sphericity in the calculations. Also in mean field calculations, e.g., $^{20}$Ne is not only deformed but shows a clear $^{16}$O $+ \alpha$ structure \cite{Ebran}. Therefore, the gain in energy with a cluster approach, as will be discussed in section 18, will not be very significant. However, at least for spherical nuclei smaller than $^{40}$Ca, such as Oxygen isotopes the explicit consideration of such correlations could, in principle, improve present mass tables which always show their greatest uncertainties precisely for lighter nuclei.\\
These extra $\alpha$-like correlations in the ground state also help to excite those nuclei to the Hoyle or Hoyle analog states, since the groundstates contain already the seeds of the $\alpha$'s.

\section{Alpha cluster states and monopole excitations in $^{13}$C}

It is a very intriguing issue to study what kinds of structures appear in $^{13}$C when an extra neutron is added into $^{12}$C, which has the shell-model-like ($0^+_1$), $3\alpha$-gas-like ($0^+_2$), higher-nodal $^8$Be($0^+$)+$\alpha$ cluster ($0^+_3$), and linear-chain-like ($0^+_4$) states as well as the $2^{+}_2$, $4^+$, $3^-$, and $1^-$ states etc., where the $0^+_3$, $0^+_4$, $2^+_2$, and $4^+$ states have been recently observed above the Hoyle state \cite{Freer, It13, It11}.
How do we identify cluster states in $^{13}$C? 
Isoscalar (IS) monopole transition strengths are very useful to search for cluster states in the low-energy region of 
light N $\equiv$ Z nuclei as well as in neutron rich nuclei~\cite{kawabata07,yamada08_monopole,yamada12}. 
The IS monopole excitations to cluster states in light nuclei are in general strong, comparable with single particle strengths.~\cite{yamada08_monopole,yamada12}. 
Their experimental strengths share about 20\% of the sum rule value in the case of $^{4}$He, $^{11}$B, $^{12}$C, and $^{16}$O etc.
They are very difficult to be reproduced by mean-field calculations~\cite{gambacurta10,ma97,bender03}.
Quite recently the enhanced monopole strengths in $^{12}$Be, predicted by the cluster model~\cite{ito08,ito12,ito14}, 
have been observed in the breakup-reaction experiment using a $^{12}$Be beam~\cite{yang14}, and the enhanced monopole state observed corresponds to the $0^{+}$ state at $E_x=10.3$~MeV with an $\alpha$+$^{8}$He cluster structure.   
Thus the IS monopole transition strengths indicate to be a good physical quantity to explore cluster states in light nuclei.
In the case of $^{13}$C, there is a long-standing problem that the {\rm C0} transition matrix elements to the $1/2^{-}_{2,3}$ states measured by the $^{13}$C $(e,e')$ experiments~\cite{wittwer69}, which are of the same order as that of the Hoyle state~\cite{ajzenberg93}, are very difficult to be reproduced within the shell-model framework~\cite{millener89}, where {\rm C0} denotes a longitudinal electric {\it monopole} transition.
There are no papers reproducing the experimental {\rm C0} matrix elements with the $(0+2)\hbar\omega$ shell-model calculations as far as we know.
In addition to the experimental {\rm C0} matrix elements, the IS monopole transition rates of $^{13}$C for the lowest three excited $1/2^-$ states have been reported with  inelastic $\alpha$ scattering on the target of $^{13}$C by Kawabata {\it et al}.~\cite{kawabata08}, and their experimental values are comparable to the single particle one~\cite{yamada08_monopole}. 
These experimental facts suggest that the two $1/2^{-}_{2,3}$ states have  cluster structure. 

\begin{figure}[h]
\begin{center}
\includegraphics[scale=0.36]{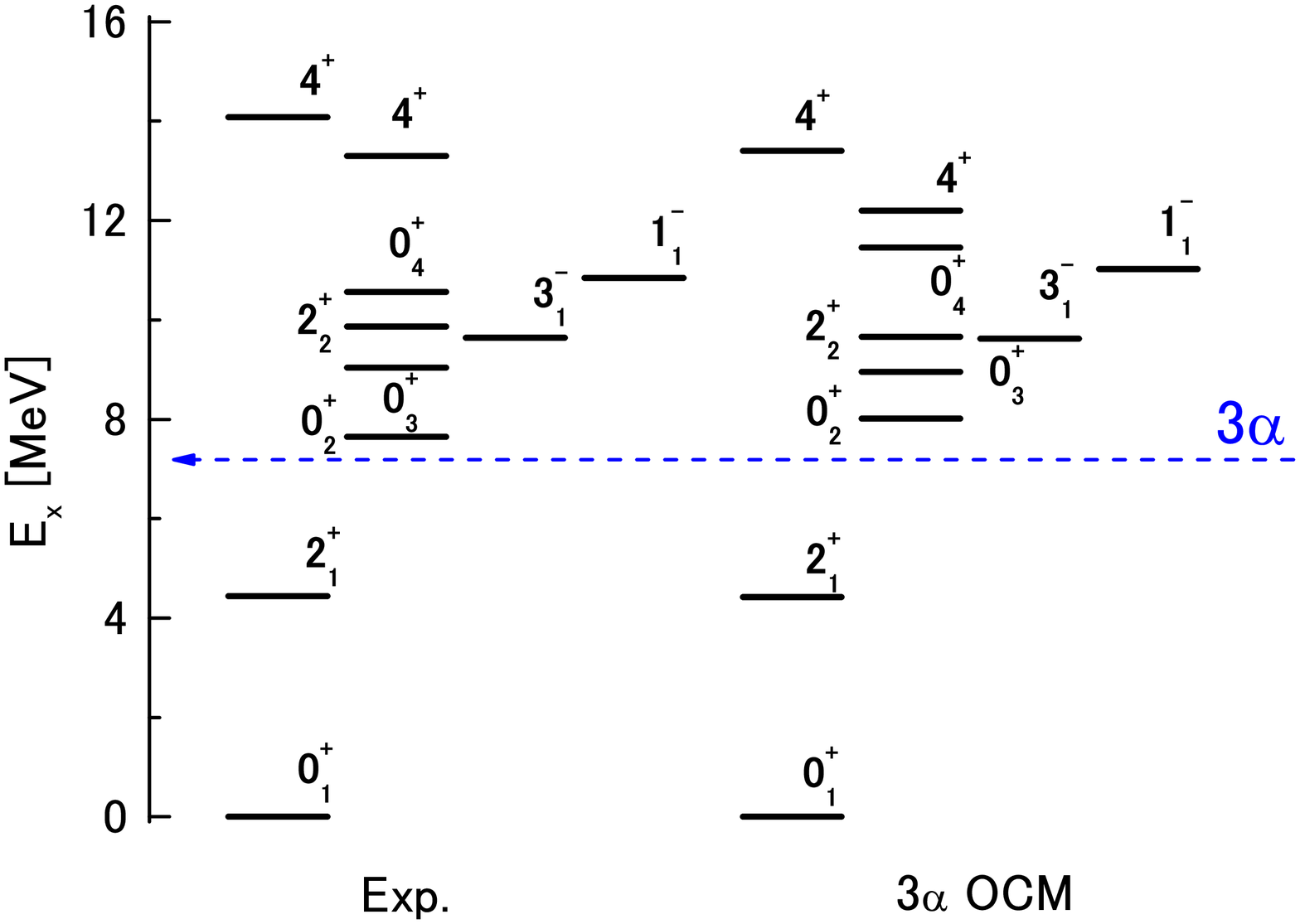}
\caption{Energy spectra of $^{12}$C obtained from the 3$\alpha$ OCM calculation~\cite{yamada15, kawabata08} with the 3-body force $V_{3\alpha}$, compared with the experimental data.}
\label{fig:3a_levels}
\end{center}
\end{figure}

\begin{figure}[h]
\begin{center}
\includegraphics[scale=0.46]{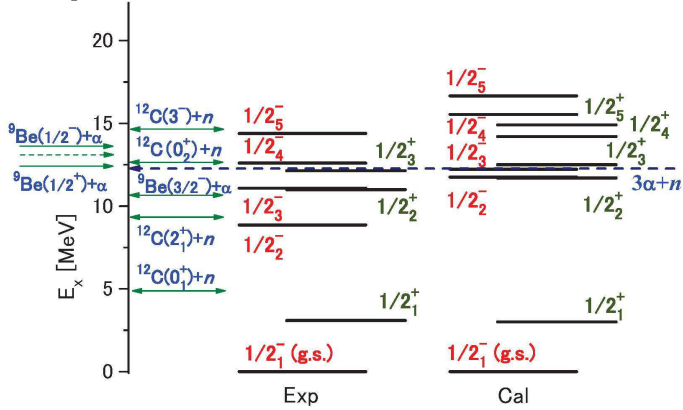}
\caption{Energy levels of the $1/2^-$ and $1/2^+$ states of $^{13}$C obtained from the 3$\alpha + n$ OCM calculation~\cite{yamada15}, compared with the experimental data. The threshold of the $^9$Be($5/2^-) + \alpha$ channel at $E_x = 13.1$ MeV, located between the $^9$Be($5/2^-) + \alpha$ and $^9$Be($1/2^-) + \alpha$ channels, is presented by the dashed arrow on the left hand side of the panel. }
\label{fig:3a+n_levels}
\end{center}
\end{figure}

\begin{table*}[htbp]
\begin{center}
\caption{Excitation energies ($E_{x}$), r.m.s.~radii ({$R$}), ${\rm C0}$ transition matrix elements [${\mathcal M}({\rm C0})$], isoscalar monopole transition matrix elements [${\mathcal M}({\rm IS})$] of the excited $1/2^-$ states in $^{13}$C obtained by the $3\alpha+n$ OCM calculation, in units of MeV, fm, fm$^2$, and fm$^2$, respectively. 
The experimental data are taken from Refs.~\cite{ajzenberg93,wittwer69} and from Ref.~\cite{kawabata08} for the $1/2^-_4$ state. }
\label{tab:12minus_ocm}
\begin{tabular}{cccccccccc}
\hline\hline
  & \multicolumn{4}{c}{Experiment} & \hspace*{10mm} & \multicolumn{4}{c}{$3\alpha+n$ OCM} \\
  & \hspace{1mm}{$E_{x}$}\hspace{1mm} & \hspace{1mm}{{$R$}}\hspace{1mm} & \hspace{1mm}${\mathcal M}({\rm C0}) $ \hspace{1mm} &  \hspace{1mm}${\mathcal M}({\rm IS})$\hspace{1mm} & & \hspace{1mm}{$E_{x}$}\hspace{1mm} & \hspace{1mm}{{$R$}}\hspace{1mm} & \hspace{1mm}{${\mathcal M}({\rm C0})$}\hspace{1mm} & \hspace{1mm}{${\mathcal M}({\rm IS})$}\hspace{1mm} \\
\hline
\hspace{1mm}$1/2_1^-$\hspace{1mm} &  $ \ 0.00 $  & $2.4628$ &  &  & & 0.0 &  $2.4$  &  & \\
$1/2_2^-$ & \ 8.86 & & $2.09\pm0.38$   & $6.1$ & & 11.7 & 3.0 & 4.4 & 9.8 \\
$1/2_3^-$ & 11.08  & & $2.62\pm0.26$   & $4.2$ & & 12.1 & 3.1 & 3.0 & 8.3 \\
$1/2_4^-$ & 12.5~  & & ${\rm No~data}$ & $4.9$ & & 15.5 & 4.0 & 1.0 & 2.0 \\
$1/2_5^-$ & 14.39\  & & ${\rm No~data}$ &  ${\rm No~data}$ & & 16.6 & 3.7 & 2.0 & 3.3 \\
\hline\hline
\end{tabular}
\end{center}
\end{table*}

The structure of the $1/2^{\pm}$ states of $^{13}$C up to around $E_x \sim 16$~MeV has been investigated with the full four-body $3\alpha$+$n$ OCM~\cite{yamada15}.
The $3\alpha$ OCM, the model space of which is the subspace of the $3\alpha+n$ model, describes well the structure of the low-lying states of $^{12}$C including the $2^{+}_2$, $0^{+}_3$, and $0^{+}_4$ states above the Hoyle state, as shown in Fig.~\ref{fig:3a_levels}.
According to the $3\alpha$ OCM calculations~\cite{kurokawa05,yamada15,Ohtsubo}, the $0^+_3$ state of $^{12}$C has a prominent $^8$Be($0^+$)+$\alpha$ structure with a higher nodal behavior, while the $0^+_4$ state is characterized by a linear-chain-like structure having the dominant configuration of ${^8}{\rm Be}(2^+)$+$\alpha$ with a relative $D$-wave motion. 
On the other hand, the low-lying states of $^9$Be, $^8$Be, and $^5$He are also described well by the $2\alpha$+$n$, $2\alpha$, and $\alpha$+$n$ OCM\rq{}s, respectively (see Ref.~\cite{yamada15}).
It should be recalled that their model spaces are also  subspaces of the $3\alpha+n$ OCM calculation.

Figure~\ref{fig:3a+n_levels} shows the calculated energy levels of the $1/2^{\pm}$ states with the $3\alpha+n$ OCM.
The five $1/2^{-}$ states and three $1/2^{+}$ states  observed up to $E_x \sim 16$~MeV are reproduced successfully.
The $1/2^{-}_1$ state, located at $E=-12.3$ MeV measured from the $3\alpha+n$ threshold, is the ground state of $^{13}$C, which has a shell-model-like structure.
Its calculated r.m.s.~radius is $R_N=2.4$ fm (see Table~\ref{tab:12minus_ocm}), the value of which agrees with the experimental data ($2.46$ fm).
The calculated Gamow-Teller transition rates $B({\rm GT})$ between the $^{13}$C ground state ($1/2^-_1$) and $^{13}$N states ($1/2^-_1,3/2^-_1$), together with the E1 transition rate $B(E1)$ between the ground state and first $1/2^+$ state of $^{13}$C are given as follows:~$B^{\rm cal}({\rm GT})=0.332$ vs.~$B^{\rm exp}({\rm GT})=0.207\pm0.002$ for the transition from the $^{13}$C($1/2^{-}_1$) state to $^{13}$N($1/2^{-}_1$), $B^{\rm cal}({\rm GT})=1.27$ vs.~$B^{\rm exp}({\rm GT})=1.37\pm0.07$ from  $^{13}$C($1/2^{-}_1$) to $^{13}$N($3/2^-_1$).
The calculated results are in agreement with the experimental data within a factor of $1.5$. 
On the other hand, the calculated value of $B({\rm E1}:{1/2^{-}_{1}} \rightarrow {1/2^+_1})$ is $2.0 \times 10^{-3}$~fm$^2$ in the present study, while the experimental value is $14 \times 10^{-3}$~fm$^2$.
This enhanced E1 transition rate has been pointed out by Millener et al.~\cite{millener89}, where the result of the shell model calculation is $B({\rm E1})=9.1 \times 10^{-3}$~fm$^2$, which is about two-third of the experimental value. This discrepancy between cluster and shell model can be solved in the future with a mixed cluster-shell-model approach, see also \cite{yamada15}.

The four excited $1/2^{-}$ states, $1/2^{-}_{2}$, $1/2^{-}_{3}$, $1/2^{-}_{4}$, and $1/2^{-}_{5}$, have larger nuclear radii ($3.0$, $3.1$, $4.0$ and $3.7$ fm, respectively) than that of the ground state (see Table~\ref{tab:12minus_ocm}).
It was  found that the $1/2^-_{2}$ and $1/2^-_{3}$ states have mainly $^9$Be($3/2^-$)+$\alpha$ and $^9$Be($1/2^-$)+$\alpha$ cluster structures, respectively. 
The $1/2^-_{4}$ and $1/2^-_{5}$ states are characterized by the dominant structures of $^9$Be($3/2^-$)+$\alpha$ and $^9$Be($1/2^-$)+$\alpha$ with higher nodal behaviors, respectively.

The present $3\alpha+n$ OCM calculations for the first time have provided reasonable agreement with the experimental data on the ${\rm C0}$ matrix elements ${\mathcal M}({\rm C0})$ of the $1/2^{-}_{2}$ ($E_x=8.86$~MeV) and $1/2^-_3$ ($E_x=11.08$~MeV) states obtained by the ${^{13}{\rm C}}(e,e')$ reaction~\cite{wittwer69}, and isoscalar monopole matrix elements ${\mathcal M}({\rm IS})$ of the $1/2^{-}_{2}$ ($E_x=8.86$~MeV), $1/2^{-}_{3}$ ($E_x=11.08$~MeV), and $1/2^-_4$ ($E_x=12.5$~MeV) states by the ${^{13}{\rm C}}(\alpha,\alpha')$ reaction~\cite{kawabata08}.
As mentioned above, they are very difficult to be reproduced in the shell model framework~\cite{millener89}.
The mechanism why the $^9$Be+$\alpha$ cluster states are populated by the isoscalar monopole transition and ${\rm C0}$ transition from the shell-model-like ground state is common to those in $^{16}$O, $^{12}$C, $^{11}$B, and $^{12}$Be etc., originates from the dual nature of the ground state~\cite{yamada08_monopole,yamada12,yamada10}:~The ground states in light nuclei have in general both the mean-field degree of freedom and cluster degree of freedom, the latter of which is activated by the monopole operator and then cluster states are excited from the ground state. 
The present results indicate that the $\alpha$ cluster picture is unavoidable to understand the structure of the low-lying states of $^{13}$C. The {C0} transitions together with the isoscalar monopole transitions are also useful to explore cluster states in light nuclei.

From the analyses of the spectroscopic factors and overlap amplitudes of the $^{9}$Be+$\alpha$ and $^{12}$C+$n$ channels in the $1/2^-$ states, dominant $^{12}$C(Hoyle)+$n$ states do not appear in the $1/2^-$ states in the present study.
This is mainly due to the effect of the enhanced $^9$Be+$\alpha$ correlation induced by the attractive odd-wave $\alpha$-$n$ force:~When an extra neutron is added into the Hoyle state, the attractive odd-wave $\alpha$-$n$ force reduces the size of the Hoyle state with the $3\alpha$ gas-like structure and then the $^9$Be+$\alpha$ correlation is significantly enhanced in the $3\alpha$+$n$ system.
Consequently the $^9$Be($3/2^-$)+$\alpha$ and $^9$Be($1/2^-$)+$\alpha$ states come out as the excited states, $1/2^-_{2}$ and $1/2^-_3$, respectively.
On the other hand, higher nodal states of the $1/2^-_{2,3}$ states, in which the $^9$Be-$\alpha$ relative wave function has one node higher than that of the $1/2^-_{2,3}$ states, emerge as the $1/2^-_{4}$ and $1/2^-_5$ states, respectively, in the present study.
It is recalled that the $0^+_3$ state of $^{12}$C has a $^8$Be+$\alpha$ structure with higher nodal behavior. 
Thus, the $^9$Be+$\alpha$ cluster states with higher nodal behavior,  $1/2^-_{4,5}$, are regarded as the counterpart of the $0^+_3$ state in $^{12}$C.

\begin{figure}[h]
\begin{center}
\includegraphics[scale=0.33]{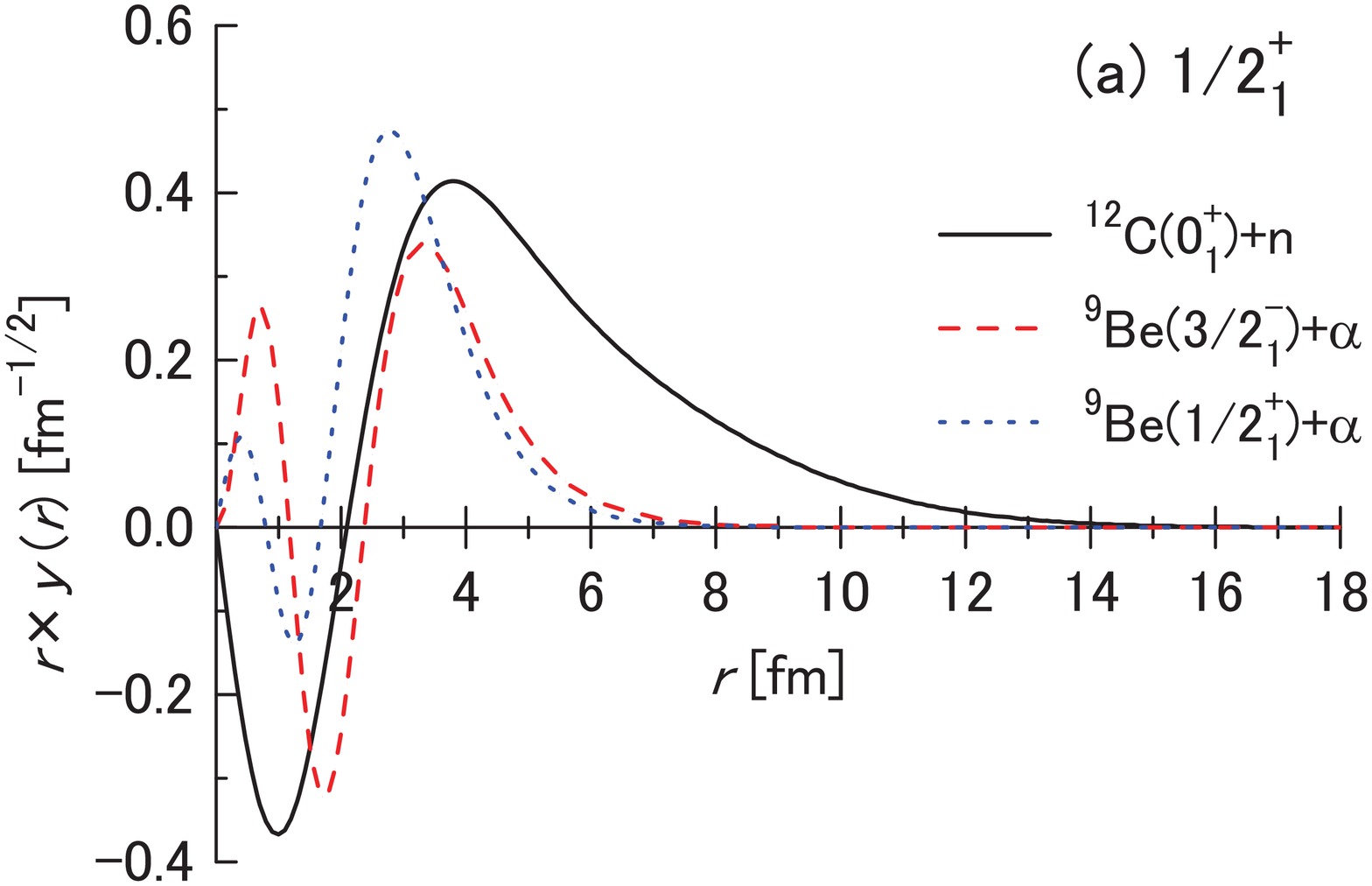}
\includegraphics[scale=0.33]{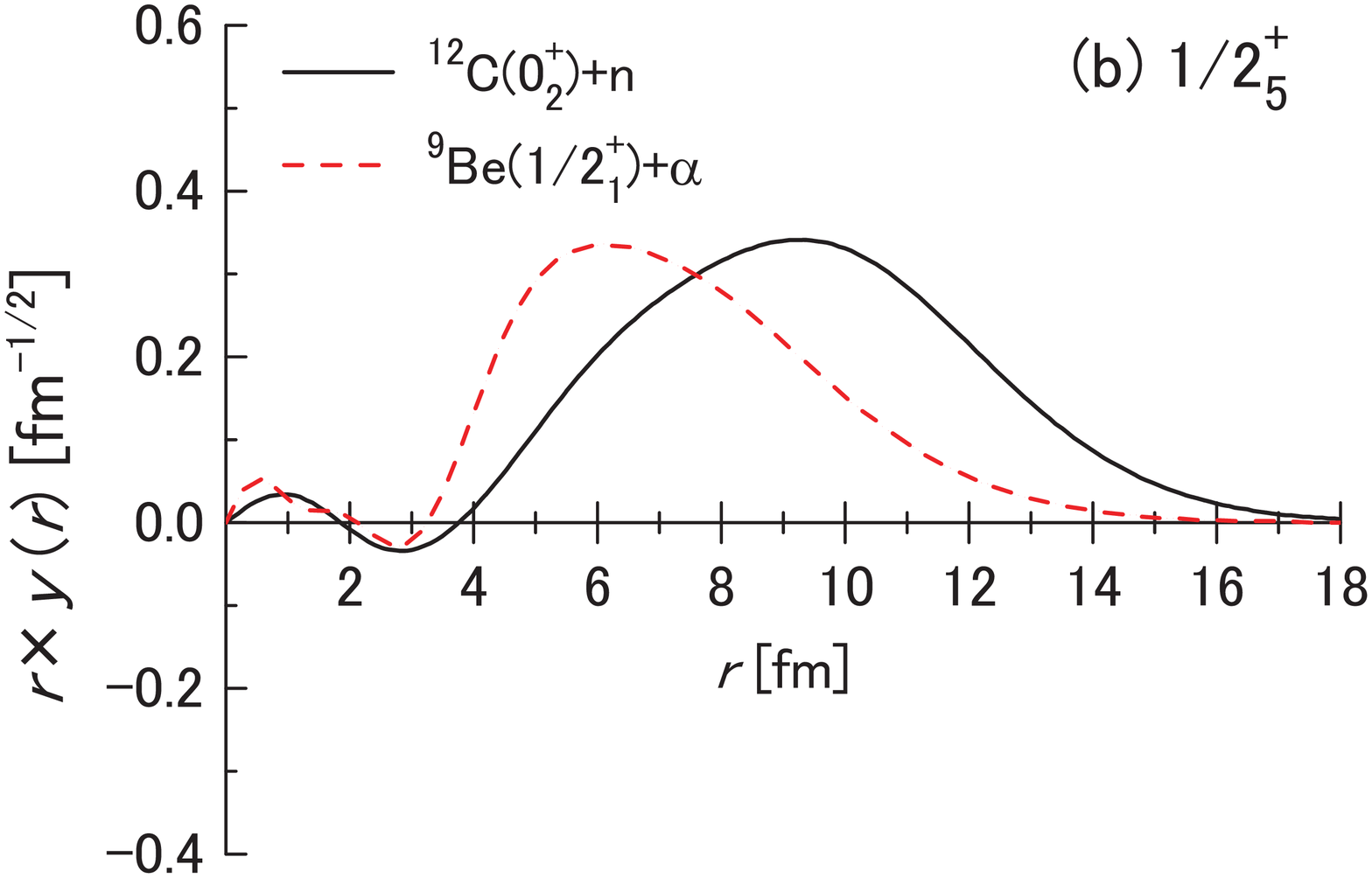}
\caption{Overlap amplitudes of the $^{12}$C+$n$ channels and $^{9}$Be+$\alpha$ channels for (a)~the $1/2^+_1$ state of $^{13}$C with a loosely bound neutron structure and (b)~the $1/2^+_5$ state with an $\alpha$-condensate-like structure ~\cite{yamada15}. In the panels we present only the overlap amplitudes with the $S^2$ factor larger than $0.2$. }
\label{fig:rwa_1st_12plus_3a-n_OCM}
\end{center}
\end{figure}

As for the $1/2^+$ states, the $1/2^+_1$ state appears as a bound state lying $1.9$~MeV below the $^{12}$C($0^+_1$)+$n$ threshold (see Fig.~\ref{fig:3a+n_levels}).
This state dominantly has a loosely bound neutron structure, in which the extra neutron moves around the $^{12}$C($0^+_1$) core in a $1S$ orbit, shown in Fig.~\ref{fig:rwa_1st_12plus_3a-n_OCM}(a).
The calculated radius of the $1/2^+_1$ state ($R=2.6$~fm), slightly larger than that of the ground state ($R=2.4$~fm), is consistent with the experimental suggestion~\cite{ogloblin11}.
It was found that the $1/2^+_2$ and $1/2^+_3$ states have mainly $^9$Be($3/2^-$)+$\alpha$ and $^9$Be($1/2^+$)+$\alpha$ structures, respectively, and their radii are around $R=3$~fm.
These two states are characterized by strong isoscalar monopole excitations from the $1/2^+_1$ state.
On the other hand, we found that the $1/2^+_4$ and $1/2^+_5$ states have dominantly a $^9$Be($3/2^-$)+$\alpha$ structure with higher nodal behavior and $3\alpha$+$n$ gas-like structure, respectively, although experimentally the two states have not been identified so far. 
The $1/2^+_5$ state with a larger radius ($R\sim 4$~fm) has the dominant configurations of $^{12}$C(Hoyle)+$n$ and $^9$Be($1/2^+$)+$\alpha$ as shown in Fig.~\ref{fig:rwa_1st_12plus_3a-n_OCM}(b).
According to the analyses of the single-cluster density matrix for $\alpha$ clusters with an extra neutron, this state is described by the product states of constituent clusters, having a configuration of $(0S)^3_\alpha(S)_n$, with the probability of $52~\%$.
Thus, the $1/2^+_5$ state can be regarded as an $\alpha$-condensate-like state with one extra loosely bound neutron.
The probability of $52~\%$ is comparable to that of the Hoyle state, $(0S)^3_\alpha$  ($70~\%$)~\cite{Yam05} and that of the $1/2^+_2$ state of $^{11}$B just above the $2\alpha+t$ threshold,  $(0S)^2_\alpha(0S)_t$ ($60~\%$)~\cite{yamada10}.

\section{Experimental evidences?}

\begin{figure*}
\begin{center}
\includegraphics[scale=0.4]{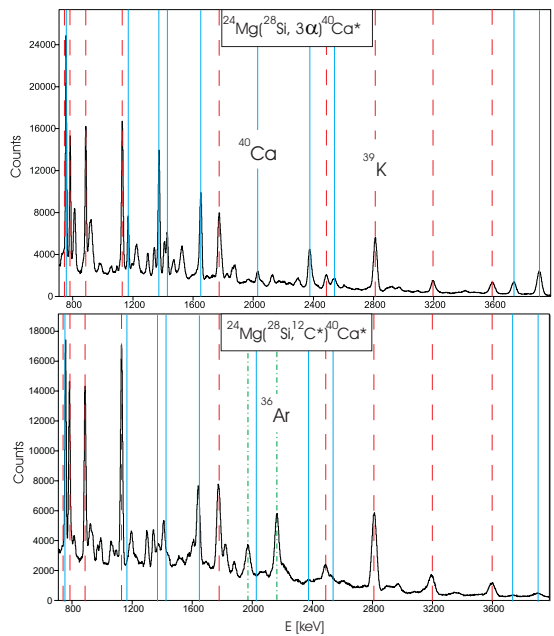}
\caption{ Coincident $\gamma$-spectra gated with the $\alpha$ particles hitting randomly three different detectors (upper panel) in comparison with the case where three $\alpha$'s hit same detector (lower panel). Note the additional lines for $^{36}$Ar in the lower panel.}
\label{36Ar}
\end{center}
\end{figure*}

Unfortunately, contrary to pairing, the experimental evidences for $\alpha$ condensation are  rare and, so far, only indirect. We, nevertheless, want to elaborate on this issue here, even though the situation is far from being satisfying. However, this may incite experimenters to perform more extensive and more accurate measurements.\\

The most prominent feature is the inelastic form factor which, as stated above, is very sensitive to the extension of the Hoyle state and shows that the Hoyle state has a volume 3-4 times larger than the one of the ground state of $^{12}$C. A state at low density is, of course, very favorable to $\alpha$ condensation as we have seen from the infinite matter study. Nevertheless, this does not establish a direct evidence for an $\alpha$ condensate. 
Other attempts to search for signatures of $\alpha$ condensate structures are heavy ion collisions around the Fermi energy where a condensate structure may be formed as intermediate state and correlations between the final $\alpha$ particles may reveal this structure.\\

For example von Oertzen {\it et al.} re-analized old data ~\cite{Ko05} of the $^{28}$Si $+ ^{24}$Mg $\rightarrow ^{52}$Fe $\rightarrow ^{40}$Ca $+ 3\alpha$ reaction at 130 MeV which, at that time, could not be explained with a Hauser-Feshbach approach for the supposedly statistical decay of the compound nucleus $^{52}$Fe. Analysing the spectrum of the decaying particles via $\gamma$-decay, obtained in combination with a multi-particle detector, it was found that the spectrum is dramatically different for events where the three $\alpha$'s are emitted randomly hitting various detectors under different angles from the ones where the three $\alpha$ were impinging on the same detector. This is shown in Fig.~\ref{36Ar} where the upper panel corresponds to the case of the 3$\alpha$'s in different detectors and lower panel, 3$\alpha$'s in same detector. A spectacular enhancement of the $^{36}$Ar line is seen in the lower panel. This is then explained  by a strong lowering of the emission barrier, due to the presence of an $\alpha$ gas state, for the emission of $^{12}$C($0^+_2$). This fact explains that the energies of the $^{12}$C($0^+_2$) are concentrated at much lower energies as compared to the summed energy of 3$\alpha$ particles under the same kinematical conditions ~\cite{oertzen-beck}. In this way, the residual nucleus ($^{40}$Ca) attains a much higher excitation energy which leads to a subsequent $\alpha$ decay and to a pile up of $^{36}$Ar in the $\gamma$ spectrum. One could also ask the question whether four $\alpha$'s have not been seen in the same detector. However, this only will happen at somewhat higher energies, an important experiment to be done in the future. \\
The interpretation of the experiment is, thus, the following, we cite v. Oertzen: {\it due to the coherent properties of the threshold states consisting of $\alpha$ particles with a large de-Broglie wave length, the decay of the compound nucleus $^{52}$Fe did not follow the Hauser-Feshbach assumption of the statistical model: a sequential decay and that all decay steps are statistically independent. On the contrary, after emission of the first $\alpha$ particle, the residual $\alpha$ particles in the nucleus contain the phase of the first emission process. The subsequent decays will follow with very short time delays related to the nuclear reaction times. Actually, a simultaneous decay can be considered. Very relevant for this scenario is, as mentioned, the large spacial extension  of the Bose condensate states, as discussed in} ~\cite{oertzen-beck}.\\

\begin{figure}
\begin{center}
\includegraphics[scale=0.33]{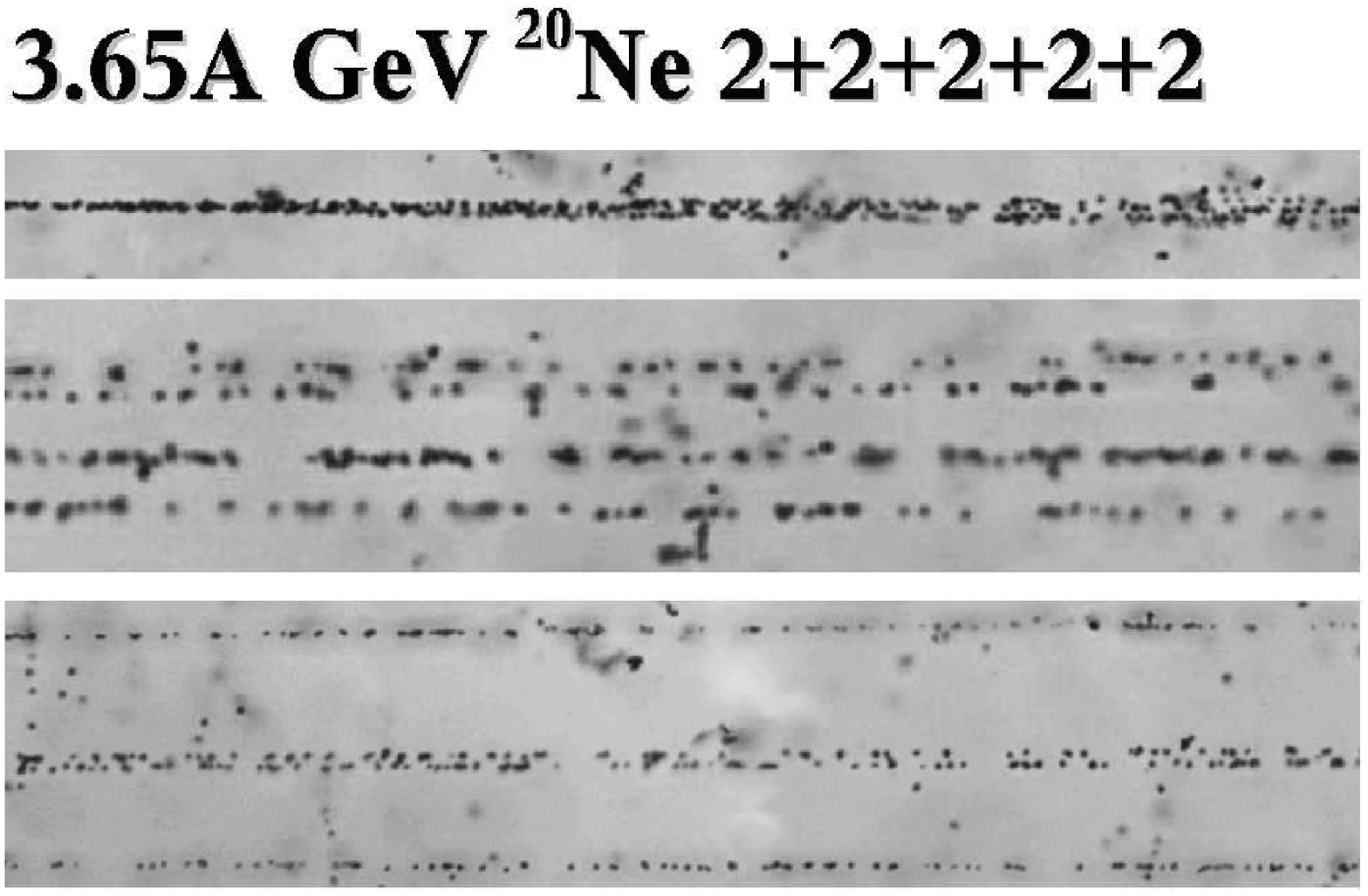}
\caption{Break-up of $^{20}$Ne at 3.65 GeV/nucleon with the emission of 5 $\alpha$'s (again partially as $^8$Be), registered in an emulsion. Different stages of the decay, registered down stream in the emulsion are shown in three panels on top of each other. P. Zarubin, private communication, see also \cite{Bradnova}.}
\label{Zarubin2}
\end{center}
\end{figure}

However, the above view of von Oertzen concerning a sign of $\alpha$ particle condensation may be too optimistic. The difference of the two events shown in Fig.\ref{36Ar} may stem from trivial energy conservation arguments and, thus, cannot be advanced as a firm evidence of Bose condensation. Probably the anomaly lies in the branching ratio of such events: if it is higher than predicted by Hauser Feshbach, this might be due to the reduced Coulomb barrier associated to the large extent of the Hoyle state (as stated by von Oertzen himself), which is probably the doorway state of the three $\alpha$'s. A similar deviation from Hauser-Feshbach in the multiple $\alpha$ channel has also been observed in a much more quantitative way in \cite{Gulmi} but no conclusion of $\alpha$ condensation is given there. A very intriguing paper appeared recently \cite{Bonasera} where the authors selected from the fragments of heavy ion reactions only the boson-like ones (deuterons, $\alpha$'s, etc.) versus only fermion like ones (p, n, tritons, $^3$He, etc.). It is found that the bosons, for instance the $\alpha$'s occupy a much denser, i.e., smaller space than the fermions. This would then be in analogy to what has been seen in Fermi-Bose mixtures of cold atoms \cite{atoms1, atoms2} and is, thus, interpreted as a sign of Bose condensation. It is, however, clear that this result has to be confirmed by other, independent experiments with also more refined analyses.

\begin{figure}
\begin{center}
\includegraphics[scale=0.33]{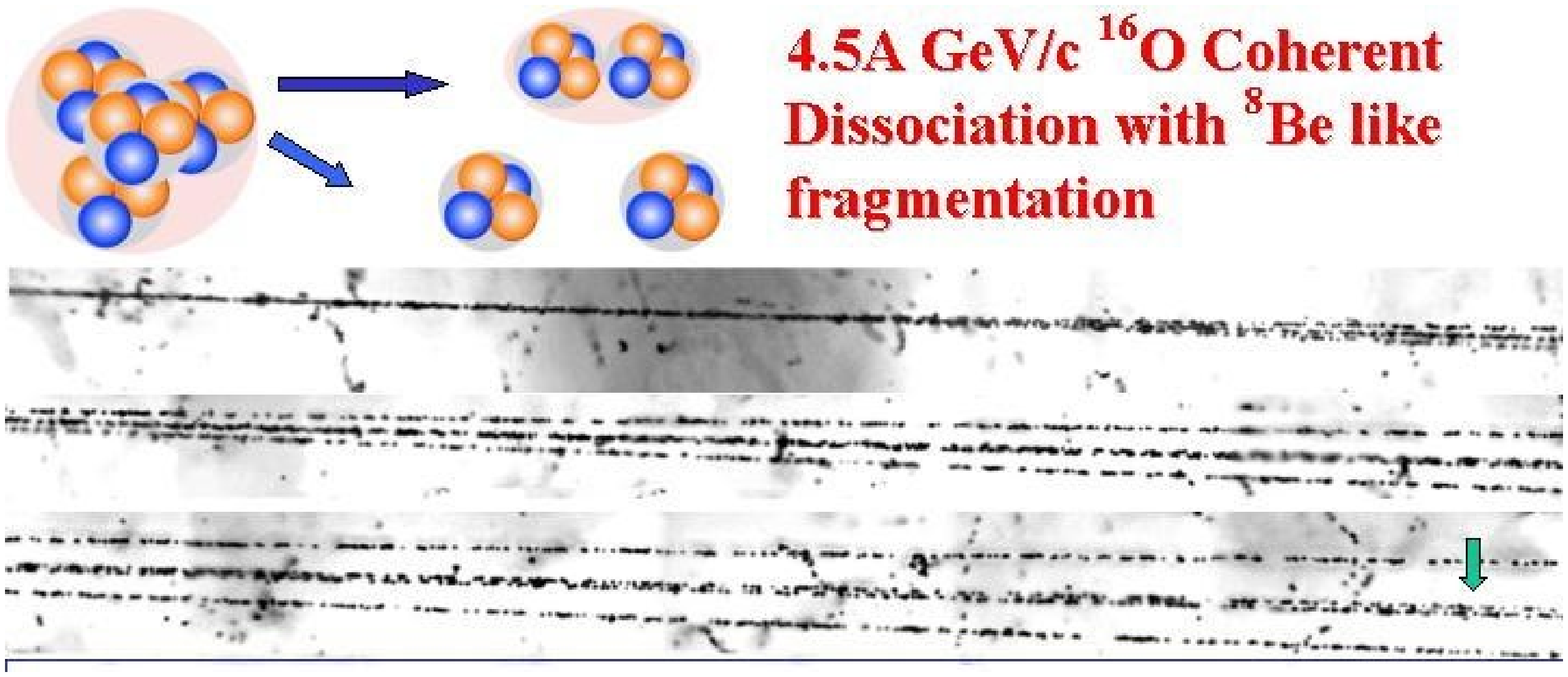}
\caption{Break-up of $^{16}$O at 4.5 GeV/nucleon with the emission of 4 $\alpha$'s, registered in an emulsion. Details of the decay can be seen, e.g., with the narrow cone of two $\alpha$'s, due to the emission of $^8$Be. Different stages of the decay, registered down stream in the emulsion are shown in consecutive panels. P. Zarubin, private communication, see also \cite{Bradnova}.}
\label{Zarubin1}
\end{center}
\end{figure}

Despite of all these uncertainties, it may become a rewarding research field to analyse heavy ion reactions more sytematically for non-statistical, coherent $\alpha$ decays.\\
A promising route may also be Coulomb excitation. In Fig.\ref{Zarubin2}, we show emulsion images of coherent $\alpha$ decay of $^{20}$Ne into three $\alpha$'s and one $^8$Be, or into 5 $\alpha$'s with remarkable intensity from relativistic Coulomb excitation at the Dubna Nucletron accelerator \cite{oertzen-beck}, see also \cite{Bradnova}. The Coulomb break-up being induced by  heavy target nuclei, Silver (Ag).
The break-up of $^{16}$O into 4$\alpha$'s, or into 2$\alpha$'s and one $^8$Be is shown in Fig.\ref{Zarubin1}. The presence of $^8$Be in the two reactions shows that the $\alpha$'s travel coherently, otherwise the $^8$Be-resonance could not be formed.

\begin{figure}
\begin{center}
\includegraphics[scale=0.5]{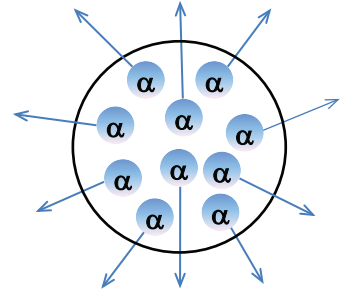}
\caption{Artist's view of a Coulomb explosion of $^{40}$Ca}
\label{40Ca-expl}
\end{center}
\end{figure}

A dream could be to Coulomb excite $^{40}$Ca to over 60 MeV and observe a slow coherent $\alpha$ particle Coulomb explosion, see Fig. \ref{40Ca-expl} for an artist's view.

Coulomb explosions have been observed in highly charged atomic van der Waals clusters, see ~\cite{explosion}. Coulomb excitation is insofar an ideal excitation mechanism as it transfers very little angular momentum and the projectile essentially gets into radial density expansion mode.

Next, we want to argue that the $^8$Be decay of the 6th $0^+$ state at 15.1 MeV in $^{16}$O can eventually show Bose enhancement, if the 15.1 MeV state is an $\alpha$ condensate. 

We know that a pick-up of a Cooper pair out of a superfluid nucleus is enhanced if the remaining nucleus is also superfluid \cite{Broglia}. For example $^{120}$Sn $\rightarrow$ $^{118}$Sn + Cooper pair. Of course same is true for pick up of 2 Cooper pairs simultaneously. We want to make an analogy between this and $^8$Be-decay of the 15.1 MeV state. In the decay probability of coincident two $^8$Be, the following spectroscopic factor should enter

\begin{equation}
S= \langle ^8\mbox{Be} + ^8\mbox{Be}|15.1 \mbox{MeV}\rangle 
\label{sqrt(6)}
\end{equation}

The reduced width amplitude $y$ is roughly related to the spectroscopic factor as  
 $y = 2^{-1/2} (4!/2!2!)^{1/2} S$.
Adopting the condensation approximation of $^8$Be
and 15.1 MeV states, this yields
\[   S=\langle B^2 B^2|(B^+)^4\rangle/(2!2!4!)^{1/2} = (4!/2!2!)^{1/2} = 6^{1/2}\]
entailing $y=6/(2^{1/2}) (y^2=18)$. In above expression for $S$, $B^+ (B)$ stands for an ideal boson creator (destructor), representing the $\alpha$ particle.

When we say that $S$ is large, we need to compare this $S$ with
some standard value.  So we consider the case that the 15.1 MeV state is
a molecular state of $^8$Be-$^8$Be.  We have
\[   S=\langle ^8\mbox{Be(I)} ^8\mbox{Be(II)}|^8\mbox{Be(I)}^8\mbox{Be(II)}\rangle=1 \]
and, therfore, $y=3^{1/2} (y^2=3)$.\\
This result shows that the condensation character of the 15.1 MeV state
gives an $^8$Be decay width which is 6 times larger than the molecular
resonance character.\\

We should be aware that above estimate is extremely crude and one rather should rely on a microscopic calculation of the reduced width amplitude $y$ what seems possible to do in the future.\\ 
One may also formulate above question somewhat differently:
For example, Ishikawa has recently calculated the probability of direct three $\alpha$ decay out of the Hoyle state \cite{Ishikawa}. It was also found that the Hoyle state is to 80\% an $\alpha$ Bose condensate \cite{Ishikawa}. 
Would the decay probability of the Hoyle (as well as of the 15.1 MeV state) be the same as if one neglected the bosonic aspect? In other words, is the decay probability (or tunnelling rate) out of Bose condensate influenced by the condensate, similar to what happens with pairing?
As just mentioned, only a realistic numerical investigation can give a final answer.

 Anyway, this example shows that the decay of the 15.1 MeV state into two $^8$Be's may be a very rewarding subject, experimentally as well as theoretically, in order to elucidate further its $\alpha$ cluster structure.\\

\section{Parity doublet rotational bands in $^{20}$Ne with the THSR wave function and its $\alpha +$$ ^{16}$O cluster structure.}

In Section 2., we have treated the $^8$Be nucleus and have seen that the THSR wave function yields a very different picture of the intrinsic deformed two $\alpha$ cluster state than the one from the traditional Brink approach. Though in both cases a clear two $\alpha$ cluster structure is seen, the Brink solution resembles the classical dumbbell picture of $^8$Be which was prevailing since the early days of cluster physics. On the other hand, the THSR approach yields a much more diffuse quantal image of $^8$Be. We should be aware of the fact that it is in the intrinsic state where the physics lies. On the other hand, we have seen in Sect.7 that Brink-GCM and THSR give practically the same {\it spherical} density after angular momentum projection, that is in the laboratory frame. However, the intrinsic state is a superposition of many eigenstates of angular momentum and projection means to filter out of this wave packet the component which has the angular momentum of interest, that is $J=0$ of the ground state in our case. So,  different wave functions of $^8$Be  may  contain  different superpositions of practically same eigenstates of angular momentum. \\

After this short recap of the situation in $^8$Be, we will now turn to other two cluster systems where the above considerations about $^8$Be may again be useful. In particular we want to study the $\alpha$ + $^{16}$O cluster structure of $^{20}$Ne in this section. It is, indeed, well known since long that despite of the fact that $^{20}$Ne can be described with the well known mean field approaches of, e.g., Skyrme or Gogny types, one nevertheless sees a relatively well pronounced $^{16}$O $+ \alpha$ structure. This may not be entirely surprising, since we have seen already in the Introduction that mean field theory can reveal clustering. The only question is whether mean field correctly describes the cluster features. That $^{20}$Ne shows a $^{16}$O $+ \alpha$ cluster structure even in the ground state may be due to the fact that both, $^{16}$O as well as $^4$He are very stiff doubly magic nuclei. So, the melting into one spherical Fermi gas state can be strongly hindered. The situation likely is similar for all doubly magic nuclei plus an $\alpha$ particle. We will later treat the situation of $^{212}$Po $= \alpha +$$^{208}$Pb but nuclei like $^{44}$Ti $= \alpha +$$^{40}$Ca or $^{100}$Sn $+ \alpha$ may show similar features.\\

Since the two clusters in $^{20}$Ne have different masses, the intrinsic cluster configuration has no good parity and one has to consider even and odd parity configurations for the $^{16}$O plus $\alpha$ system. Let us write down the THSR ansatz for $^{20}$Ne generalising in a rather obvious way the one for $^8$Be in (\ref{thsr-2})
\begin{equation}
\Psi^{\rm THSR}({^{20}\mbox{Ne}}) \propto {\mathcal A} e^{-\frac{8}{5B^2}r^2}\phi(^{16}\mbox{O})\phi_{\alpha}
\label{Ne}
\end{equation}
where ${\vek r} = {\vek R}_{16} - {\vek R}_4$ is the relative distance between the c.o.m. positions of $^{16}$O and the $\alpha$ particle and $\phi(^{16}\mbox{O})$ and $\phi_{\alpha}$ are the intrinsic translationally invariant mean field wave functions for $^{16}$O and $\alpha$. Usually, one takes, e.g., for $\phi(^{16}\mbox{O})$ a harmonic oscillator Slater determinant where the c.o.m. part has been eliminated for translational invariance. The $\alpha$ particle wave function $\phi_{\alpha}$ is the same as in (\ref{a-wf}).\\

\begin{figure}
\begin{center}
\includegraphics[scale=0.6]{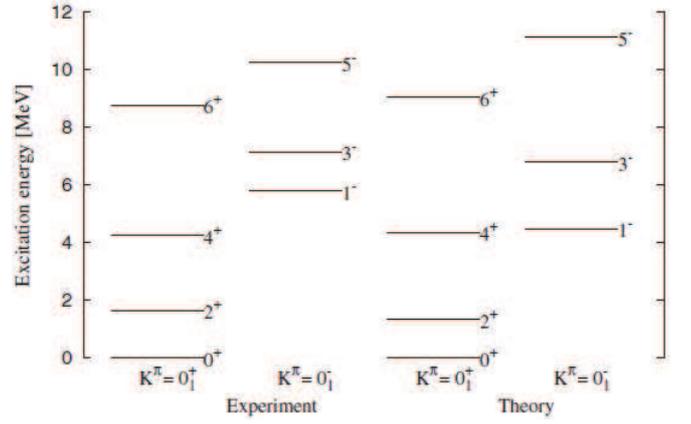}
\end{center}
\caption{Parity doublet spectrum of $^{20}$Ne.}
\label{fig:20Ne_levels}
\end{figure}

It is clear that the THSR function (\ref{Ne}) has positive parity as all other THSR wave functions treated so far. This also holds if one generalizes (\ref{Ne}) as in the $^8$Be case to the deformed intrinsic situation. How to generate a parity odd THSR wave function? The answer comes from applying the same trick as is used to demonstrate that a two particle state consisting of an antisymmetrised product of two Gaussians can lead to the correct P-wave harmonic oscillator wave function. For this one displaces the two Gaussians slightly from one another, normalises the wave function and takes the limit of the displacement going to zero. Translated to our situation here this leads to
\begin{equation}
\Psi^{\rm THSR}_{\mbox{hyb}}({^{20}\mbox{Ne}}) = N_{S_z} {\mathcal A} e^{-\frac{8}{5B^2}({\vek r} - S_z{\vek e}_z)^2}\phi(^{16}\mbox{O})\phi_{\alpha}
\label{Ne-hybrid}
\end{equation}
where $N_{S_z}$ is the normalisation constant and ${\vek e}_z$ the unit vector in $z$-direction. This ansatz means that the two clusters are displaced by the amount $S_z$. The cross term in the exponential containing the displacement vector can be expanded into partial waves. Projecting on even or odd parities, that is taking even or odd angular momenta and the limit $S_z \rightarrow 0$, one obtains THSR functions with good angular momenta and good $\pm$ parities (more details can be found in ~\cite{Bo}). Proceeding now exactly as in the case of $^8$Be, we obtain the following rotational spectrum of $^{20}$Ne for even and odd parity states as shown in Fig.~\ref{fig:20Ne_levels}. We see that the parity doublet spectrum is very nicely reproduced demonstrating again the flexibility and efficiency of the single THSR wave function.\\

\begin{figure}[h]
\begin{center}
\includegraphics[scale=0.45]{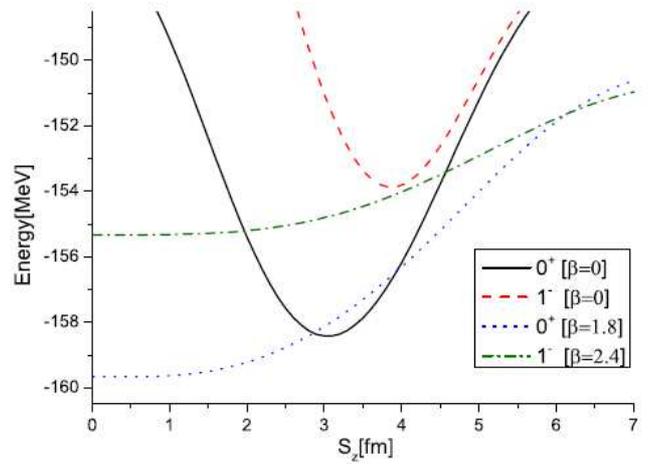}
\caption{Energy curves of $J^\pi=0^+,1^-$ states with different widths $B^2 = b^2 + 2\beta^2$ of Gaussian relative wave functions in the hybrid model.}
\label{fig:hybenene}
\end{center}
\end{figure}

Let us analyse the content of $\alpha$ clustering in $^{20}$Ne comparing again the Brink and THSR approaches. As a matter of fact, if in (\ref{Ne-hybrid}) one takes $B=b$, one obtains a single Brink wave function, see Eq. (\ref{thsr-2}) for the $^8$Be case. In Fig.~\ref{fig:hybenene}, we compare the energies for the $0^+$ and $1^-$ states obtained with this single Brink wave function as a function of $S_z$ with corresponding THSR wave functions but with optimised width parameters $B$. We see that in the latter case the minimum of energy is obtained for $S_z=0$ whereas with the Brink wave function the minimum is obtained with a finite value of $S_z$ lying higher in energy. We, thus, conclude that $B$ is a more efficient variational parameter than $S_z$. It is to be pointed out that for $S_z=0$ the wave function in (\ref{Ne-hybrid}) is just the THSR one. The one with $S_z \ne 0$ is called the hybrid Brink-THSR wave function, since it contains the wave functions of Brink and THSR as specific limits. \\

What about the cluster structure of the ground state of $^{20}$Ne ? To this end, we consider the following deformed intrinsic state
\begin{equation}
\Phi^{\mbox{hyb}}_{20} \propto {\mathcal A}\bigg [ \exp \bigg (-\frac{8(r_x^2 +r_y^2)}{5B_x^2} -\frac{8(r_z - S_z)^2}{5B_z^2} \bigg )\phi(^{16}\mbox{O})\phi_{\alpha} \bigg ]
\label{hyb-def}
\end{equation}

\begin{figure}[h]
\begin{center}
\includegraphics[scale=0.45]{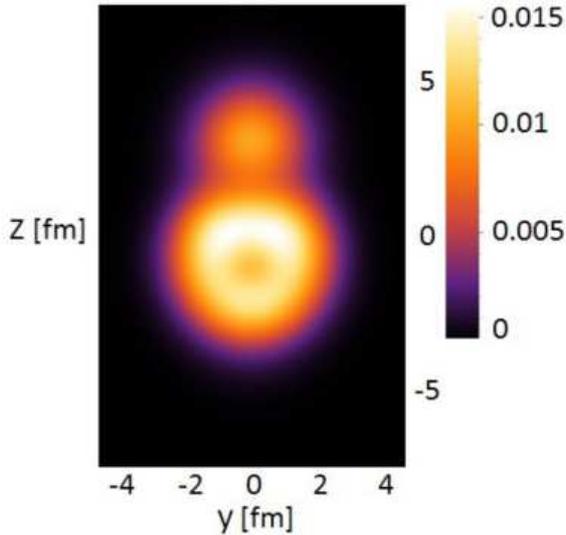}
\caption{Density distribution of the ${^{16}{\rm O}}+\alpha$ hybrid-Brink-THSR wave function with $S_z= 0.6$ fm and $(\beta_x,\beta_y,\beta_z)=(0.9, 0.9, 2.5\ {\rm fm})$.}
\label{fig:alodens}
\end{center}
\end{figure}

In Fig.~\ref{fig:alodens} we show the density distribution of $^{16}$O $+ \alpha$ corresponding to the wave function in (\ref{hyb-def}) with $S_z = 0.6$ fm and $(\beta_z, \beta_y, \beta_z) = (0.9, 0.9, 2.5$ fm) where $B^2_k = b^2 + 2\beta^2_k$. For numerical convenience a small but finite value of 0.6 fm for $S_z$ was taken which has to be compared to the large inter-cluster distance of about 3.6 fm. Clearly, this large inter-cluster distance cannot be attributed to the small value of $S_z$, rather the density distribution in Fig.~\ref{fig:alodens} reflects the amount of $\alpha$ clustering on top of the doubly magic nucleus, $^{16}$O contained in the THSR approach.

The situation of $^{20}$Ne has some similarity with the $^8$Be case but is nevertheless quite distinct, see Fig.~\ref{densities8Be}. Apparently the $^{16}$O nucleus attracts the $\alpha$ much stronger than this is the case in $^8$Be, so that $^{20}$Ne is quite compact in its ground state with more or less usual saturation density. One may also argue that $^{16}$O is much less stiff than an $\alpha$ in $^8$Be. The first excited state in $^4$He is at about 20 MeV whereas in $^{16}$O it is the $0^+_2$ state at 6.33 MeV.

One can consider this as the preformation of an $\alpha$ particle, a notion which is for instance used in spontaneous $\alpha$ decay in heavy nuclei. We will precisely discuss this issue in Sect. 20.\\

So far we have discussed two different binary cluster situations, one with two equal clusters ($^8$Be) and one with two un-equal ones ($^{20}$Ne). We may suspect that the features which have been revealed for these two cases essentially will repeat themselves in other binary cluster systems. This may, e.g., be the case with the molecular resonances found in $^{32}$S$=$$^{16}$O $+ ^{16}$O~\cite{Oh02,Ki04} or in other hetero-binary cluster systems.\\

At the end of this section, let us also mention that for light doubly open shell nuclei as, e.g., $^{20}$Ne, there is also recent progress using the symplectic NCSM and ab-initio shell model approaches, see \cite{Tobin, Stroberg, Jansen}.

\section{More on localised versus delocalised cluster states}

Several times in this review, we alluded already to the fact that the physical picture of cluster motion delivered by the THSR ansatz is very different from the Brink approach. Let us elaborate somewhat more on this aspect. As already mentioned, it was and, may be, still is the prevailing opinion of the cluster community that clusters in nuclear systems are localised in space. This opinion stems from Brink's ansatz for his cluster wave function ( see (\ref{Ne-hybrid}) for $B=b$) where, e.g., the $\alpha$ particle is explicitly placed at a definite position in space. Even though, later, the fixed position is smeared in the Brink-GCM approach, the picture of an essentially localised $\alpha$ remained. In the THSR wave function, there is a priori absolutely no spacial localisation visible. The parameter $B$ which enters the THSR wave function is a quantal width parameter which makes the c.o.m. distribution, centered around the origin, of an $\alpha$ particle wider or narrower. Nevertheless, one sees localisation of $\alpha$ particles as, e.g., in $^8$Be or in $^{20}$Ne as seen in Fig.\ref{fig:alodens}. This localisation can only come from the Pauli principle, i.e., the antisymmetriser ${\mathcal A}$ in the THSR wave function. So localisation, like seen in $^8$Be is entirely an effect of kinematics, that is the $\alpha$'s cannot be on top of one another because of Pauli repulsion. In an $\alpha$ chain state, and $^8$Be is the smallest chain state, the $\alpha$'s are, therefore, always quite well localised, see Fig. 7. In other spatial configurations of $\alpha$'s like in the Hoyle state, the freedom of motion of the $\alpha$ particles is much greater, in
spite of the fact that they mutually avoid each other. The emerging picture of an $\alpha$ gas state is then the following: the $\alpha$'s freely move as bosons in a large container (the mean field of the clusters) but avoid to come too close to one another due to the Pauli principle in spite of the fact that there is also some attraction between two $\alpha$'s at not so close distance. This picture is well born out in an $\alpha-\alpha$ correlation function study of the Hoyle state ~\cite{Suzuki}.  The situation is similar to the more phenomenological one of the excluded volume.\\ 


\section{$^{212}$Po seen as  $^{208}$Pb $+ \alpha$ .}

The formation of $\alpha$ particle-like clusters in nuclear systems and the description of its possible condensate properties is a challenge to present many-body theory. Whereas in the case of two-nucleon correlations efficient approaches to describe pairing are known, no first-principle formalism is available at present to describe quartetting in heavy nuclei. The THSR approach may be considered as an important step in this direction but is restricted to light nuclei. A more general approach should also describe $\alpha$-like clustering in arbitrary nuclear systems.

An interesting example where the formation of $\alpha$ particle-like clusters is relevant is radioactive decay by  $\alpha$ particle emission. 
The radioactive $\alpha$ decay occurs, in particular, near the doubly magic nuclei
$^{100}$Sn and $^{208}$Pb, and in superheavy nuclei. The standard approach to 
the decay width considers the transition probability for the $\alpha$ decay 
as product of the preformation probability
$P_{\alpha}$, a frequency factor, and an exponential factor.
Whereas the tunneling of
an $\alpha$ particle across the Coulomb barrier is well described
in quantum physics, the problem in understanding the $\alpha$
decay within a microscopic approach is the preformation $P_{\alpha}$ of the
$\alpha$ cluster in the decaying nucleus. 

In the case of four nucleons moving in a nuclear environment,
we obtain from a Green function approach \cite{Gerd4} an in-medium wave equation which reads in position space
\begin{eqnarray}
&&[E_4-\hat h_1 -\hat h_2- \hat h_3 - \hat h_4]\Psi_4({\bf r}_1 {\bf r}_2 {\bf r}_3{\bf r}_4)\nonumber \\ && =
\int d^3 {\bf r}_1'\,d^3 {\bf r}_2' \langle {\bf r}_1{\bf r}_2|B \,\,V_{N-N}| {\bf r}_1'{\bf r}_2'\rangle
\Psi_4({\bf r}_1'{\bf r}_2'{\bf r}_3{\bf r}_4)\nonumber \\ && +
\int d^3 {\bf r}_1'\,\,d^3{\bf r}_3'  \langle {\bf r}_1{\bf r}_3|B \,\,V_{N-N}|
{\bf r}_1'{\bf r}_3'\rangle \Psi_4({\bf r}_1'{\bf r}_2{\bf r}_3'{\bf r}_4)\nonumber \\ &&
+~~ \mbox{four further permutations.}
\label{15}
\end{eqnarray}
The single-particle Hamiltonian $\hat h_1$ contains the mean field that may be dependent on position, 
in contrast to our former considerations for homogeneous matter. The six  nucleon-nucleon interaction terms contain besides the nucleon-nucleon potential $V_{N-N}$ also the 
 blocking operator $B$ which can be given in quasi-particle state representation. 
For the first term on the r.h.s. of Eq. (\ref{15}), the expression
\begin{eqnarray}
 B(1,2)=[1-f_1(\hat h_1)-f_2(\hat h_2)]  
\label{15a}
\end{eqnarray}
results which is the typical blocking factor of the so-called particle-particle 
Random-Phase Approximation  (ppRPA) \cite{RS}.
The phase space occupation (we give the  internal quantum state $\nu={\sigma,\,\tau}$ explicitly)
\begin{equation}
\label{occ}
f_\nu(\hat h) =\sum_n^{\mbox{occ.}}| n,\nu \rangle \langle n,\nu |
\end{equation}
indicates the phase space which according to the Pauli principle is not available for an interaction process of a nucleon 
with internal quantum state $\nu$.

As worked out in the previous chapters, the formation of a well-defined 
$\alpha$-like bound state is possible only at low density of nuclear matter
because Pauli blocking suppresses the in medium four particle level density at the Fermi level and, thus,  the interaction strength necessary for 
the bound state formation. For homogeneous symmetric matter, $\alpha$-like bound states
can exist if the nucleon density is
comparable or below 1/5 of saturation density $\rho_0 = 0.16$
fm$^{-3}$. With the density dependence of Pauli blocking at zero temperature
given in \cite{Po}, a value for the Mott density $n_B^{\rm Mott}=0.03$ fm$^{-3}$
results for the critical density. For $n_B>n_B^{\rm Mott}$, the four nucleons 
which may form the $\alpha$-like particle are in continuum states which are 
approximated by independent single-nucleon quasiparticle states, as known from
shell model calculations.
Adopting a local-density approach, this argument confines the 
preformation of $\alpha$-like bound states to the tails of the
density distribution of the heavy nuclei.

As a typical example, $^{212}$Po has been considered in \cite{Po} which
decays into the doubly magic $^{208}$Pb core nucleus and an $\alpha$ particle,
the half life being 0.299 $\mu$s and $Q_\alpha = 8954.13$ keV. The proton
density as well as the total nucleon density have been measured \cite{Tarbert2014}. 
Outside of a critical radius $r_{\rm cluster}=7.44$ fm, the baryon density $n_B(r)$
falls below the critical value, $n_B(r)<n_B^{\rm Mott}$, so that $\alpha$ particle
preformation is possible \cite{XuPo}.

Another issue we discussed in the previous chapters is the treatment of the 
c.o.m. motion of the $\alpha$-like clusters, in contrast to the localized Brink states.
It is trivial that the state of the preformed $\alpha$ particle and its 
decay process demands the treatment of the c.o.m. motion like in the gas-like 
motion in the Hoyle state. However, this is not a simple task because only in 
homogeneous matter the c.o.m. motion can be separated from the intrinsic motion
describing the four nucleons inside the $\alpha$ particle. For inhomogeneous systems 
such as the case of $^{212}$Po, the intrinsic wave function of the $\alpha$-like clusters
depends on the nuclear matter density $n_B(r)$ and, consequently, on the position $r$,
the distance from the center of the core nucleus. In the previous chapters this
problem was not analyzed any further and, within a variational approach, a rigid 
internal structure of the $\alpha$ particle was assumed with fixed rms point radius 1.36 fm.

At short distances between the $\alpha$ particles, the antisymmetrization 
of the nucleon wave function within the THSR approach gives also 
the transition to single-nucleon shell model states. However, the treatment of heavy
nuclei like $^{212}$Po is not possible within THSR at present so that we use
a hybrid approach where the core nucleus $^{208}$Pb is described by an independent nucleon
approach (Thomas-Fermi or shell model) whereas the full antisymmetrization with the 
additional $\alpha$ particle is realized by the Pauli blocking terms. It is a challenge to future
research to formulate a full consistent approach where also the four-nucleon correlations in the 
core nucleus $^{208}$Pb are taken into account. As a step in this direction, we 
can consider the THSR treatment of $^{20}$Ne as a system where an $\alpha$ particle
is moving on top of the double magic $^{16}$O core \cite{Bo,Bo2}.

We shortly review the treatment of $^{212}$Po within a quartetting wave function approach \cite{Po}.
Similar to the case of pairing, we derive  
an effective $\alpha$ particle equation 
\cite{Po} for cases where an $\alpha$ particle is bound to the $^{208}$Pb.
Neglecting recoil effects, we assume that the core nucleus is fixed at
${\bf r} = 0$. The core nucleons are distributed with the baryon density $n_B(r)$ and produce
a mean field $V_\tau^{\rm mf}(r)$ acting on the two neutrons ($\tau = n$)
and two protons ($\tau = p$) moving on top of the lead core.
We give not a microscopic description
of the core nucleons (e.g., Thomas-Fermi or shell model calculations)
but consider both $n_B(r)$ and $V_\tau^{\rm mf}(r)$ as phenomenological inputs.
Of interest is the wave function of the four nucleons on top of the core nucleus
which can form an $\alpha$-like cluster.

The four-nucleon wave function (quartetting state) 
\begin{equation}
\Psi({\bf R},{\bf s}_j)=\varphi^{\mbox{intr}}({\bf s}_j,{\bf R})\,\Phi({\bf R}) 
\end{equation}
can be subdivided in a unique way in the (normalized) center of mass (c.o.m.)
part $\Phi({\bf R})$ depending only on the c.o.m. coordinate ${\bf R}$
and the intrinsic part $\varphi^{\mbox{intr}}({\bf s}_j,{\bf R})$
which depends, in addition, on the relative coordinates ${\bf
s}_j$ (for instance, Jacobi-Moshinsky coordinates)
\cite{Po}. The respective c.o.m. and intrinsic Schr\"odinger
equations are found from a variational principle, in particular the wave equation for the c.o.m. motion
\begin{eqnarray}
\label{9}
&&-\frac{\hbar^2}{2Am} \nabla_R^2\Phi({\bf R})-\frac{\hbar^2}{Am}\int ds_j \varphi^{\mbox{intr},*}({\bf s}_j,{\bf R})\nonumber 
\\ && 
[\nabla_R \varphi^{\mbox{intr}}({\bf s}_j,{\bf R})][\nabla_R\Phi({\bf R})]
\nonumber \nonumber  \\ &&
-\frac{\hbar^2}{2Am}\int ds_j \varphi^{\mbox{intr},*}({\bf s}_j,{\bf R}) 
[ \nabla_R^2 \varphi^{\mbox{intr}}({\bf s}_j,{\bf R})] \Phi({\bf R}) \nonumber \\ &&
+\int dR'\,W({\bf R},{\bf R}')  \Phi({\bf R}')=E\,\Phi({\bf R})\,
\end{eqnarray}
with the c.o.m. potential
\begin{eqnarray}
\label{9c}
 &&W({\bf R},{\bf R}')=\int ds_j\,ds'_j\,\varphi^{\mbox{intr},*}({\bf s}_j,{\bf R}) [T[\nabla_{s_j}] \nonumber \\
&&\delta({\bf R}-{\bf R}')\delta({\bf s}_j-{\bf s}'_j)+V({\bf R},{\bf s}_j;{\bf R}',{\bf s}'_j)]
\varphi^{\mbox{intr}}({\bf s}'_j,{\bf R}')\,. \nonumber \\
\end{eqnarray}
with $T$ the kinetic energy operator.\\
A similar wave equation is found for the intrinsic motion, see Ref. \cite{Po}.

The c.o.m. and intrinsic Schr\"odinger
equations are  coupled by contributions containing the expression
$\nabla_R \varphi^{\mbox{intr}}({\bf s}_j,{\bf R})$ which will be
neglected in the present discussion. In contrast to homogeneous matter
where this expression disappears, in finite nuclear systems such
as $^{212}$Po this gradient term will give a contribution to the
wave equations for $\Phi({\bf R})$ as well as for
$\varphi^{\mbox{intr}}({\bf s}_j,{\bf R})$. Up to now, there are
no investigations of such gradient terms.

The intrinsic wave equation describes in the zero density limit
the formation of an $\alpha$ particle with binding energy $B_\alpha= 28.3$
MeV. For homogeneous matter, the binding energy will be reduced
because of Pauli blocking. In the zero temperature case considered
here, the shift of the binding energy is determined by the baryon
density $n_B=n_n+n_p$, i.e. the sum of the neutron density $n_n$
and the proton density $n_p$. Furthermore, Pauli blocking depends on the asymmetry given by
the proton fraction $n_p/n_B$ and the c.o.m. momentum ${\bf P}$ of the
$\alpha$ particle. Neglecting the weak dependence on the
asymmetry, for ${\bf P}=0$ the density dependence of the Pauli
blocking term
\begin{equation}
\label{Paul}
W^{\rm Pauli}(n_B)=4515.9\, n_B -100935\, n_B^2+1202538\, n_B^3
\end{equation}
was found in \cite{Po}.
In particular, the bound state is dissolved and merges with the continuum
of scattering states at the Mott density $n_B^{\rm Mott}= 0.02917$ fm$^{-3}$. The
intrinsic wave function remains nearly  $\alpha$-particle like up
to the Mott density (a small change of the width parameter $b$ of
the four-nucleon bound state is shown in Fig. 2 of Ref.
\cite{Po}), but becomes a product of free nucleon wave functions
(more precisely the product of scattering states) above the Mott
density. This behavior of the intrinsic wave function will be used
below when the preformation probability for the  $\alpha$ particle
is calculated. Below the Mott density the intrinsic part of the
quartetting wave function has a large overlap with the intrinsic
wave function of the free $\alpha$ particle.
In the region where the $\alpha$-like cluster
penetrates the core nucleus, the intrinsic bound state wave
function transforms at the critical density $n_B^{\rm Mott}$ into an unbound
four-nucleon shell model state.

In the case of $^{212}$Po considered here, an $\alpha$ particle is
moving on top of the doubly magic $^{208}$Pb core. The tails of
the density distribution of the Pb core where the baryon density
is below the Mott density $n_B^{\rm Mott}$, is relevant for
the formation of  $\alpha$-like four-nucleon correlations. Simply
spoken, the $\alpha$ particle can exist only at the surface of the
heavy nucleus. This peculiarity has been considered since a long
time for the qualitative discussion of the preformation of
$\alpha$ particles in heavy nuclei \cite{Xu}.

Using the empirical results for the nucleon densities obtained recently
\cite{Tarbert2014} which are parametrized by Fermi functions,  
the Mott density $n_B^{\rm Mott}= 0.02917$ fm$^{-3}$ occurs
at $r_{\rm cluster} =7.4383$ fm, $n_B(r_{\rm cluster})=n_B^{\rm Mott}$.
This means that $\alpha$-like
clusters can exist only at distances $r > r_{\rm cluster}$, for
smaller values of $r$ the intrinsic wave function is characterized
by the nearly uncorrelated motion of the four nucleons.
Note that this transfer of results
obtained for homogeneous matter to finite nuclei is based on a
local density approach. In contrast to the weakly  bound
di-nucleon cluster, the $\alpha$ particles are more compact so
that a local-density approach seems to be better founded. However,
the Pauli blocking term is non-local. As shown in \cite{Po}, the
local density approach can be improved systematically. It is
expected that non-local  interaction terms and gradient terms will
make the sudden transition  at $r_{\rm cluster}$ from the
intrinsic $\alpha$-like cluster wave function to an uncorrelated four-nucleon
wave function more smooth but will not change the general picture.

Our main attention is focussed on the c.o.m. motion  $\Phi({\bf R})$
of the four-nucleon wave function (quartetting state of four nucleons
$n_\uparrow, n_\downarrow, p_\uparrow, p_\downarrow$). Because
the lead core nucleus is very heavy, we replace the c.o.m.
coordinate ${\bf R}$ by the distance $r$ from the center of the
$^{208}$Pb core. Neglecting the gradient terms, 
the corresponding Schr\"odinger equation (\ref{9}) contains
the kinetic part $-\hbar^2 \nabla_r^2/8 m$ as well as the
potential part $W({\bf r},{\bf r}')$ which, in general, is
non-local but can be approximated by an effective c.o.m. potential
$W(r)$. The effective c.o.m. potential 
\begin{equation}
 W(r)=W^{\rm intr}( r)+W^{\rm ext}(r)
\end{equation}
 consists of two
contributions, the intrinsic part $W^{\rm intr}( r)=E_\alpha^{(0)} + W^{\rm Pauli}(r)$ and the external
part $W^{\rm ext}(r)$ which is determined by the mean-field
interactions.


The intrinsic part $W^{\rm intr}(r)$ approaches for large
$r$ the bound state energy $E_\alpha^{(0)} =- B_\alpha= -28.3$ MeV of the $\alpha$ particle.
In addition, it contains the Pauli
blocking effects $W^{\rm Pauli}(r)$.
Since the distance from the center of the lead
core is now denoted by $r$, we have for $r >  r_{\rm cluster}$ the
shift of the binding energy of the $\alpha$-like cluster.
Here, the Pauli blocking part has the form ${ W}^{\rm Pauli}[n_B(r)]$ given above (\ref{Paul}).
For $r < r_{\rm cluster}$, the density of the core nucleus is larger than
the Mott density so that no bound state is formed. As lowest
energy state, the four nucleons of the quartetting state are added
at the edge of the continuum states which is given by the chemical potential.
In the case of the Thomas-Fermi model, not accounting
for an external potential, the chemical potential coincides with
the sum of the four constituting Fermi energies. For illustration,
the intrinsic part $W^{\rm intr}( {\bf r})$ in Thomas-Fermi
approximation, based on the empirical density distribution, is
shown in Fig. \ref{W-int}.
The repulsive contribution of the Pauli exclusion principle is
clearly seen.

\begin{figure}
\begin{center}
\includegraphics[scale=0.33]{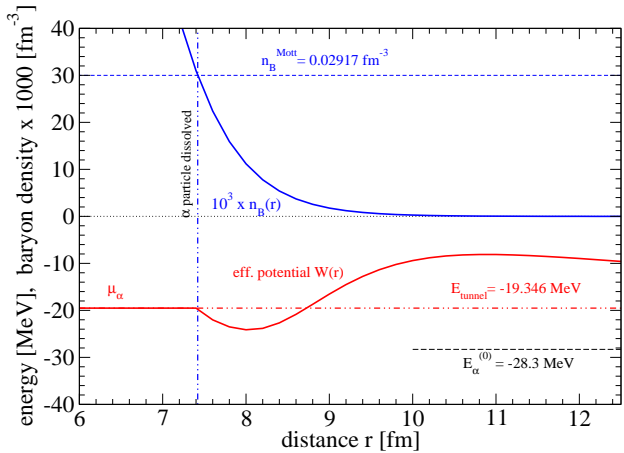}
\end{center}
\caption{Effective c.o.m. potential $W(r)$. The empirical baryon density distribution \cite{Tarbert2014} for the $^{208}$Pb core is also shown. The chemical potential $\mu_{\alpha}$ coincides with the binding energy $E_{\mbox{tunnel}}$.}
\label{W-eff}
\end{figure} 

\begin{figure}
\begin{center}
\includegraphics[scale=0.33]{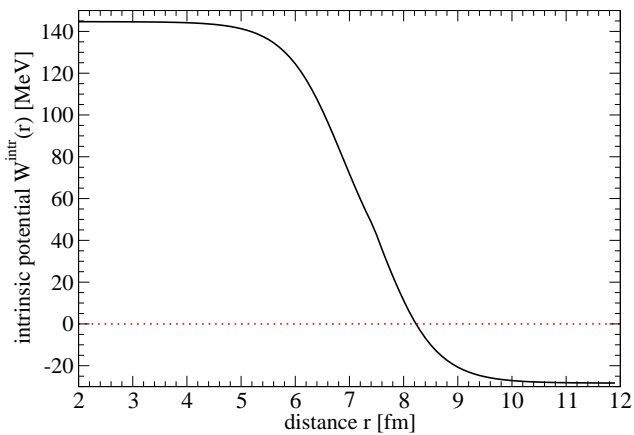}
\end{center}
\caption{
Intrinsic part $W^{\mbox{intr}}({\bf r})$ of the effective potential $W({\bf r})$. The empirical density distribution \cite{Tarbert2014} for the $^{208}$Pb core has been used. The four-nucleon Fermi energy for $r < r_{\mbox{cluster}}$ is taken in Thomas-Fermi approximation and $r_{\rm cluster} = 7.44$ fm.}
\label{W-int}
\end{figure}

The external part $W^{\rm ext}( {\bf r})$ is given by the mean
field of the surrounding matter acting on the four-nucleon system.
It includes the strong nucleon-nucleon interaction as well as the
Coulomb interaction. It is
given by a double-folding potential using the intrinsic
$\alpha$-like cluster wave function, see \cite{Po}. For $r >  r_{\rm cluster}$
the simple Woods-Saxon potential used in \cite{Po} can be improved 
\cite{XuPo} using the M3Y double-folding potential
\cite{M3YReview}. This M3Y potential contains in addition to the
Coulomb interaction the direct nucleon-nucleon interaction $V_N(
r)$ and the exchange terms $V_{\rm ex}( r)+V_{\rm Pauli}( r)$
\cite{M3YReview}.


The Coulomb interaction is calculated as a  double-folding potential
using the proton density $n_p(r )$ of the $^{208}$Pb core \cite{Tarbert2014}
and a Gaussian density distribution for the $\alpha$ cluster,
with the charge r.m.s. radius 1.67 fm. The direct
nucleon-nucleon interaction is
obtained by folding the measured nucleon density distribution of the
$^{208}$Pb core $n_B( r)$ \cite{Tarbert2014} and the Gaussian density distribution for the $\alpha$ cluster
(point r.m.s. radius 1.36 fm)  with a parameterized  nucleon-nucleon
effective interaction
\begin{equation}
\label{nnpot}
 v(s)=c \exp(-4s)/(4s)-d \exp(-2.5s)/(2.5s)
\end{equation}
describing a short-range repulsion and a long-range attraction,
$s$ denotes the nucleon-nucleon distance.

In principle, the nucleon mean field should
reproduce  the empirical densities of the $^{208}$Pb core. For $r
<  r_{\rm cluster}$  a local-density (Thomas-Fermi) approach will give a
constant chemical potential $\mu_{4}$ which is the sum of the
mean-field potential and the Fermi energy of the four nucleons,
\begin{equation}
\mu_{4}=W^{\rm ext}( {\bf r})+2E_{{\rm F},n}(n_n)+2E_{{\rm F},p}(n_p)
\end{equation}
with
\begin{equation}
 E_{{\rm F},\tau}(n_\tau)=(\hbar^2/2 m_\tau) (3 \pi^2 n_\tau)^{2/3}.
\end{equation}
The chemical potential $\mu_{4}$ is not depending on position. Additional four nucleons must be
introduced at the value $\mu_4$.
We consider this property as valid for any local-density approach, the continuum edge for adding
quasiparticles to the core nucleus is given by the chemical potential, not depending on position.
In a rigorous Thomas--Fermi approach for the core nucleus, this chemical potential coincides with the bound state
energy $E_{\rm tunnel}$ of the four-nucleon cluster, $E_{\rm tunnel}=\mu_4$.
For $r <  r_{\rm cluster}$, the effective c.o.m. potential $W(r)$ describes the edge of the
four-nucleon continuum where the nucleons can be introduced into the core nucleus.
Note that we withdraw this relation for shell model calculations where all states below the Fermi energy
are occupied, but the next states (we consider the states above the Fermi energy as "continuum states"
with respect to the intrinsic four-nucleon motion) are separated by a gap so that $E_{\rm tunnel}>\mu_4$.
We come back to this issue below.

The effective interaction $v(s)$ is designed according to this simple local-density approach, see Fig. \ref{W-eff}. 
The two parameter
values $ c=13866.30$ MeV and $d=4090.51$ MeV fm in Eq. (\ref{nnpot}) are determined by the conditions $\mu_4=E_{\rm tunnel}=-19.346$ MeV,
see Fig. \ref{W-eff}. The tunneling energy is identical with the energy at which the four nucleons
are added to the core nucleus. The total c.o.m. potential is continuous at $r=  r_{\rm cluster}$
and is constant for $r<  r_{\rm cluster}$, where the effective c.o.m. potential is
$W( { r}) =\mu_{4}$. We solve the effective Schr\"odinger equation for the c.o.m. potential $W( r)$
to find the the c.o.m. wave function $\Phi(r)$. In simplest approximation we assume that the intrinsic 
four-nucleon wave function coincides with the free $\alpha$-particle wave function for matter density below
the Mott density $n_B^{\rm Mott}$ but the overlap is zero for $n_B(r)>n_B^{\rm Mott}$ where the intrinsic wave function is a product of single-nucleon states,
\begin{eqnarray}
P_{\alpha}=\int_0^\infty  d^3r |\Phi(r)|^2 \Theta
\left[n_B^{\rm Mott}-n_B(r)\right]
\end{eqnarray}
with the step function $\Theta(x)=1$ for $x>0$ and = 0 else.
From the solution of the the effective Schr\"odinger equation follows the tunneling energy 
$E_{\rm tunnel}=-19.346$ MeV. The corresponding values for the preformation factor $P_\alpha=0.367$ 
and the decay half-life $2.91 \times 10^{-8}$ s are found \cite{XuPo}.

In a better approximation, the simple local-density (Thomas-Fermi) approach
for the $^{208}$Pb core nucleus has to be replaced by a shell model calculation. Then, the
single-particle states are occupied up to the Fermi energy, and additional nucleons
are introduced on higher energy levels according to the discrete
structure of the single energy level spectrum. The condition   $E_{\rm tunnel}=\mu_4$ is withdrawn.
If the nucleon-nucleon potential Eq. (\ref{nnpot}) is determined to reproduce not only the correct energy $-19.346$ MeV
of the $\alpha$ decay but also the decay half-life $2.99 \times 10^{-7}$ s,
the value $E_{\rm tunnel}-\mu_4=0.425$ MeV results \cite{XuPo} so that the additional four nucleons
forming the $\alpha$-like cluster are above the Fermi energy of the $^{208}$Pb core.

Clearly these calculations have to be improved with the aim of the THSR ansatz where all nucleons are treated 
on the same footings. Shell model
calculations are improved by including four-particle
($\alpha$-like or BCS-like) correlations that are of relevance when the matter
density becomes low. A closer relation of the calculations
 to the THSR calculations is of great interest, see
the calculations for $^{20}$Ne \cite{Bo,Bo2}. Related calculations
are performed in Ref. \cite{Horiuchi}. Note that the problem with the gradient terms 
in inhomogeneous nuclear systems can also be treated this way. The comparison with THSR
calculations would lead to a better understanding of the
microscopic calculations, in particular the c.o.m. potential, the
c.o.m. wave function, and the preformation factor.\\

\section{Outlook and Conclusions}

In this review, we tried to summarize where we presently stand with $\alpha$ particle clustering and $\alpha$ particle condensation in lighter nuclei. We mostly considered $N=Z$ nuclei but also studied successfully the case of $^{13}$C, that is 3 $\alpha$'s plus one neutron. We have seen that the $\alpha$ condensate in this even-odd nucleus is only born out around the threshold energy for 3 $\alpha$'s plus one neutron. At lower energies we identified cluster states with $^5$He bound states due to the strong neutron-$\alpha$ attraction. The study of analog states in $^{12}$C + proton and cluster states of $^{12}$C plus two nucleons may be an interesting subject for the future. First successful investigations of $^{14}$C have already been performed \cite{Lyu} employing a generalized THSR wave function. However, the more phenomenological OCM approach as applied here to $^{13}$C also turns out to be very useful for the description of clustering.\\
Concerning the $\alpha$ particle condensation aspect, we concluded that there are no serious objections which would invalidate this novel and exciting aspect of nuclear physics, in spite of the fact that direct proofs of condensation are difficult to obtain experimentally. Analysis of heavy ion collisions (HIC's) and relativistic Coulomb excitations may be promising fields of future investigations both theoretically as well as experimentally. An interesting study in this respect may be the $^8$Be decay of the 6-th $0^+$ state at 15.1 MeV in $^{16}$O where one eventually may see an enhancement if the 15.1 MeV state is an $\alpha$ condensate state as predicted from our studies. Extremely important future investigations concern the Hoyle excited states and and Hoyle analog states in $^{16}$O and heavier $N=Z$ nuclei. Already some experimental result show that the situation in $^{16}$O may have some similarity with the $^{12}$C case \cite{It14}. Confirmation of this fact would further give credit to the  $\alpha$ condensate and $\alpha$ gas hypothesis for states around the $\alpha$ disintegration threshold. The condensate states may be considered as the ground states on top of which $\alpha$ particles may be lifted into higher nodal states. Whether $\alpha$ gas states can be deformed or not will need further studies, experimentally and theoretically. For this, it may be useful to consider cranked THSR wave functions corresponding to a Hamiltonian of the form

\[ H_{\Omega} = H - \Omega L_X \]

\noindent
with $L_X = [{\bf R}\times {\bf P}]_X$ where ${\bf R}, {\bf P}$ are c.o.m. positions and corresponding momenta of the $\alpha$ particles. In this case the c.o.m. part of one $\alpha$ in the THSR wave function is of the following form

\begin{equation}
|\mbox{THSR}_{\rm rot}\rangle ={ \cal A}[e^{-\frac{R_x^2}{2b_x^2}} e^{-\frac{R_2^2}{2B_2^2}} 
e^{-\frac{R_3^2}{2B_3^2}} \phi_{\alpha_1}\phi_{\alpha_2}\phi_{\alpha_3}]
\end{equation}

\noindent
where $b_x=1/\sqrt{\hbar m \omega_x}; ~~~B_{2/3} = 1/\sqrt{\hbar m\Omega_{2/3}}$ with $\Omega_{2/3}^2 = \omega_r^2 + \omega_{+}^2\pm \sqrt{\omega_-^4 + 4\omega_r^2 
\omega_+^2}$ and $\omega_{\pm} = \frac{1}{2}(\omega_y^2 \pm \omega_z^2)$ as well as $R_2 = \alpha_2(R_y + \beta P_z); ~~~ R_3 = \alpha_3(R_z + \beta P_y)$.
Cranking a condensate state fast may align the $\alpha$'s into a chain state \cite{Bo14, Maruhn}. There have been speculations that up to 6 $\alpha$ chain states may exist \cite{Merchant}. One should be aware of the fact that  one dimensional Bose condensates do not exist and that, when the bosons are not interpenetrable (hard core bosons), a so-called Tonks-Girardeau boson gas forms where the bosons act like fermions because they cannot be at the same spacial point (as spinless fermions) \cite{Girardeau, Anna}. Since our $\alpha$ particles can practically not penetrate one another, it would be interesting to investigate how much linear $\alpha$ chain states resemble a Tonks-Girardeau Bose gas.\\

Other time dependent cluster processes could qualitatively be studied with time-dependent THSR wave functions where the width parameters $B(t)$ and $b(t)$ are considered time-dependent. For example from a time-dependent variational principle $\delta \langle \rm THSR(t)|i\partial/ \partial t - H|\rm THSR(t)\rangle =0$, one could let start a compressed $n\alpha$ nucleus and follow its expansion as a function of time, that is its clusterisation into $\alpha$ particles. This may give some qualitative insight into the dynamics of $\alpha$ particle clustering when the nuclei reach a low density phase during their expansion. It may numerically within reach to perform such a study with the THSR wave function.\\

The THSR approach also has shed a completely new light on the question of the spacial localisation of the alpha particles in states like $^8$Be or the Hoyle state. Since the much employed Brink and Brink-GCM wave function suggest a crystal arrangement of the $\alpha$'s, a dumbbell picture for $^8$Be or an equilateral triangle for the Hoyle state was in the past the common idea. However, the THSR wave function provides quite some other image of the situation. The $\alpha$ particles move freely in their common cluster mean field potential, except for mutual overlaps which are prevented by the Pauli principle. Therefore, in linear chain states still some localisation can be seen, even with the THSR wave function, see Fig.\ref{densities8Be}, this being a purely kinematical effect. However, in essentially spherical condensates, as, e.g., the Hoyle state, localisation is much suppressed and the $\alpha$'s most of their time (over 70\%) move as free bosons, that is they perform a delocalised motion. The situation is actually not far from the old phenomenological idea of the excluded volume. This can nicely be seen with the $\alpha-\alpha$ correlation function in the Hoyle state studied by Suzuki {\it et al.} in \cite{Suzuki}.\\
It is tempting to try to treat in the future nucleons as clusters of three quarks with a similar THSR ansatz

\begin{equation}
\Psi_{\rm THSR, Fermi} \propto {\mathcal A} \psi_1...\psi_n \equiv {\mathcal A}|F\rangle
\label{thsr-fermi}
\end{equation}
with $\psi_i \propto \exp[-2({\bf R}_i - {\bf X}_G)^2/B^2 \phi_{i,\rm nucleon}$ where $\phi_{i,\rm nucleon}$ is the intrinsic translationally invariant three quark wave function of the i-th nucleon. Of course, the nucleons form a Fermi gas of three quark clusters, opposite to the case of $\alpha$'s which are bosons. The antisymmetriser ${\mathcal A}$ should take care of the fermionic aspect of the quark clusters.  Of course, this fermion-THSR approach is very hypothetical and much will depend on whether an effective quark-quark interaction can be modeled which describes well the nucleon-nucleon scattering data. This in analogy to the effective nucleon-nucleon force which was invented by Tohsaki \cite{Tohsaki} to successfully describe $\alpha-\alpha$ scattering data.  For heavier nuclei where the THSR approach is inapplicable, one also could think of some fermionic OCM description which eases the solution of the many fermion systems. Gases of trimers in atomic traps \cite{Ottenstein} or triton and $^3$He gases may be other fields of applications of (\ref{thsr-fermi}).\\

We also discussed and showed that the THSR wave function not only is well adopted for the description of condensate states. It also is apt to describe $\alpha$-type of correlations in ground states. A paradigmatic case is $^{20}$Ne where two doubly magic nuclei quite unsuccessfully try to fuse completely. Indeed in Fig. 33 we show the ground state density distribution of $^{20}$Ne where still a pronounced $^{16}$O + $\alpha$ structure can be seen. As well known, this left-right asymmetric shape gives raise to the so-called parity doublet rotational spectrum. It is quite remarkable that a {\it single} THSR wave function can account for the experimental situation.\\
Another case of binding of two doubly magic nuclei is $^{212}$Po = $^{208}$Pb + $\alpha$. This case is much more difficult to treat than $^{20}$Ne because of its high mass. Indeed the THSR wave function which needs explicit antisymmetrisation could, so far, not handle nuclei beyond $^{20}$Ne because antisymmetrisation of heavy systems engenders very small differences of huge numbers, as is well known. However, in the case of $^{212}$Po, just because of its high mass, it actually shows also an advantage: the underlying core $^{208}$Pb can be considered as a fixed center, i.e., recoil effects, still essential in $^{20}$Ne, can be neglected here. Because of the compact size of the $\alpha$ particle, barely larger than the surface thickness of $^{208}$Pb, we then calculated the effective $^{208}$Pb + $\alpha$ potential with the Local Density Approximation (LDA). A very genuine effect, revealed in our study of $\alpha$ condensation in infinite matter, comes into play here. This concerns the fact that no BEC to BCS like continuous cross-over exists for four fermion clusters, see Sec.2. Therefore, when the $\alpha$ particle is approaching the $^{208}$Pb core from the outside, it first, at very low density, feels the strong attraction from the usual fermionic mean field. However, very soon, still at very low density of about a 5-th of saturation density  in the tail of the $^{208}$Pb density distribution the $\alpha$ particle dissolves into 2 neutron-2 proton shell model states on top of the $^{208}$Pb core. Thus, a potential pocket forms at the point where the $\alpha$ dissolves revealing the preformation of the $\alpha$ particle. This is contrary to what happens for example for a two fermion cluster on top of the $^{208}$Pb core. One could, for example, think of $^{210}$Bi
with a deuteron as the cluster. The effective deuteron-Pb potential should reveal a substantially different behavior from the $\alpha$-Pb case. Of course, the LDA has its limitations. However, since we have the quantal THSR solution for the analogue situation of $^{20}$Ne at hand, we can investigate the effective $\alpha$-Oxygen potential fully microscopically. Again, one may study the difference with the deuteron + Oxygen case of $^{18}$F and see in how much this case is different from the $\alpha$ cluster case. These very important and interesting studies remain for the future.\\

In Sec.12, we also shortly discussed promising progress of ab initio approaches (EFT plus lattice QMC and symplectic NCSM) to nuclear clustering.\\

All in all, we feel that nuclear cluster physics will still make tremendous progress in the future. It may be a fore-runner of other cluster systems, like they are already produced with trimers in cold atoms or as one suspects to exist with bi-excitons in semi-conductors. Other yet unexpected cluster problems may pop up in the future.


\section{Appendix}

Let us try to set up a BCS 
analogous procedure for quartets. Obviously we should write for the wave 
function

\begin{equation}
\label{quartet-gs}
|Z\rangle = e^{\frac{1}{4!}\sum_{k_1k_2k_3k_4}Z_{k_1k_2k_3k_4}
c^+_{k_1}c^+_{k_2}c^+_{k_3}c^+_{k_4}}|\mbox{vac}\rangle
\end{equation}

\noindent
where the quartet amplitudes $Z$ are fully antisymmetric (symmetric) 
with respect to 
an odd (even) permutation of the indices. 
The task will now be to find a killing operator for this quartet condensate 
state. Whereas in the pairing case the partitioning of the pair 
operator into a linear combination of a fermion creator and a fermion 
destructor is unambiguous, in the quartet case there exist two ways to 
partition the quartet operator, that is into a single plus a triple or into 
two doubles. Let us start with the superposition of a single and a triple. As 
a matter of fact it is easy to show that ( in the following, we always will assume that all amplitudes are real)

\begin{equation}
\label{q-killer}
q_{\nu} = u^{\nu}_{k_1}c_{k_1} -\frac{1}{3!}\sum v^{\nu}_{k_2k_3k_4}c^+_{k_1}c^+_{k_2}c^+_{k_3}
\end{equation}

\noindent
kills the quartet state under the condition

\begin{equation}
\label{Z=v/u}
Z_{k_1k_2k_3k_4} = \sum_{\nu}(u^{-1})^{\nu}_{k_1}v^{\nu}_{k_2k_3k_4}
\end{equation}

\noindent
However, so far, we barely have gained anything, since above quartet destructor
contains a non-linear fermion transformation which, a priory, cannot be 
handled. Therefore, let us try with a superposition of two fermion pair 
operators which is, in a way, the natural extension of the Bogoliubov 
transformation in the pairing case, i.e. with $Q = \sum [XP - YP^+]$ 
where $P^+ = 
c^+c^+$ is a fermion pair creator. We will, however,  find out that such an 
operator cannot kill the quartet state of Eq.~(\ref{quartet-gs}). 
In analogy to the so-called Self-Consistent RPA (SCRPA) approach 
\cite{Jemai}, we will 
introduce a slightly more general operator, that is

\begin{eqnarray}
\label{pair-bogo}
Q_{\nu} &=& \sum_{k<k'}[X^{\nu}_{kk'}c_kc_{k'} - Y^{\nu}_{kk'}c^+_{k'}c^+_k]\nonumber \\
&-&\sum_{k_1<k_2<k_3k_4}\eta^{\nu}_{k_1k_2k_3k_4}c^+_{k_1}c^+_{k_2}c^+_{k_3}c_{k_4}
\end{eqnarray}

\noindent
with $X,Y$ antisymmetric in $k,k'$.
Applying this operator on our quartet state, we find $Q_{\nu}|Z\rangle = 0$ 
where the relations between the various amplitudes turn out to be

\begin{equation}
\label{XYZ-relations}
\sum_{k<k'}X^{\nu}_{kk'}Z_{kk'll'} = Y^{\nu}_{ll'}~~~~~~\mbox{and}
~~~~~~~~\eta^{\nu}_{l_2l_3l_4;k'} = \sum_kX^{\nu}_{kk'}Z_{kl_2l_3l_4}
\end{equation}

\noindent
These relations are quite analogous to the ones which hold in the case of the 
SCRPA approach \cite{Jemai}. One also notices that the 
relation between $X, Y, Z$ amplitudes is similar in structure to the 
one of BCS theory for
pairing. As with SCRPA, in order to proceed, we have to approximate the 
additional
$\eta$-term. The quite suggestive recipe is to replace in the $\eta$-term of Eq.~(\ref{pair-bogo}) the density operator $c^+_{k'}c_k$ by 
its mean value $\langle Z|c^+_{k'}c_k|Z\rangle/\langle Z|Z\rangle \equiv 
\langle c^+_{k'}c_k\rangle  =
\delta_{kk'}n_k$, i.e.  $c^+_{k_1}c^+_{k_2}c^+_{k_3}c_{k_4} \rightarrow c^+_{k_1}c^+_{k_2}n_{k_3}\delta_{k_3k_4}$  where we supposed that we work in the basis where the single 
particle density matrix is diagonal, that is, it is given by the occupation 
probabilities $n_k$. This approximation, of course, violates the Pauli 
principle but, 
as it was found in applications of SCRPA \cite{Jemai}, we suppose that 
also here 
this violation will be quite mild (of the order of a couple of percent). With 
this approximation, we see that the $\eta$-term only renormalises the $Y$ 
amplitudes and, thus, the killing operator boils down to a linear super
position of a fermion pair destructor with a pair creator. This can then be 
seen as a Hartree-Fock-Bogoliubov (HFB) transformation of fermion pair 
operators, i.e., pairing of 'pairs'. Replacing the pair operators by 
ideal bosons as done in RPA, would 
lead to a standard bosonic HFB approach \cite{RS}, ch.9 and Appendix. 
Here, however, we will stay with the 
fermionic description and elaborate an HFB theory for fermion pairs. For this, 
we will suppose that we can use the killing property $Q_{\nu}|Z\rangle = 0$ 
even with the approximate $Q$-operator. As with our experience from SCRPA, 
we assume that this violation of consistency is weak.\\

Let us continue with elaborating our just defined frame. We will then use for the 
pair-killing operator 

\begin{equation}
\label{approxi-Q}
Q_{\nu} = \sum_{k<k'}[X^{\nu}_{kk'}c_kc_{k'} 
- Y^{\nu}_{kk'}c^+_{k'}c^+_k]/N^{1/2}_{kk'} 
\end{equation}

\noindent
with (the approximate) property $Q|Z\rangle=0$ and the first 
relation in (\ref{XYZ-relations}). The normalisation factor $N_{kk'}=
|1-n_k-n_{k'}|$ has been introduced so that $<[Q,Q^+]>= 
\frac{1}{2}\sum (X^2-Y^2) =1$, 
i.e., the quasi-pair state $Q^+|Z\rangle$ and the $X, Y$ amplitudes 
being normalised to one. We now 
will minimise the following energy weighted sum rule 

\begin{equation}
\label{pair-sum-rule}
\Omega_{\nu} = \frac{\langle Z| [Q_{\nu},[H - 2\mu \hat N,Q^+_{\nu}]]|Z\rangle}
{\langle Z| [Q_{\nu},Q^+_{\nu}]|Z\rangle}
\end{equation}

\noindent
The minimisation with respect to $X, Y$ amplitudes leads to

\begin{equation}
\label{f-pair-gap}
\left (
\begin{array}{cc}
{\mathcal {\bf H}}       &  {\bf \Delta}^{(22)}    \\
-{{\bf \Delta}^{(22)}}^+ & -{\mathcal {\bf H}}^*   
\end{array}
\right)
\left(
\begin{array}{c}
X^{\nu}  \\
Y^{\nu}  
\end{array}
\right)
=\Omega_{\nu}
\left(
\begin{array}{c}
X^{\nu}  \\
Y^{\nu}  
\end{array}
\right)
\end{equation}

\noindent
with (we eventually will consider a symmetrized double commutator in ${\bf H}$)

\begin{eqnarray}
\label{A-matrix}
 &&{\mathcal {\bf H}}_{k_1k_2,k'_1k'_2} \nonumber \\
&&= \langle [c_{k_2}c_{k_1},[H-2\mu \hat N,c^+_{k'_1}c^+_{k'_2}]]\rangle/(N^{1/2}_{k_1k_2}N^{1/2}_{k'_1k'_2})\nonumber\\
&& = (\xi_{k_1} + \xi_{k_2})\delta{k_1k_2,k'_1k'_2}\nonumber \\
&&  +N^{-1/2}_{k_1k_2}N^{-1/2}_{k'_1k'_2}  \{N_{k_1k_2}\bar v_{k_1k_2k'_1k'_2}N_{k'_1k'_2}\nonumber\\ 
&& +  [ (\frac{1}{2}\delta_{k_1k'_1}\bar v_{l_1k_2l_3l_4}C_{l_3l_4k'_2l_1}+\bar v_{l_1k_2l_4k'_2}C_{l_4k_1l_1k'_1} )\nonumber \\
&& -(k_1 \leftrightarrow k_2) ] - [k'_1 \leftrightarrow k'_2] \}
\end{eqnarray}

\noindent
where 

\begin{equation}
\label{corr-fct}
C_{k_1k_2k'_1k'_2}=\langle c^+_{k'_1}c^+_{k'_2}c_{k_2}c_{k_1}\rangle -n_{k_1}n_{k_2}[\delta_{k_1k'_1}\delta_{k_2k'_2}-\delta_{k_1k'_2}\delta_{k_2k'_1}]
\end{equation}

\noindent
is the two body correlation function and

\begin{eqnarray}
\label{B-matrix}
&&{\bf \Delta}^{(22)}_{k_1k_2,k'_1k'_2} \nonumber \\
&& = - \langle [c_{k_2}c_{k_1},[H-2\mu \hat N ,c_{k'_1}c_{k'_2}]]\rangle/(N^{1/2}_{k_1k_2}N^{1/2}_{k'_1k'_2})\nonumber\\
&&= N^{-1/2}_{k_1k_2}[( \Delta_{k_1k'_2;k'_1k_2} - k_1 \leftrightarrow k_2 ) - (k'_1 \leftrightarrow k'_2) ]N^{-1/2}_{k'_1k'_2} \nonumber \\
\end{eqnarray}
with
\begin{equation}
\label{a-gap}
 \Delta_{k_1k'_2;k'_1k_2}=
\sum_{l<l'}\bar v_{k_1k'_2ll'}\langle c_{k'_1}c_{k_2}c_{l'}c_l\rangle 
\end{equation}
In (\ref{f-pair-gap}) the matrix multiplication is to be understood  
as $\sum_{k'_1<k'_2}$ 
for restricted summation (or as 
$\frac{1}{2}\sum_{k'_1k'_2}$ for unrestricted summation ) . We see 
from (\ref{B-matrix}) and (\ref{a-gap}) that the bosonic gap 
${\bf \Delta}^{(22)}$ involves the quartet order parameter quite in 
analogy to the 
usual gap field in the BCS case. The ${\bf H}$ operator in (\ref{f-pair-gap}) 
has already been discussed in \cite{NPA628} in connection with SCRPA in 
the particle-particle channel. Equation (\ref{f-pair-gap}) has the typical 
structure of a bosonic HFB equation but, here, for fermion pairs, instead of 
bosons. It remains the task to close those HFB equations in expressing all 
expectation values involved in the ${\bf H}$ and ${\bf \Delta}^{(22)}$ fields 
by the 
$X, Y$ amplitudes. This goes in the following way. Because of the HFB 
structure of (\ref{f-pair-gap}), the $X, Y$ amplitudes obey the usual 
orthonormality relations, see \cite{RS}. Therefore, one can invert 
relation (\ref{approxi-Q}) 
to obtain

\begin{equation}
\label{inversion}
c^+_{k'}c^+_k = N^{1/2}_{kk'}\sum_{\nu}[X^{\nu}_{kk'}Q^+_{\nu} + Y^{\nu}Q_{\nu}]~~~~~~~~~~~(k<k')
\end{equation}

\noindent
and by conjugation the expression for $cc$.
With this relation, we can calculate all two body correlation functions 
in (\ref{B-matrix}) and (\ref{A-matrix}) in terms of $X, Y$ amplitudes. 
This is achieved in commuting the 
destruction operators $Q$ to the right hand side and use the killing property. 
For example, the quartet order parameter in the gap-field (\ref{a-gap}) is 
obtained as $\langle c_{k'_1}c_{k_2}c_{l'}c_l\rangle = 
N^{1/2}_{k'_1k_2}\sum_{\nu}X^{\nu}_{k_2k'_1}Y^{\nu}_{ll'}N^{1/2}_{ll'}$. 
Remains the task how to link the occupation numbers 
$n_k = \langle c^+_kc_k\rangle$ to the $X, Y$ amplitudes. Of course, that is 
where our partitioning of the quartet operator into singles and triples comes 
into play. Therefore, let us try to work with the operator (\ref{q-killer}). 
First, as a side-remark, let us notice that if in (\ref{q-killer}) we replace 
$c^+_{k_1}c^+_{k_2}$ by its expectation value which is the pairing tensor, we are 
back to the standard Bogoliubov transformation for pairing. Here we want to 
consider quartetting and, thus, we have to keep the triple operator fully. 
Minimising, as in (\ref{pair-sum-rule}) an average single particle energy, we 
arrive at the following equation for the amplitudes $u, v$ in (\ref{q-killer})

\begin{equation}
\label{sp-eq}
\left(
\begin{array}{cc}
\xi & {\bf \Delta}^{(13)}  \\
{{\bf \Delta}^{(13)}}^+ & -{\mathcal N}{\mathcal H}^*
\end{array}
\right)
\left(
\begin{array}{c}
u \\
v 
\end{array}
\right)
=E
\left(
\begin{array}{cc}
 1 &  0             \\
 0 & {\mathcal N}
\end{array}
\right)
\left(
\begin{array}{c}
u \\ 
v 
\end{array}
\right)
\end{equation}

\noindent
with (we disregard pairing, i.e., $\langle cc\rangle$ amplitudes)

\begin{equation}
\label{13-gap}
{\bf \Delta}^{(13)}_{k;k_1k_2k_3}= \Delta_{kk_3;k_2k_1} -
[(k_2 \leftrightarrow k_3) - (k_1 \leftrightarrow k_2)]
\end{equation}
and
\begin{eqnarray}
&&\hspace{-0.5cm} ({\mathcal N}{\mathcal H}^*)_{k_1k_2k_3;k'_1k'_2k'_3}=\langle \{ c^+_{k_3}c^+_{k_2}c^+_{k_1},[H-3\mu\hat N,
c_{k'_1}c_{k'_2}c_{k'_3}]\}\rangle \label{3body-H} \nonumber \\
&& \\
&&\hspace{-0.5cm} {\mathcal N}_{k_1k_2k_3;k'_1k'_2k'_3}=\langle \{ c^+_{k_3}c^+_{k_2}c^+_{k_1},
c_{k'_1}c_{k'_2}c_{k'_3}\}\rangle \label{3body-N} \nonumber \\
\end{eqnarray}
with $\{..,..\}$ an anticommutator.
 We will not give ${\mathcal H}$ in full because it is a very complicated 
expression involving self-consistent determination of three-body densities. 
To lowest order in the interaction it is given by
\begin{eqnarray}
\label{H33}
&&{\mathcal H}_{k_1k_2k_3;k'_1k'_2k'_3}= (\xi_{k_1} +\xi_{k_2}+\xi_{k_3})\delta_{k_1k_2k_3,k'_1k'_2k'_3} \nonumber \\ 
&& + [(1 - n_{k_1}-n_{k_2})\bar v_{k_1k_2k'_1k'_2}\delta_{k_3k'_3} + ~~\mbox{permutations}] \nonumber \\
\end{eqnarray}

\noindent
where $\delta_{k_1k_2k_3,k'_1k'_2k'_3}$ is the fully antisymmetrised three fermion Kronecker symbol.
Even this operator is still rather complicated for numerical applications and 
mostly one will replace the correlated occupation numbers by their free Fermi-
Dirac steps, i.e., $n_k \rightarrow n^{0}_k$. To this order the three body 
norm in (\ref{3body-N}) is given by
\begin{eqnarray}
&&\hspace{-0.3cm}{\mathcal N}_{k_1k_2k_3;k'_1k'_2k'_3} \simeq [\bar n^0_{k_1}\bar n^0_{k_2}\bar n^0_{k_3}
+n^0_{k_1} n^0_{k_2} n^0_{k_3}]\delta_{k_1k_2k_3,k'_1k'_2k'_3} \nonumber \\
\end{eqnarray}
with $\bar n^0 = 1-n^0$.
In principle this effective 
three-body Hamiltonian leads to three-body bound and scattering states. In our 
application to nuclear matter given below, we will make an even more drastic 
approximation and completely neglect the interaction term in the three-body 
Hamiltonian. Eliminating under this condition the $v$-amplitudes from 
(\ref{sp-eq}), one can write down the following effective single particle 
equation
\begin{eqnarray}
\label{eff-sp-field}
&& \xi_ku_k^{(\nu)} + \nonumber \\
&&\hspace{-0.6cm} \sum_{k_1<k_2<k_3k'}\hspace{-0.2cm}\frac{{\bf \Delta}^{(13)}_{k,k_1k_2k_3}(\bar n^{0}_{k_1}\bar n^{0}_{k_2}\bar n^{0}_{k_3} + n^{0}_{k_1}n^{0}_{k_2}n^{0}_{k_3}){\bf \Delta}^{{(13)}^*}_{k_3k_2k_1k'}}{E_{\nu} + \xi_{k_1}+\xi_{k_2}+\xi_{k_3}}u_{k'}^{(\nu)} \nonumber \\
&&= E_{\nu}u_k^{(\nu)}
\end{eqnarray}
The occupation numbers are given by
\begin{equation}
\label{occ-numbers}
n_k =1-\sum_{\nu}|u^{(\nu)}_k|^2
\end{equation}


\noindent
The effective single particle field in (\ref{eff-sp-field}) is grapphically interpreted in Fig.~\ref{fig:Sogomassoperator}, lower panel.
The gap-fields in (\ref{eff-sp-field}) are then to be calculated as 
in (\ref{13-gap}) and (\ref{a-gap}) 
with (\ref{inversion}) and the system of equations is fully closed. 
This is quite in parallel to the pairing case. 
In cases, 
where the quartet consists out of four different fermions and in addition is 
rather strongly bound, as this will be the case for the $\alpha$ particle in 
nuclear physics, one still can make a very good but drastic simplification: 
one writes the quartic order parameter as a translationally invariant product 
of four times the same single particle wave function in momentum space. We 
have seen, how this goes  when we apply our theory to $\alpha$ particle 
condensation in nuclear matter (Sect.2). Comparing the effective single particle field 
in (\ref{eff-sp-field}) with the one of standard pairing ~\cite{FW}, we 
find strong analogies but also several 
structural differences. The most striking is that in the quartet case Pauli 
factors figure in the numerator of (\ref{eff-sp-field}) whereas this is not the case for pairing. 
In principle in the pairing case, they are also there, but since 
$\bar n_k +n_k = 1$, they drop out. This difference has quite dramatic 
consequences between the pairing and the quartetting case. Namely when the 
chemical potential $\mu$ changes from negative (binding) to positive, the 
implicit three hole level density

\begin{equation}
\label{eq:3h-level}
g_{3h}(\omega)=\sum_{k_1k_2k_3}(\bar n^0_{k_1}\bar n^0_{k_2}\bar n^0_{k_3} + n^0_{k_1}n^0_{k_2}n^0_{k_3})\delta(\omega+\xi_{k_1}+\xi_{k_2}+\xi_{k_3})
\end{equation}

\noindent
passes through zero at $\omega-3\mu=0$ because phase space constraints and 
energy conservation cannot be fullfilled simultaneously at that point. 



\begin{ack}
Our collaboration on $\alpha$ clustering and $\alpha$ condensation has profited over the years from discussions with many protagonists in the field. Let us cite W. von Oertzen, M. Freer, M. Itoh, T. Kawabata, Y. Kanada-En'yo, M. Girod,  B. Borderie, E. Khan, J.-P. Ebran, and many more. 
More recently our collaboration was joined by Z. Ren, Chang Xu, M. Lyu, B. Zhou who contributed with important works. We are very grateful.

\end{ack}


\end{document}